\documentclass[]{aa}  

\usepackage{graphicx}
\usepackage{natbib} 
\bibpunct{(}{)}{;}{a}{}{,} 

\usepackage{txfonts}
\usepackage{amssymb}
\def\gapprox{\;\rlap{\lower 2.5pt            
 \hbox{$\sim$}}\raise 1.5pt\hbox{$>$}\;}
\def\lapprox{\;\rlap{\lower 2.5pt            
 \hbox{$\sim$}}\raise 1.5pt\hbox{$<$}\;}

\begin{document} 

\title{Resolution and calibration effects in high contrast polarimetric imaging of circumstellar scattering regions}
    \author{
      H.M.~Schmid\inst{1},
      \and
      J. Ma\inst{2}
}

\institute{
  ETH Zurich, Institute for Particle Physics and Astrophysics,
  Wolfgang-Pauli-Strasse 27, CH-8093 Zurich, Switzerland\label{instch1}
  \\
  \email{schmid@astro.phys.ethz.ch}
  \and
  Univ. Grenoble Alpes, CNRS, IPAG, F-38000 Grenoble, France\label{instf1}\\
}
           \date{Received ...; accepted ...}

\abstract
{Many circumstellar dust scattering regions
have been detected and investigated with polarimetric
imaging.
However, the quantitative determination of the intrinsic
polarization and of dust properties is 
difficult because of complex observational effects.}
{This work investigates instrumental convolution and polarimetric
calibration effect for high contrast imaging polarimetry with the aim
to define measuring parameters and calibration procedures
for accurate measurements of the circumstellar polarization.}
{We simulate
the instrumental convolution and polarimetric cancellation effects
for two axisymmetric point spread functions (PSF),
a Gaussian PSF$_{\rm G}$ and an extended PSF$_{\rm AO}$ typical for a modern
adaptive optics system, both with the same
diameter {$D_{\rm PSF}$} for the PSF peak. Further, polarimetric
zero-point corrections (zp-corrections) are simulated for different cases
like coronagraphic observations or systems with barely resolved
circumstellar scattering regions.}
{The PSF convolution reduces the integrated azimuthal polarization
$\Sigma Q_\phi$ for the scattering region while the net Stokes
signals $\Sigma Q$ and $\Sigma U$ are not changed. For non-axisymmetric
systems a spurious $U_\phi$-signal is introduced. These effects are
strong for compact systems and
for the convolution with an extended PSF$_{\rm AO}$. 
Compact scattering regions can be detected down to an inner working
angle of $r\approx D_{\rm PSF}$ based on the presence of a net
$\Sigma Q_\phi$ signal.
Unresolved central scattering regions can introduce
a central Stokes $Q,\, U$ signal which can be used to constrain
the scattering geometry even at separations $r< D_{\rm PSF}$.
The smearing by the halo of the PSF$_{\rm AO}$ produces an extended, low
surface brightness polarization signal.
These effects change the angular distribution of the azimuthal polarization
$Q_\phi(\phi)$, but the initial $Q'_\phi(\phi)$-signal can be
recovered partly with the analysis of measured Stokes $Q$ and $U$
quadrant pattern.
A polarimetric zp-correction applied for the removal of offsets
from instrumental or interstellar polarization
depends on the selected reference region and can introduce
strong bias effects for $\Sigma Q$ and $\Sigma U$
and the azimuthal distribution of $Q_\phi(\phi)$.
Strategies for the zp-correction are described for different data types,
like coronagraphic data or observations of partly unresolved systems.
These procedures provide polarization parameter which can
be reproduced easily with model simulations.}
{The simulations describe observational effects for imaging polarimetry
in a systematic way, and they show when these effects are
strong, and how they can be considered in the analysis. This defines
suitable measuring parameters and procedures for the quantitative
characterization of the intrinsic
polarization of circumstellar scattering regions.}

\authorrunning{H.M. Schmid \& J. Ma}

\titlerunning{Resolution and calibration effects in polarimetric imaging}

\maketitle

\section{Introduction} 
\label{SectIntro}

Dust in circumstellar disks around young stars or dust in 
shells around red giants has been recognized in many systems
as infrared excess in the spectral energy distribution. This dust
plays an important role in the formation of stars and planetary
systems \citep{Birnstiel24},
or the mass loss of evolved stars
and the dust enrichement of the interstellar medium \citep{Ferrarotti06}.

In the last decade a lot of new information on circumstellar dust
has been obtained with imaging observations of scattered
stellar light in the near-IR and visual wavelength range.
High-contrast instruments at large ground based telescopes
achieve an angular resolution of up to 20~mas and this provides
important information about the geometric distribution and the
scattering properties of the circumstellar dust. In particular, 
the new information from scattered light observations is complementary
to the extensive data already available 
from IR observations of the thermal emission of the dust around
protoplanetary disks \citep[e.g.,][]{Woitke19},  
debris disk \citep[e.g.,][]{Hughes18,Chen14},  
and red giants \citep[e.g.,][]{Hoefner18}.

The detection of the scattered intensity from circumstellar dust 
is difficult because of the strong glare of the central star.
This requires a careful separation of the stellar signal from
the circumstellar signal, which can be achieved with
high contrast techniques using space telescopes like HST
\citep[e.g.,][]{Kalas13,Schneider14,Zhou23} or JWST
\citep[e.g.,][]{Gaspar23,Lawson23}
or with ground based telescopes using adaptive optics (AO) systems
\citep[e.g.,][]{Milli17,Ren23}.

High resolution imaging polarimetry is a very powerful method
to separate the circumstellar polarization signal introduced by dust scattering
from the direct light of the bright central source, 
which has usually no or only a small net linear polarization
\citep[e.g.,][]{Kuhn01,Quanz11}.   
Large progress has been achieved with new instruments for imaging
polarimetry in combination with AO
systems at large ground based telescopes, including Subaru
CIAO, HiCIAO and SCeXAO instruments
\citep{Murakawa04,Hodapp08,Lucas24}, Gemini GPI    
\citep{Perrin15}, and VLT NACO and SPHERE \citep{Lenzen03,Beuzit19}.
Polarimetric imaging of
many objects has been obtained, including protoplanetary disks
\citep[e.g.,][and references therein]{Avenhaus18,Benisty23},
debris disks
\citep[e.g.,][]{Esposito20,Crotts24},
circumstellar shells around red giants 
\citep[e.g.,][]{Ohnaka16,Khouri20,Montarges23},
and post-AGB stars \citep[e.g.,][]{Andrych23}.

High resolution imaging polarimetry is a quite new and not so simple
observing technique because of PSF smearing and polarization
cancellation effects, and noise in the
weak differential signal \citep{Schmid06,Schmid22}. In addition
the polarimetric calibration of the complex high contrast imaging systems
has to be considered
\citep[e.g.][]{Tinbergen07,Schmid18,deBoer20,vanHolstein20}
including the temporal PSF variations for ground based observations using
AO systems \citep{Tschudi21,Ma23}. Depite this,
many impressive results
on the dust scattering geometry has been obtained,
but for most objects no detailed analysis of
the polarized intensity was attempted, mostly because this was not
the focus of the study. For some special cases the polarized flux is derived
more accurately to constrain the properties of the scattering dust, for
example for extended circumstellar disks
\citep{Stolker16,Monnier19,Milli19,Arriaga20,Hunziker21},
axisymmetric systmes \citep{Pinilla18,Tschudi21,Ma23}, edge-on debris disks
\citep[e.g.,][]{Graham07,Engler17}.
It is highly desirable that such detailed studies on the
circumstellar polarization are applied to a larger
sample and are further improved for some key objects.

Accurate polarimetric measurements for circumstellar
scattering regions require a good understanding of
the observational aspects. Convolution effects
degrade the measurable polarization signal as 
described for some cases in \citet{Schmid06}, \citet{Avenhaus17},
\citet{Heikamp19}, \citet{Tschudi21}, and \citet{Ma23,Ma24b}.
Additional issues are polarization offsets introduced by
instrumental effects, interstellar polarization,
or an intrinsically polarized central star. These offsets
can be eliminated with a polarimetric normalization, which sets the
polarization for a certain region to zero
\citep{Quanz11,Avenhaus14}.
We call this procedure hereafter polarimetric zero-point
correction (zp-correction).

A very useful and frequently used way to described the circumstellar
signal is the azimuthal polarization $Q_\phi$ \citep{Schmid06,Quanz13}.
A convolution or an offset by a zp-correction introduces
for $Q_\phi$ quite complex changes and some of these effects have not been
described in detail or have not been investigated at all. 
For example, it is clear that polarimetric offsets, or a zp-correction, change
the polarization signal, but this work describes for the first time how
this affects the azimuthal distribution of the circumstellar $Q_\phi$ signal.
Because of the lack of a systematic study, it is also difficult to extrapolate
findings from the papers cited above to a more general picture.

For compact scattering regions near the star the mentioned effects
are much stronger and they must even be considered for a
qualitative interpretation of the polarization signal.
Therefore, the scattering regions at small separations to the star
$r\lapprox 5\,\lambda/D$, or $r\lapprox 200~$mas for large telescopes
with AO systems, are often disregarded
despite the fact that these regions are scientifically very interesting.
They correspond for protoplanetary disks at $\approx 100$~pc 
to the innermost $<20~$AU, where 
planets form \citep[e.g.,][]{Birnstiel24} and for nearby red giant stars
to a separation of a few stellar radii where dust
particles condensate
\citep[e.g.,][]{Hoefner18}. In non-coronagraphic and non-saturated data,
this region is sometimes masked in the data reduction process,
because it looks "noisy", or shows an unexplained polarization pattern.
However, these features might just be caused by a polarization offset
or a convolution effect.

In coronagraphic observations the central region is hidden behind
the focal plane mask to achieve more sensitive observations for
the polarization signal further out. Also for this type of data, it is
important to understand the impact of the convolution and of
polarization offsets.

This work presents model simulations and a parameter framework
for a systematic description of the PSF convolution effects and
an investigation of the polarimetric offsets introduced by the
instrument, the interstellar polarization, the central star, or by the 
zp-correction applied to the data. The results should provide
a better understanding of these effects for the interpretation
of observational results, and be useful for the definition of
observing and measuring strategies for polarimetric imaging of
circumstellar scattering regions.

This paper is organized as follows. The next section describes the model
simulation and introduces the used polarimetric parameters.
Section 3 describes the degradation of
the scattering polarization by the PSF convolution for axisymmetric
circumstellar scattering regions and for inclined
disk models. The polarimetric calibration and zp-correction 
effects are investigated in Sect.~4, and discussed for coronagraphic
observation and for data of partially resolved circumstellar
scattering region. In Sect.~5 the
results are summarized and discussed in the context of
observational data. Detailed information, in particular numerical
results for the presented simulations, are given in the appendix.

\begin{figure}
\includegraphics[trim=-0.4cm 0.2cm 0.2cm 0.0cm, width=8.0cm]{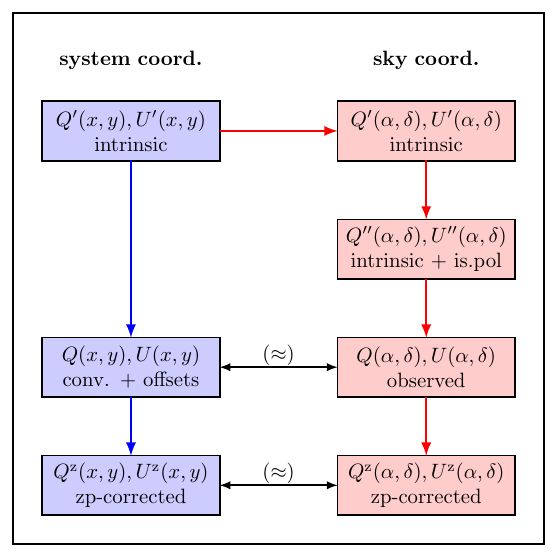}
\caption{Block diagram with the simplified description of the simulated
  imaging polarimetry given in blue. The red arrows show the full 
  imaging process from the intrinsic model, the on-sky model including
  interstellar polarization, to the observed and possibly
  zp-corrected polarization signal.
  }
\label{BlockDiag}
\end{figure}

\section{Model calculations}
\label{SectModel}

The block diagram in Fig.~\ref{BlockDiag} gives an overview
of the steps involved in imaging polarimetry using the
Stokes $Q=I_0-I_{90}$ and $U=I_{45}-I_{135}$ polarization
parameters as example. The full process
is described by the boxes connected with red arrows, from the
intrinsic polarization model defined in system coordinates $Q'(x,y),U'(x,y)$,
to the model in sky coordinates $Q'(\alpha,\delta),U'(\alpha,\delta)$,
including the contribution of interstellar polarization
$Q''(\alpha,\delta),U''(\alpha,\delta)$, the observed signal
$Q(\alpha,\delta),U(\alpha,\delta)$,
and the result after a possible polarimetric zp-correction
$Q^{\rm z}(\alpha,\delta),U^{\rm z}(\alpha,\delta)$.
This is simplified in the simulations according to the blue
path in Fig.~\ref{BlockDiag} by calculating the observed
polarization signal in system coordinates $Q(x,y),U(x,y)$
considering only convolution and
polarimetric offsets, and a zp-correction
$Q^{\rm z}(x,y),U^{\rm z}(x,y)$ for the calibration of the data.
This approach still considers many of the key aspects of polarimetric
imaging, but disregards second order effects and
particular problems of individual instruments or data sets.
The parameters for the $x,y$-coordinates are used for
the description of the model simulations the $x,y$-coordinates, while
some general polarimetric principle are discussed
for on sky parameters using $(\alpha,\delta)$-coordinates. 
However, it is important to be aware of simplifying assumptions outlined
in Fig.~\ref{BlockDiag} for the interpretation of the model results.

\subsection{Intrinsic scattering models}

The intrinsic models for the circumstellar
dust scattering region and a point like central star
at $x_0,y_0$ are described
by 2-dimensional (2D) maps for each component of the Stokes vector
\begin{equation}
\mathbf{I}'(x,y) = (I'(x,y),Q'(x,y),U'(x,y))\,,
\end{equation}
describing the intensity $I'(x,y)$, and the linear
polarization $Q'(x,y)$, $U'(x,y)$ in x,y-coordinates aligned with the
object. The circular polarization signal $V'$ is expected to be
much weaker for circumstellar scattering and is neglected. 
In principle, multiple scattering by dust can produce
a circular polarization signal, in particular for (magnetically)
aligned dust grains. Measurements for circumstellar
circular polarization exist since many decades
\citep[e.g.,][]{Kwon14,Bastien89,Angel73}, but the signals are weak, or
originate from regions far away from the star, where interstellar
magnetic fields may play a role. Considering this in our models
is beyond the scope of this paper.

The models
use $\mathbf{I}'(x,y)=\mathbf{I}'_s + \mathbf{I}'_d(x,y)$ 
consisting of a star
$\mathbf{I}'_s(x,y)=\mathbf{I}'_s(0,0)=\mathbf{I}'_s$, with or without
an intrinsic polarization $Q'_s$, $U'_s$ and an extended dust scattering
region $\mathbf{I}'_{\rm d}(x,y)$ according to the vector components
\begin{eqnarray}
  I'(x,y) & = & I'_s + I'_d(x,y), \\
  Q'(x,y) & = & Q'_s + Q'_d(x,y), \label{QIntr} \\
  U'(x,y) & = & U'_s + U'_d(x,y)\,. \label{UIntr} 
\end{eqnarray}  
In most models the star is not polarized and $Q'_s$ and $U'_s$ are set
to zero or $Q'(x,y) = Q'_d(x,y)$ and $U'(x,y) = U'_d(x,y)$.

\begin{figure}
  \includegraphics[trim=-0.5cm 0.5cm 1.0cm 0.0cm,width=8.8cm]{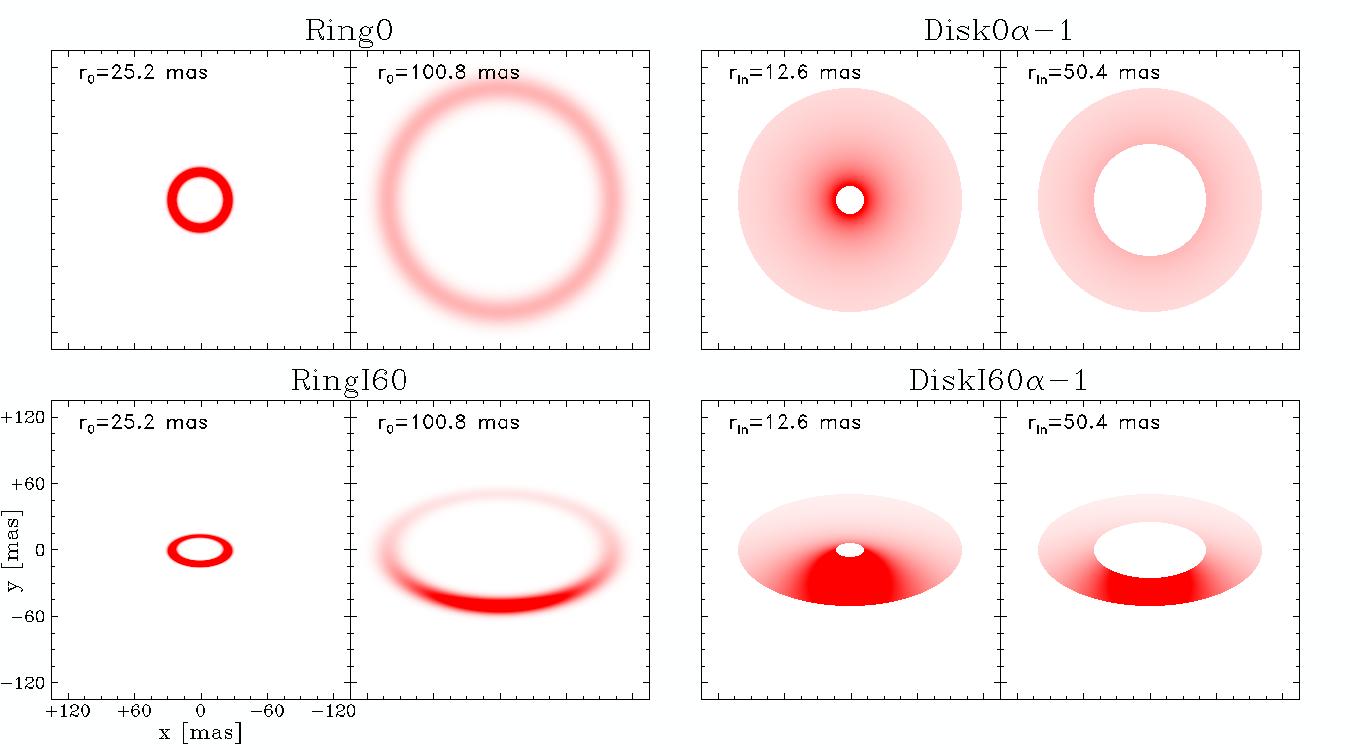}
  \caption{Maps for the intrinsic disk intensity $I'_d(x,y)$ for the four
    used types of circumstellar scattering models.}
\label{Fig.IntrModel}
\end{figure}

\paragraph{Scattering geometries.}
We consider the axisymmetric models
Ring0 and Disk0 representing circumstellar disks
seen pole-on or spherical dust shells as illustrated by the
intensity maps $I'_d(x,y)$ in the upper row of 
Fig.~\ref{Fig.IntrModel}. The intrinsic polarization flux $P'_d$ is
proportional to the
intensity $P'_d(x,y)=0.25 \cdot I'_d(x,y)$.

The Ring0 models have a mean radius $r_0$ and a
Gaussian radial profile with full width at half maximum
(FWHM) of $\Delta r=0.2~r_0$ and
$r_0$ is varied from $3.15~$mas to 806.4~mas.
The Disk0 models are axisymmetric scattering regions extending
from an inner $r_{\rm in}$ to an outer radius $r_{\rm out}$, with
a surface brightness (SB) described by the power law 
\begin{equation}
I'_d(r)=A_I\,(r/r_{\rm ref})^\alpha\,,
\label{Eq.SBdisk}  
\end{equation}  
with reference radius $r_{\rm ref}$. Three cases, Disk0$\alpha 0$,
Disk0$\alpha$-1, and Disk0$\alpha$-2, are considered with $\alpha=0$ for a
constant SB and $\alpha=-1$ and $-2$ for a brightness
decreasing with distance. For all cases the outer radius is
$r_{\rm out}=100.8~$mas while the radius of the inner cavity is varied between
$r_{\rm in}=3.15$~mas and 50.4~mas to investigate the differences
between fully resolved and partially resolved scattering regions.
Figure~\ref{Fig.IntrModel} shows $I'_d(x,y)$ maps for Disk0$\alpha$-1
with $r_{\rm in}=12.6$~mas and 50.4~mas. 

Simulations for non-axisymmetric scattering geometries are
obtained by adopting the Ring0 and Disk0 geometries for inclined, flat disks
with $i=60^\circ$. The dust density in 
the disk plane $\rho_d(r_d)$ of the inclined models RingI60 and
DiskI60$\alpha$0, DiskI60$\alpha$-1, DiskI60$\alpha$-2 are
described by the same radial parameters as
for the axisymmetric models. The scattering angle in the inclined
disk model varies as function of azimuthal angle $\phi$ on the disk,
and therefore also the scattered intensity and polarization depend
on $\phi$. This is simulated as in
\citet{Schmid21} for flat, optically thin disks with a
dust scattering phase function described by a Henyey-Greenstein
function for the intensity with asymmetry parameter $g=0.6$ and
a Rayleigh scattering like dependence for the fractional polarization
with $p_{\rm max}=0.25$.
With these settings the resulting model maps $\mathbf{I}'(x,y)$
depend only on $r_0$ for the RingI60 model
and on $\alpha$ and the radius of the central cavity
$r_{\rm in}$ for the DiskI60 models. Figure \ref{Fig.IntrModel}
shows the maps for RingI60 and DiskI60$\alpha$-1 with the
same $r_0$ and $r_{\rm in}$ parameters as for the pole-on models.
The major and minor axes of the inclined disks are aligned
with the $x$ and $y$ coordinates and the backside of the disk is in the
$+y$ direction. Because of the strong forward scattering ($g=0.6$) the
intensity signal is enhanced on the front-side, and this is
a frequently observed property for
inclined disks \citep[e.g.,][]{Ginski23}

\subsection{Sky coordinates and interstellar polarization}
Observations are obtained in sky coordinates 
$(\alpha,\delta)$ and the intrinsic maps in system coordinates
$Q'(x,y),U'(x,y)$ can be transformed into the sky coordinates
$Q'(\alpha,\delta),U'(\alpha,\delta)$ by
a rotation of the geometry and the polarization vector
\citep[e.g.,][]{Schmid21}.
The signal reaching Earth $Q''(\alpha,\delta),U''(\alpha,\delta)$ is
often affected by interstellar polarization introduced
by dichroic absorption of magnetically aligned interstellar grains
\citep[e.g.][]{Draine03}.
This introduces a fractional polarization offset, which can be
described for the usual, low polarization case
($Q',U'\ll I'$ and $q,u\ll 1$) by
\begin{eqnarray}
Q''(\alpha,\delta)\approx Q'(\alpha,\delta)
         +q_{\rm is}I'(\alpha,\delta)\,, \label{Eq.qis}\\
U''(\alpha,\delta)\approx U'(\alpha,\delta)
           +u_{\rm is}I'(\alpha,\delta)\,, \label{Eq.uis}  
\end{eqnarray}
where $q_{\rm is}, u_{\rm is}$ are the components of the
fractional interstellar polarization
$p_{\rm is}=(q_{\rm is}^2+u_{\rm is}^2)^{-1/2}$ with position angle
$\theta_{\rm is} = 0.5\cdot {\rm atan2}(u_{\rm is},q_{\rm is})$
\footnote{Defined as in the FORTRAN function atan2(y,x) for Cartesian to
polar coordinate conversions.}
defined in sky coordinates. We use flux ratios to quantify
the resulting intensity or polarization so that the interstellar
transmission losses can be neglected. Extreme dichroic 
extinction by the interstellar medium can convert linear
polarization partly to circular polarization \citep[e.g.,][]{Kwon14},
but this is not considered in this work. 

\subsection{Signal degradation by ground-based AO observations.}
The turbulence in the Earth atmosphere leads to a strong seeing
convolution of the incoming signal $(I'',Q'',U'')$. With an
AO system the seeing can be strongly reduced
but there remain smearing and polarimetric cancellation
effects which can change strongly the spatial distribution
of the observed signal $I(\alpha,\delta),Q(\alpha,\delta),U(\alpha,\delta)$
when compared to the incoming signal
\citep[e.g.,][]{Perrin03,Fetick19}. The effects
are particularly strong for not well
resolved structures near the star. The AO correction
depends on the atmospheric turbulence and instrument properties
and therefore the observational point spread function (PSF)
changes with time and shows various types of non-axisymmetric
structures \citep[e.g.,][]{Cantalloube19}.
In addition, AO instruments are complex and they
usually introduce instrumental polarization offsets which are
difficult to calibrate accurately \citep{Tinbergen07}. These
observational effects are investigated in this work.

\begin{figure}
\includegraphics[trim=1.3cm 13cm 2.2cm 5.0cm, width=8.5cm]{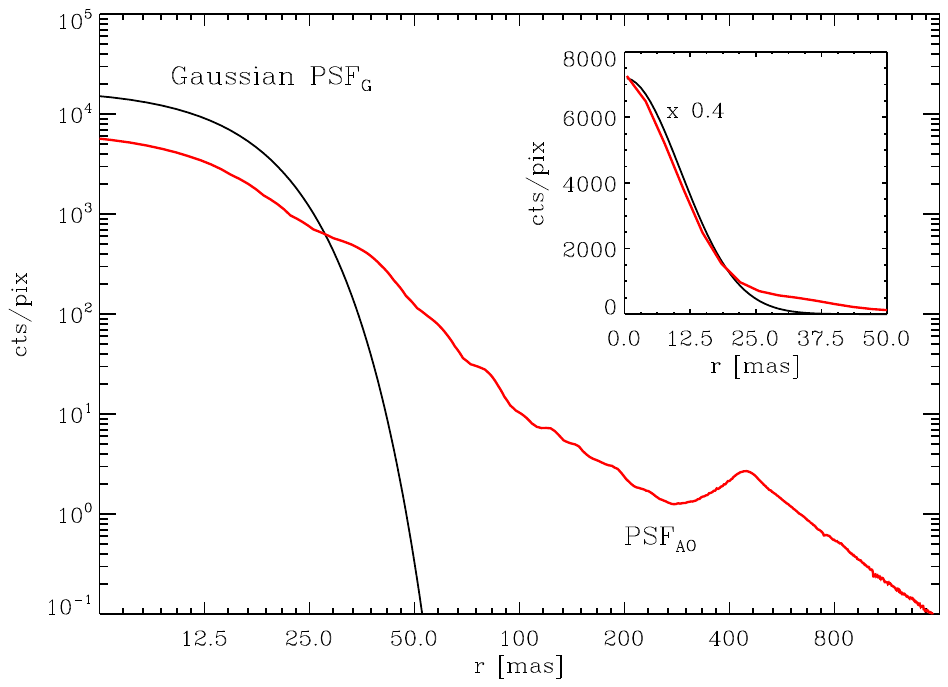}
\caption{Radial profiles for the extended PSF$_{\rm AO}$ (red) and the Gaussian
  PSF$_{\rm G}$ (black). In the main panel the total flux is normalized
  to $10^6$ counts. In the inset PSF$_{\rm G}$
  is reduced by a factor 0.4 for a comparison of the PSF cores.
  The pixel size is $3.6\,{\rm mas}\times 3.6\,{\rm mas}$.
  }
\label{ProfPSFs}
\end{figure}

\paragraph{PSF convolution.}
Axisymmetric PSFs are adopted so that the
convolved scattering signal does not depend 
on the orientation of the observed system. Therefore we
can apply the convolution directly to the models
described in disk coordinates $(x,y)$.
Real PSFs are variable and deviate from axisymmetry but
assuming a stable, axisymmetric PSF is a reasonable
approximation for PSFs with high Strehl ratio or for
the averaged PSF obtained after a number of
polarimetric cycles taken in field rotation mode.

We use PSF$_{\rm AO}(x,y)$
representing an AO system with a narrow core with FWHM or diameter of
$D_{\rm PSF}=25.2$~mas
and an extended halo. This profile
is obtained by averaging azimuthally the PSF described in
\cite{Schmid18} for the standard star HD 161096 taken under excellent
condition with the AO instrument VLT/SPHERE/ZIMPOL in the
N$\_$I filter ($\lambda_c=817$~nm) with a Strehl ratio of about 0.4.
The radial profile of PSF$_{\rm AO}$ is plotted in Fig.~\ref{ProfPSFs}
together with a Gaussian profile PSF$_{\rm G}$ with the same diameter
$D_{\rm PSF}$, which
is used to investigate the impact of
the extended halo in PSF$_{\rm AO}$ on the convolved signal.
Convolution with a Gaussion PSF with $D_{\rm PSF}\approx \lambda/D$
could also represent roughly diffraction limited imaging polarimetry
with a space telescope with diameter $D$ or ground based seeing
limited imaging polarimetry with $D_{\rm PSF}$ of the
order $\approx 1000$~mas.
  
\paragraph{Polarimetric calibration.}
Many types of instrumental effects are introduced in
observations taken with high-contrast imaging polarimetry
and there exist established procedures 
to calibrate the data, for example for SPHERE/IRDIS
\citep{deBoer20,vanHolstein20} and SPHERE/ZIMPOL polarimetry 
\citep{Schmid18,Hunziker20,Tschudi24}. Individual instruments
show also particular effects but
these aspects are beyond the scope of this paper.

A very general and important observational effect are the
offsets introduced by the interstellar $p_{\rm is}$ or
by the instrument polarization
$p_{\rm inst}$ of the kind described in Eqs.~\ref{Eq.qis} and \ref{Eq.uis}.
Because the intensity of the star $\Sigma I_s$ is much higher
than the polarization from the circumstellar dust $\Sigma Q_d, \Sigma U_d$,
by about a factor 100 or even more,  
already a small fractional polarization offset for the central star
of about $p\approx 0.001$ can strongly disturb the circumstellar
signal, while $p\approx 0.01$ can completely mask that signal.
In addition, there could also be a contribution from an intrinsic
polarization of the central
star. The effects of polarimetric offsets do not depend on the
orientation of the selected coordinate system 
and therefore they can also be treated in the $(x,y)$-coordinate system. 

Unfortunately, it can be quite difficult to disentangle
the different contributions to the overall polarimetric offset and
therefore an
``ad-hoc'' zp-correction for the polarization is often
applied to the data,
based on the assumption that the polarization of the target in a
selected integration region is zero or at least
very small \citep{Quanz11,Avenhaus14}.
A polarimetric offset sets the integrated Stokes parameters
in a certain region to zero
$\Sigma Q^{\rm z}=\int Q^{\rm z}(\alpha,\delta) {\rm d}\alpha{\rm d}\delta=0$
and similar for $\Sigma U^{\rm z} =0$, but this procedure 
can introduce a bias. For coronagraphic observations the central star
cannot be included for the determination of the zp-correction,
and this is adding another complication which will also be described.

\subsection{Analysis of diagnostic polarization parameters}
The impact of the convolution and the polarimetric zp-correction,
follows from the comparisons between the
simulated observational maps $I(x,y), Q(x,y), U(x,y)$ or
the corresponding zp-corrected polarization $Q^{\rm z}(x,y),U^{\rm z}(x,y)$
with the initial maps $I'(x,y)$, $Q'(x,y)$, $U'(x,y)$.

Circumstellar scattering produces predominantly a linear
polarization in azimuthal direction $Q_\phi$ with
respect to the central star located at $x_0,y_0$.
Therefore $Q_\phi$ is a very useful polarization parameter
which is defined by
\begin{eqnarray}
Q_\phi(x,y) =  -Q(x,y)\cos(2\phi_{xy}) - U(x,y)\sin(2\phi_{x,y})\,,
      \label{EqQphiQU}\\
U_\phi(x,y) =  +Q(x,y)\sin(2\phi_{xy}) - U(x,y)\cos(2\phi_{x,y})\,,
      \label{EqUphiQU}
\end{eqnarray}  
with  $\phi_{xy} = {\rm atan2}((x-x_0),(y-y_0))$
according to the description of \citet{Schmid06} for the radial Stokes
parameters $Q_r$, $U_r$ and using $Q_\phi=-Q_r$ and $U_\phi=-U_r$.
The $U_\phi(x,y)$ parameter gives the linear polarization
component rotated by $45^\circ$ with respect
to azimuthal component $Q_\phi(x,y)$. 
The different polarized intensities for a given point $(x,y)$
in the polarization maps are related
to the polarized flux for the linear polarization
according to 
\begin{equation}
  P(x,y) = \bigl(Q^2(x,y)+U^2(x,y)\bigr)^{1/2}
          =\bigl(Q_\phi^2(x,y)+U_\phi^2(x,y)\bigr)^{1/2}\,.
\label{EqPQUQphiUphi}
\end{equation}
Figure.~\ref{FigRingPoleOn} illustrates the relation between $Q_\phi$
and the Stokes parameters $Q$ and $U$ for a pole-on disk ring.
In this model the $U_\phi$ component is zero and $P(x,y)=Q_\phi(x,y)$
because of the axisymmetric geometry. Non-axisymmetric geometries
can produce an intrinsic $U'_\phi(x,y)$ signal, for example by
multiple scattering \citep{Canovas15}, but for the simple (single
scattering) models adopted in this work the
intrinsic $U'_\phi(x,y)$-signal is also zero for inclined disks.
However, the convolution and polarimetric
offsets will introduce also for these models a $U_\phi$ signal,
which is equivalent to a non-azimuthal polarization component
(Fig.~\ref{FigRingi60}).

The observational $Q(x,y),U(x,y)$ or $Q_\phi(x,y),U_\phi(x,y)$ maps
have in high contrast imaging polarimetry often a lower signal
per pixel than the photon noise $\sigma(x,y)$ or other pixel to
pixel noise sources.
Therefore, the polarized flux $P(x,y)$ is biased, because it is
always positive, and the measured values follow a Rice probability
distribution. For a low polarization $P(x,y)\lapprox \sigma(x,y)$ the
noise will introduce on average a signal of about
$P(x,y)\approx +\sigma(x,y)$ \citep{Simmons85}, and this can add up to a
very significant spurious
signal for the polarized flux $\Sigma P$ in an integration region
$\Sigma$ containing many tens or more pixels.
This noise problem is avoided by using $Q_\phi(x,y)$ for measuring
the strength of the spatially resolved linear polarization \citep{Schmid06}.
This is a reasonable approximation, because circumstellar scattering produces
predominantly a polarization in azimuthal direction with
$Q_\phi(x,y)\gg 0$ and $U_\phi(x,y)\approx 0$ and therefore one
can consider $Q_\phi$ as rough proxy for $P$ according to
\begin{equation}
P(x,y) = \bigl(Q_\phi^2(x,y)+U_\phi^2(x,y)\bigr)^{1/2} \approx Q_\phi(x,y)\,.
\label{Eq.Polapprox}
\end{equation}
Enhanced random noise in the data does not change the mean $Q_\phi$ signal
for a pixel region, but $P$ is for observational data often very significantly
affected by the bias problem described above.
The simulations in this work do not consider statistical noise in the
data. However, there are other systematic differences between
$Q_\phi(x,y)$ and $P(x,y)$, because of the $U_\phi(x,y)$ signal introduced
by the PSF convolution and polarization offsets as will be described
by the simulation results.

The convolution and zp-correction effects are quantified using
polarization parameters integrated in well defined apertures
\begin{equation}
\Sigma X =\int X(x,y)\,{\rm d}x{\rm d}y \,,
\end{equation}
where $X$ is a place holder for the PSF convolved radiation parameters, like 
$X=\{I,Q,U,P,Q_\phi,U_\phi\}$, and similarly for $X'$ or $X^{\rm z}$
for the intrinsic models or zp-correcected models, respectively.
$\Sigma$ defines a circular aperture centered on the star for the
determination of the system integrated parameters. Later, we consider also
other axisymmetric circular and annular apertures for the parameters
of radial subregions (see Sect.~\ref{Sect.NormPolRing}). Axisymmetric
apertures are very useful for quantifying the convolution by
an axisymmetric PSF and polarization offsets for an intensity
signal dominated by the axisymmetric stellar PSF.

For a given integration region $\Sigma$ an ``aperture''
polarization ${\cal{P}}(\Sigma) = \bigl((\Sigma Q)^2 + (\Sigma U)^2\bigr)^{1/2}$
with the corresponding position angle $\theta(\Sigma_i)$ can be defines
like for aperture polarimetry. These parameters are useful
to specify the polarization of the star, or the fractional 
polarization $p={\cal{P}}(\Sigma)/\Sigma I$ and the fractional
Stokes parameters $q,u$ to quantify polarization offsets and
zp-correction factors (see Sect.~\ref{SectPar}).

The polarization parameters are usually normalized to the integrated
intrinsic polarization $\Sigma Q'_\phi$, like $\Sigma Q/\Sigma Q'_\phi$,
$\Sigma Q^{\rm z}/\Sigma Q'_\phi$, because $\Sigma Q'_\phi$ provides
a good reference for the characterization of the impact
of the convolution or of offsets on polarization parameters.

For the
characterization of the azimuthal distribution of the polarization
$Q(\phi)$ and $U(\phi)$ the four Stokes $Q$ quadrants
$Q_{xxx}=\{Q_{000},Q_{090},Q_{180},Q_{270}\}$, and the four Stokes $U$
quadrants $U_{xxx}=\{U_{045},U_{135},U_{225},U_{315}\}$ are used \citep{Schmid21},
while $X_{xxx}$ stands for all eight parameters. 
They represent the Stokes $Q$ and $U$ polarization in the four
wedges of 90$^\circ$, which form for well resolved
circumstellar scattering regions the 
positive-negative $Q$ and $U$ quadrant patterns. The signals in
these quadrants are changed by the convolution
and polarization offsets, and they are useful to quantify
the corresponding changes for the
azimuthal distribution of the polarization signal $Q_\phi(\phi)$.

An overview on the used polarization parameters is given in Sect.~\ref{SectPar}
(Tables \ref{ParXYTab} and \ref{ParXYTabzp}).

\section{Polarization degradation by the PSF convolution} 
\label{SectConv}

This section describes the basic PSF smearing and
cancellation effects for imaging polarimetry of circumstellar
scattering regions.
The convolution is always applied
to the intrinsic intensity $I'$ and Stokes $Q'$, $U'$ maps,
or from maps with a polarization offset like the $Q''$, $U''$ maps ,
and from the resulting $Q$ and $U$ one can then derive 
according to Eq.~\ref{EqQphiQU}, \ref{EqUphiQU}, and \ref{EqPQUQphiUphi}
the convolved maps for the polarization parameters
$Q_\phi$, $U_\phi$ and $P$ {\citep[e.g.,][]{Tschudi21}.
Alternatively, one could also apply the convolution to the
$I'_0$, $I'_{90}$, $I'_{45}$, and $I'_{135}$
polarization components and derive from this the convolved intensity
and polarization maps, because the convolution
operation is distributive (${\rm PSF} * A + {\rm PSF} * B = {\rm PSF} * (A+B)$).
However, a direct convolution of the intrinsic azimuthal polarization
(${\rm PSF} * Q'_\phi$) or of the polarized flux (${\rm PSF} * P'$) gives
wrong results.

\begin{figure}
\includegraphics[trim=0.0cm 0.3cm 0.5cm 1.2cm, width=8.8cm]{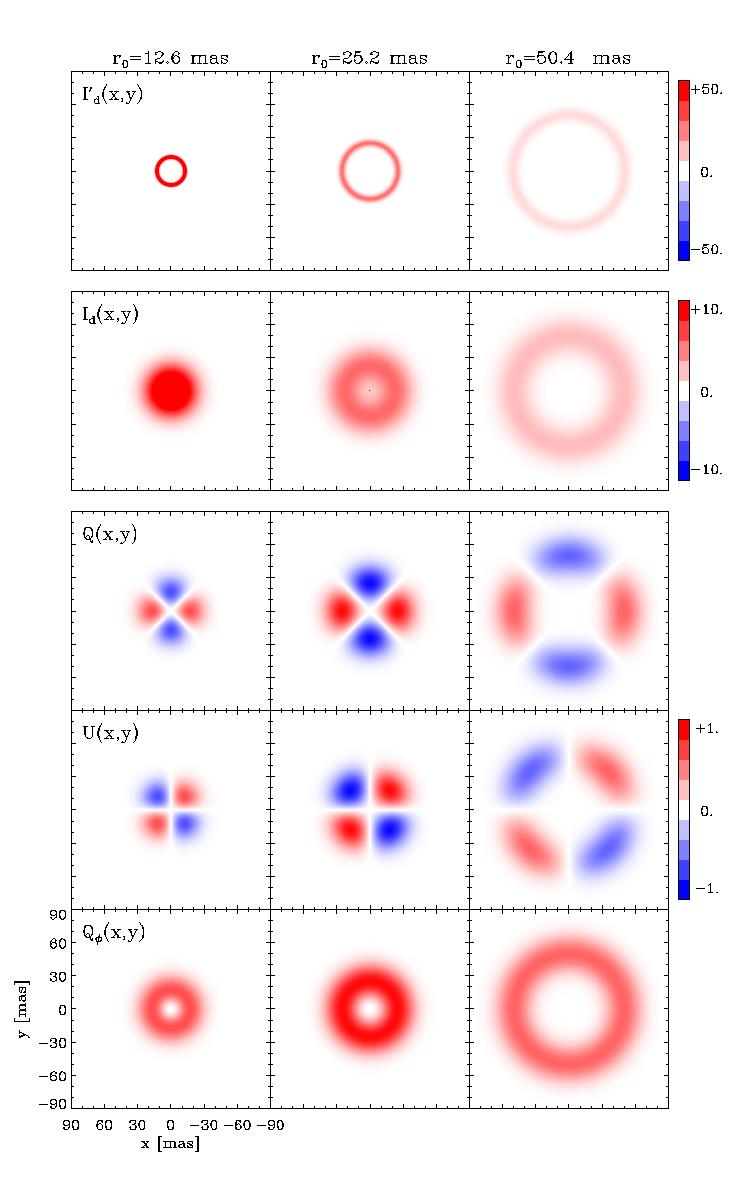}
\caption{Maps for the Ring0 models
  with  $r_0=12.6$~mas (left), 25.2~mas (middle), and 50.4~mas (right)
  for the intrinsic circumstellar intensities $I'_d(x,y)$, and the
  PSF$_G$ convolved intensity $I_d(x,y)$ and 
  polarization $Q_d(x,y)$, $U_d(x,y)$ and $Q_\phi(x,y)$.
  $\Sigma I'_d$ is the same for all models, and $\pm 1$ for the color scale
  represent the same fluxes per pixel for all panels.}
\label{FigRingPoleOn}
\end{figure}

\begin{figure}
\includegraphics[trim=2.2cm 13.2cm 8cm 5.2cm, width=8.8cm]{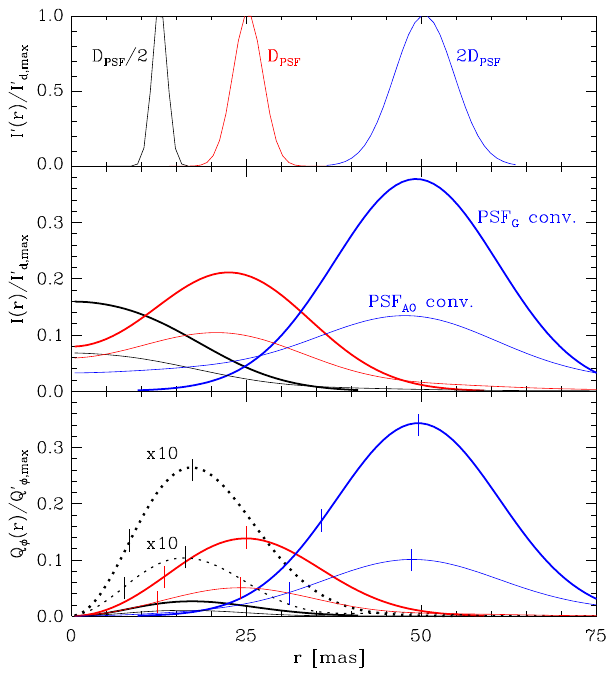}
\caption{Normalized radial profiles for the intrinsic 
  $I'_d(r)/I'_{\rm d,max}$ (top panel), and the PSF$_{\rm G}$ 
  (thick lines) and PSF$_{\rm AO}$ convolved (thin lines) intensity
  $I_d(r)/I'_{\rm d,max}$ (middle panel) for
  the Ring0 models with $r_0=12.6$~mas (black), 25.2~mas (red),
  and 50.4 mas (blue). The bottom panel shows the convolved 
  polarization $Q_\phi(r)/Q'_{\phi,max}$, where the
  black dotted lines are the black solid lines 
  multiplied by a factor of 10. The radii for the profile
  peaks $r({\rm max}Q_\phi(r))$ and the central holes $r_h(Q_\phi)$ given
  in Table~\ref{Tab.PoleOnRing} are indicated with vertical marks.}
\label{ProfRingPoleOn}
\end{figure}

\subsection{Convolution for axisymmetric scattering models}
\label{ConvAxisym}

\subsubsection{Narrow pole-on rings}
\label{Sect.RingPoleOn}

The convolution depends strongly on the spatial resolution, specifically
on the angular size of the scattering signal compared to the PSF widths.
This can be described by an axisymmetric scattering geometry 
like for a dust disk seen pole-on or a
spherical dust shell.

Simulations for the convolution of the Ring0 models are shown in
Fig.~\ref{FigRingPoleOn} for three ring sizes with
$r_0 = 12.6~$mas, 25.2~mas, and 50.4~mas and using the Gaussian
PSF$_{\rm G}$. Maps for the intrinsic dust scattering intensities
$I'_d(x,y)$, and convolved intensities $I_d(x,y)$ and the
polarization $Q_d(x,y)$, $U_d(x,y)$, $Q_{\phi,d}(x,y)$ are given.
The central star is unpolarized in this model and there is
$Q_d=Q$, $U_d=U$ and $Q_{\phi,d}=Q_\phi$.
All three models have the same intrinsic flux $\Sigma I'_d$ and
$\Sigma Q'_\phi$ and therefore the peak fluxes scale like
${\rm max}(I'_d(x,y))\propto 1/r_0^2$ and is higher for smaller $r_0$.
The star is unpolarized and therefore does not contribute
to the polarization maps.
The star $I_s$ is not included in the intensity maps because for
realistic cases it would dominate strongly the scattering intensity $I_d$.  
For the intrinsic model there is $Q'_\phi(x,y)=P'(x,y)$ and $U'_\phi(x,y)=0$.
For axisymmetric models this
property is not changed by the convolution and therefore the
maps for $P(x,y)$ and $U_\phi(x,y)$ are not shown in Fig.~\ref{FigRingPoleOn}. 

Radial profiles for the intensity and azimuthal polarization
$I_d(r)$ and $Q_\phi(r)$ are plotted in Fig.~\ref{ProfRingPoleOn}
for the Ring0 model convolved with the Gaussian PSF$_{\rm G}$ and the
extended PSF$_{\rm AO}$. The profiles are normalized to the intrinsic
peak flux $I'_{\rm d,max}={\rm max}(I_d'(r))$
and $Q'_{\phi,{\rm max}}={\rm max}(Q'_\phi(r))$ respectively,
to illustrate the signal degradation.
The intensity is smeared by the convolution and
this reduces the peak SB of the rings but the
total intensity is preserved $\Sigma I_d=\Sigma I'_d$. 
For the smallest model with $r_0=12.5$~mas the ring structure is
no more visible (Fig.~\ref{FigRingPoleOn}) and there is 
a strong maximum at the center.

The convolved polarization signal shows also 
a smearing and in addition a mutual polarimetric cancellation
between the positive and negative quadrants in the $Q$ and $U$
maps \citep{Schmid06}. This reduces strongly the polarization signal for
$Q_d(x,y)$, $U_d(x,y)$, and $Q_\phi(x,y)$ close to the star and produces
also for compact rings a central zero
(Figs.~\ref{FigRingPoleOn}, \ref{ProfRingPoleOn}). 

\begin{table}
\caption{Results for Ring0 models with different radii $r_0$.}
\label{Tab.PoleOnRing}
\begin{tabular}{lccccccc}
\noalign{\hrule\smallskip}    
\multispan{2}{\hfil $r_0$ \hfil}
       & \multispan{2}{\hfil $\Sigma Q_\phi/\Sigma Q'_\phi$ \hfil}
                       & \multispan{2}{\hfil $r({\rm max}(Q_\phi))$\hfil}
                             & \multispan{2}{\hfil $r_h(Q_\phi)$\hfil}    \\
       &     &  PSF$_{\rm G}$\hspace{-1mm} & \hspace{-2mm} PSF$_{\rm AO}$
                & PSF$_{\rm G}$\hspace{-1mm}  & \hspace{-2mm}PSF$_{\rm AO}$ 
                   & PSF$_{\rm G}$\hspace{-1mm}  & \hspace{-2mm} PSF$_{\rm AO}$ \\
\noindent
\hspace{-3mm}
${\rm [mas]}$ & \hspace{-3mm}$[D_{\rm PSF}]$\hspace{-3mm}
& [\%]        & [\%]   
                       & \hspace{-2mm}$[D_{\rm PSF}]$\hspace{-2mm}
                         & \hspace{-2mm}$[D_{\rm PSF}]$\hspace{-2mm}
                           & \hspace{-2mm}$[D_{\rm PSF}]$\hspace{-2mm}
                              & \hspace{-2mm}$[D_{\rm PSF}]$\hspace{-2mm} \\
\noalign{\smallskip\hrule\smallskip}
\hspace{-3mm} 3.15  & \hspace{-1mm}.125\hspace{-1mm}
             & 2.5       & 1.0
                          & 0.62  & 0.59 &  0.31  &  0.26 \\
\hspace{-3mm} 6.30  & .25  & $8.3$     & 3.4
                          & 0.64  & 0.59 &  0.32  &  0.27 \\
\hspace{-3mm} 12.6  & .5  & $28.2$    & 11.6
                          & 0.68  & 0.67 &  0.33  &  0.30 \\
\hspace{-3mm} 25.2  & 1  & $66.2$    & 28.4
                          & 0.99  & 0.96 &  0.52  &  0.48  \\
\hspace{-3mm} 50.4  & 2  & $90.9$    & 46.9
                          & 1.96  & 1.92 &  1.42  &  1.23  \\
\hspace{-3mm} 100.8 & 4  & $97.7$    & $60.1$
                          & 3.98  & 3.95 &  3.34  &  3.19  \\
\hspace{-3mm} 201.6 & 8  & $99.9$    & $66.1\tablefootmark{a}$ &
                         $\approx r_0$ & $\approx r_0$ \\
\hspace{-3mm} 403.2 & 16  & $99.9$    & $70.1\tablefootmark{a}$ &   \\
\hspace{-3mm} 806.4 & 32  & $100.$   & $77.9\tablefootmark{a}$ &   \\
\noalign{\smallskip}
\hspace{-3mm} error &   & \hspace{-1mm}$\pm 0.2$ \hspace{-1mm}
                        & \hspace{-1mm}$\pm 0.2$ \hspace{-1mm}
                        & \hspace{-1mm}$\pm 0.02$ \hspace{-1mm}
                        & \hspace{-1mm}$\pm 0.02$ \hspace{-1mm}
                        & \hspace{-1mm}$\pm 0.01$ \hspace{-1mm}
                        & \hspace{-1mm}$\pm 0.01$ \hspace{-1mm} \\
\noalign{\smallskip\hrule}
  \end{tabular}
\tablefoot{Values are given for the integrated azimuthal polarization
$\Sigma Q_\phi/\Sigma Q'_\phi$ for Ring0 models after convolution
with the Gaussian PSF$_{\rm G}$ and the extended PSF$_{\rm AO}$.
For compact models also the radii for the peak
flux $r({\rm max}(Q_\phi))$ and for the central hole $r_h(Q_\phi)$
are listed. The column $D_{\rm PSF}$ gives the ring radius in units of
the FWHM of the PSF.
\tablefoottext{a}{Significant amounts of polarization signal not
contained in the used aperture with $r=1.5''$.}}
\end{table}

The integrated polarization $\Sigma Q_\phi$ is reduced for the  
PSF$_{\rm G}$ convolved models
by factors $\Sigma {Q}_\phi/\Sigma {Q}'_\phi = 0.91$, 0.66, and 0.28
for ring radii of $r_0=50.4~$mas, 25.2~mas and 12.6~mas, respectively
(Table~\ref{Tab.PoleOnRing}). We measure the apparent ring size
using the radius for the maximum SB $r({\rm max}(Q_\phi))$
and the size of the central cancellation hole by the radius $r_h$
at half maximum $Q_\phi(r_h)=0.5\cdot ({\rm max}(Q_\phi))$.
For $r_0\lapprox 0.5\, D_{\rm PSF}$ the peak radius for the convolved ring is 
significantly larger than $r_0$. For larger rings
$r_0\gapprox D_{\rm PSF}$ these two radii agree well
($r({\rm max}(Q_\phi))\approx r_0$).

The $\Sigma Q_\phi/\Sigma Q'_\phi$ values in Table~\ref{Tab.PoleOnRing}
are significantly lower for the PSF$_{\rm AO}$ convolved models because of
the much stronger smearing effects introduced by the PSF halo. 
For $r_0 < 4~D_{\rm PSF}$
the reduction is about a factor of 0.5 lower with respect to a convolution
with PSF$_{\rm G}$, or very roughly at the level of the Strehl ratio for 
PSF$_{\rm AO}$ simulated for the AO system.

For very small rings
with $r_0=0.25~D_{\rm PSF}$ and 0.125~$D_{\rm PSF}$ only a very small amount of the
$Q_\phi$ polarization remains and the radial $Q_\phi$ profile has for
these two cases roughly the same shape as the convolved Ring0 with
$r_0=0.5~D_{\rm PSF}$ with a peak at $r({\rm max}(Q_\phi))\approx (2/3)~D_{\rm PSF}$
and a hole radius $r_h\approx (1/3)~D_{\rm PSF}$.
This represents the simulated inner working angle for the detection of
a resolved polarization signal for a circumstellar scattering region.
For observational data these limits might be less good because
of PSF variation and alignment errors.

Importantly, the spatial resolution is practically
not reduced by the convolution with PSF$_{\rm AO}$ when compared
to PSF$_{\rm G}$ and the radii $r({\rm max}(Q_\phi))$ and
$r_h(Q_\phi)$ for PSF$_{\rm AO}$ are very similar for the two cases
because both PSF have the same peak width $D_{\rm PSF}$.
Therefore, the shape of the radial profiles in Fig.~\ref{ProfRingPoleOn}
are very similar for the two cases and the morphology of the
strong polarization structures in the PSF$_{\rm G}$ model maps
of Fig.~\ref{FigRingPoleOn},
would look practically the same for a convolution with PSF$_{\rm AO}$.
The halo of the PSF$_{\rm AO}$ introduces faint,
extended polarization artefacts for axisymmetric 
geometries as described in the Appendix (Sect.~\ref{Sect.artifacts}),
but they are very weak with a surface brightness of $\lapprox 1~\%$
when compared to the peak signal of the ring.

\subsubsection{Radially extended scattering regions}
\label{Sect.Exti0Models}

\begin{figure}
\includegraphics[trim=0.1cm 0.5cm 1.5cm 0.5cm,width=8.8cm]{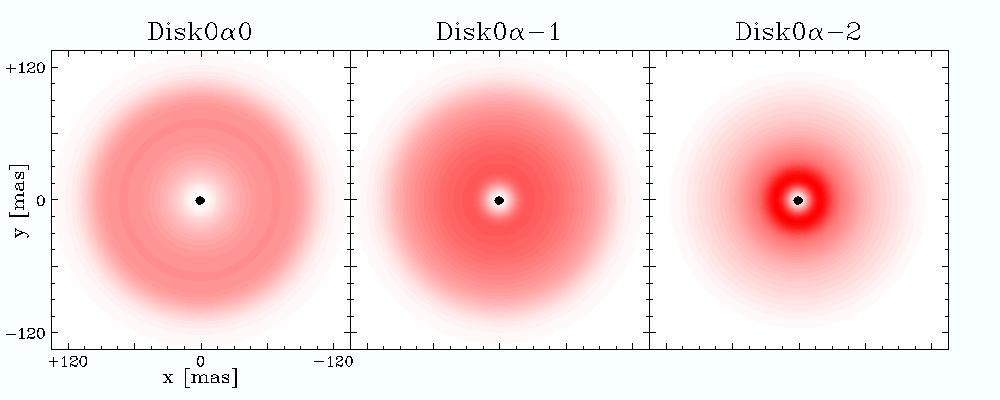}
\caption{Central cancellation holes in the $Q_\phi(x,y)$ maps
  for the models Disk0$\alpha$0, Disk0$\alpha$-1, and Disk0$\alpha$-2
  after convolution with the extended PSF$_{\rm AO}$.
  The size of the inner cavity $r_{\rm in}=0.125 D_{\rm PSF}$ (3.15~mas)
  is the same for all three models and indicated by the black dot.}
\label{FigDiskPoleOn}
\end{figure}

Axisymmetric, radially extended disks or shells models
are just superpositions of concentric ring signals, and
the polarimetric cancellation effects will be strong for
the innermost regions and much reduced further out.
Therefore, the convolved polarization
signal strongly underestimates the scattering near the central
star and can mimic the presence of a central cavity even if no such
cavity is present. The effect is illustrated
in Fig.~\ref{FigDiskPoleOn} with PSF$_{\rm AO}$
convolved $Q_\phi(x,y)$ maps for the models Disk0$\alpha$0, Disk0$\alpha$-1,
and Disk0$\alpha$-2, with radial brightness profiles $Q'_\phi(r)=A_\phi$,
$A_\phi (r/r_{\rm ref})^{-1}$, and $A_\phi (r/r_{\rm ref})^{-2}$, respectively. In all
three cases the intrinsic disk extends from $r_{\rm in}=3.15$~mas
or $0.125\,D_{\rm PSF}$ to $r_{\rm out}=100.8$~mas.
The radius of the convolution hole ($r_h(Q_\phi)$) is
slightly smaller than $0.4\,D_{\rm PSF}$ for the $\alpha=-2$ case
and slightly larger than $0.4\,D_{\rm PSF}$ for the $\alpha=-1$ disk
with a flatter brightness distribution (Table~\ref{TabPoleOnDisk}).

\begin{figure}
\includegraphics[trim=2.0cm 13cm 7cm 4.5cm, width=8.0cm]{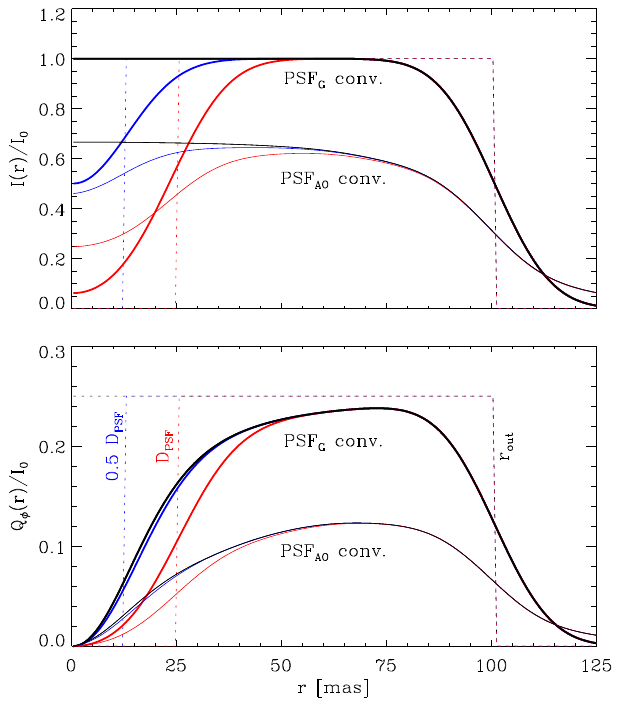}
\caption{Normalized profiles $I(r)/I'_0$ and 
  $Q_\phi(r)/I'_0$ for the Disk0$\alpha$0 models with $r_0=D_{\rm PSF}~$ (red),
  $0.5\,D_{\rm PSF}~$ (blue) and without cavity (black) for
  PSF$_{\rm G}$ (thick) and PSF$_{\rm AO}$ (thin) convolution, and for
  the intrinsic model (dotted).}
\label{ProfFlatPoleOn}
\end{figure}

\paragraph{Constant surface brightness.}
The convolution effects for Disk0$\alpha$0 are shown in
Fig.~\ref{ProfFlatPoleOn} with radial profiles for the
intrinsic parameters $I'_d(r)=I'_0$, $Q'_\phi(r)=A_\phi=0.25\,I'_0$
and the convolved intensity $I_d(r)$ and polarization 
$Q_\phi(r)$ for different inner disk
radii $r_{\rm in}$ and for PSF$_{\rm G}$ and PSF$_{\rm AO}$ convolution. 
The profiles $I_d(r)$ show for increasing
cavity size $r_{\rm in}$ an increasing central dip
depth and width. The surface brightness (SB) $I_d(r)$ is strongly reduced
after convolution with PSF$_{\rm AO}$ while the central
cavity is less pronounced. The convolution does not change $\Sigma I_d$
but for PSF$_{\rm AO}$ a lot of the signal is redistributed to radii
$r\gg 100$~mas.

The convolved $Q_\phi(r)$ profiles show for all cases a 
central zero $Q_\phi(0)=0$, even for the disk without
central cavity. Only models with $r_{\rm in}\approx D_{\rm PSF}$ 
or larger show an obvious difference when compared to the model
without cavity. The $Q_\phi(r)$ profiles do not reach the
intrinsic $Q'_\phi(r)$ level even for the convolution with a
narrow PSF$_{\rm G}$, because of the polarimetric cancellation.
For the convolution with PSF$_{\rm AO}$ there is an additional
reduction of the $Q_\phi(r)$ level by about 40~\% but despite this,
the radial shape of the profiles is very similar for the two cases
as follows also from hole radii $r_h(Q_\phi)$ given in
Table~\ref{TabPoleOnDisk}.

\begin{figure}
\includegraphics[trim=2.0cm 12.5cm 2.5cm 4.5cm, width=10.5cm]{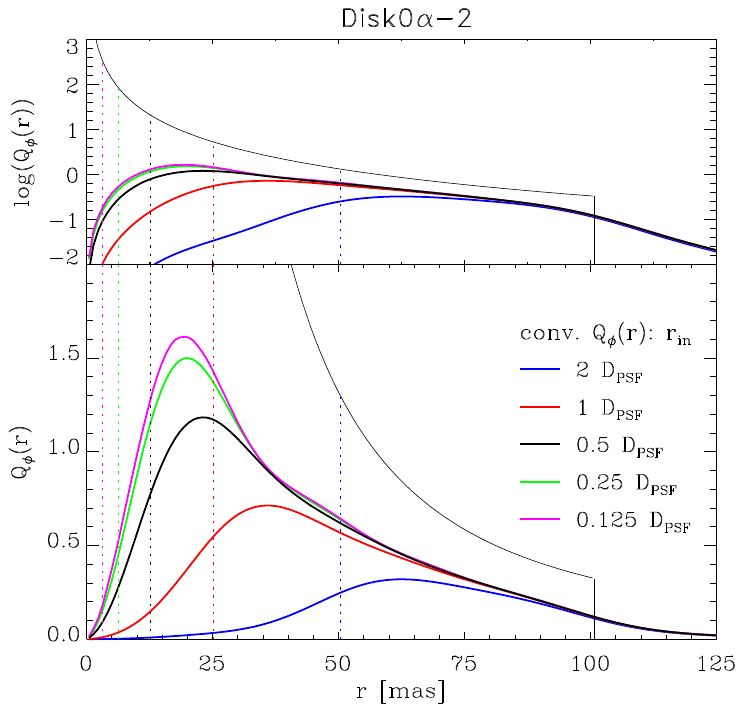}
\caption{Radial profiles $Q_\phi(r)$ for the Disk0$\alpha$-2 model
  convolved with PSF$_{\rm AO}$ on a log-scale (upper panel) and a
  linear scale (lower panel) for different inner cavities $r_0$ as
  indicated by the colors. The thin black line is the intrinsic
  SB $Q'_\phi(r)\propto r^{-2}$.}
\label{Profalpham2PoleOn}
\end{figure}

\paragraph{Centrally bright scattering regions.}
Many circumstellar disks and shells show a steep SB
profile increasing strongly towards smaller radii. Therefore,
the polarimetric cancellation suppresses efficiently
the intrinsically very bright but barely resolved central
signal \citep[e.g.,][]{Avenhaus18,Garufi20,Khouri20,Andrych23}.

Simulations of radial profiles are shown
in Fig.~\ref{Profalpham2PoleOn} for PSF$_{\rm AO}$ convolved models
DiskI60$\alpha$-2 with intrinsic SB profile
for the polarization
$Q'_\phi(r)=A_\phi (r/r_{\rm ref})^{-2}$ with $r_{\rm ref}=12.6$~mas and
$A_\phi=20.7$ similar to Eq.~\ref{Eq.SBdisk} and for different cavity
sizes $r_{\rm in}$.
The convolved profiles $Q_\phi(r)$ show clearly how the central
hole size decreases for smaller $r_{\rm in}$,
and how the measurable peak polarization signal
${\rm max}(Q_\phi(r))$ increases. Despite the strongly increasing
intrinsic $Q'_\phi$ signal at small radii the convolved $Q_\phi(r)$
curves converge to a limiting model case with $r_i\approx 0.2\,D_{\rm PSF}$
($\approx 5~$mas), because even large amounts of intrinsic signal
in the center are fully cancelled by the convolution
(Table~\ref{TabPoleOnDisk}).
The $Q_\phi(r)$ profiles are still quite sensitive for constraining the  
intrinsic $Q'_\phi(r)$ signal around 
$r\approx D_{\rm PSF}$ based on the location of the flux maximum
$r({\rm max}(Q_\phi))$ and the amount of $Q_\phi(r)$ signal
near this location.
A careful analysis of $Q_\phi(r)$ can therefore constrain an inner
cavity for a dust scattering region and this could be potentially
useful for estimates on the dust sublimation radius
for disks around young stars or the dust condensation radius
for circumstellar shells around mass losing stars.

Details of the profiles $Q_\phi(r)$
depend also on the power law index $\alpha$ for the SB.
The peak radii $r({\rm max}(Q_\phi(r))$ and
central hole radii $r_h$ for a given $r_{\rm in}$ are a bit larger
for Disk0$\alpha$-1 than Disk0$\alpha$-2 (Table~\ref{TabPoleOnDisk}),
because in this model the contributions of smeared signal
from larger separations are more important than for Disk0$\alpha$-2
and this reduces the apparent sharpness of the central cancellation
hole.

\begin{figure}
\includegraphics[trim=0.0cm 0.6cm 0.3cm 1.0cm, width=9.2cm]{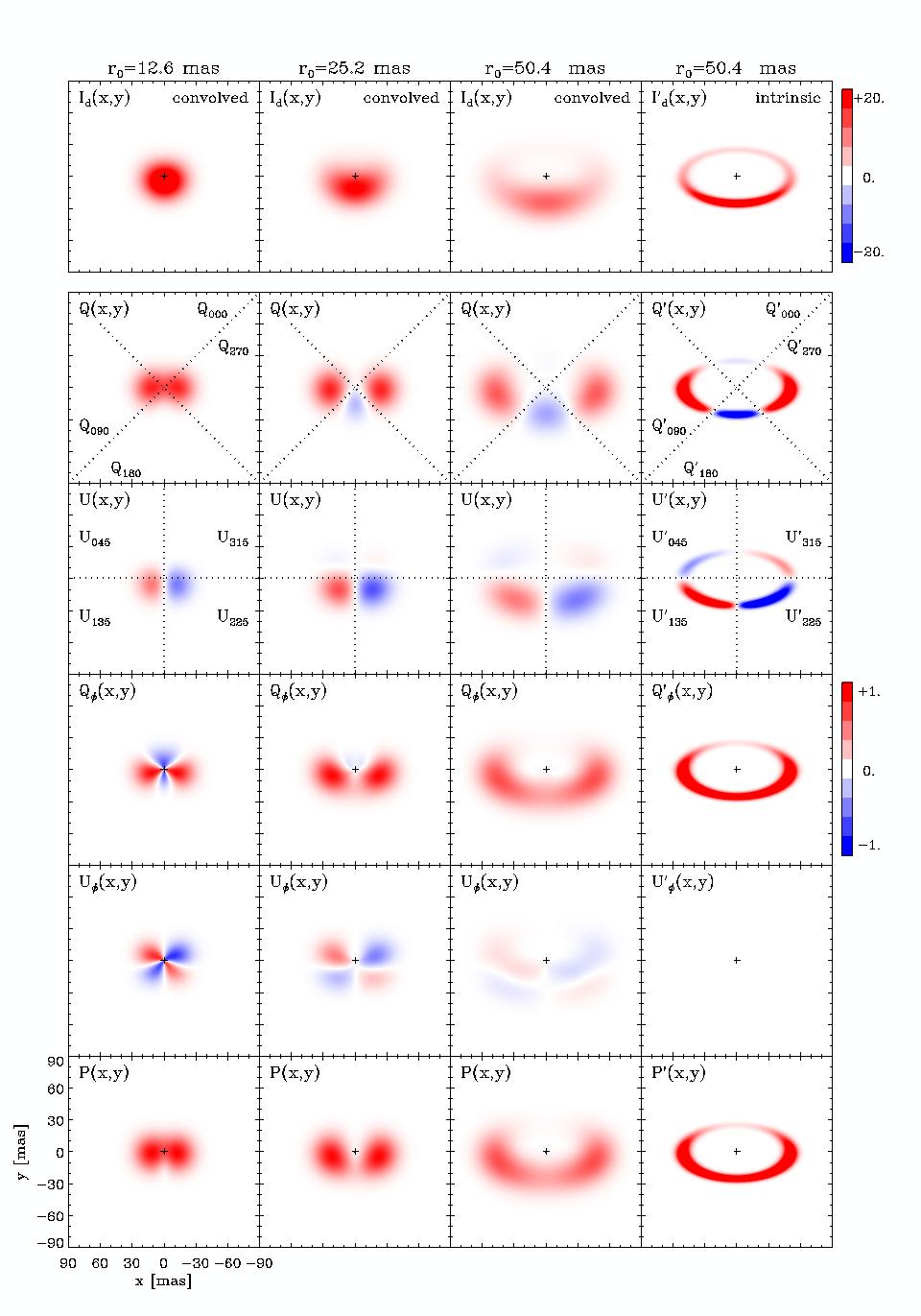}
\caption{Maps for the intensity $I_d$ and
  polarization $Q_\phi$,  $Q$, $U$, $Q_\phi$, $U_\phi$
  and $P$ (from top to bottom) for RingI60 models with $r_0=12.6$~mas,
  25.2~mas and 50.4~mas after convolution with the Gaussian
  PSF$_{\rm G}$ (first three columns). The last column gives the
  same for the intrinsic model with $r_0=50.4$~mas. Stokes quadrant
  parameters are indicated in some $Q$ and $U$ maps.}
 \label{FigRingi60}
\end{figure}

\subsection{Convolution for inclined disk ring models}
\label{Sect.RingI60}

The scattering geometry for a rotationally
symmetric system with an inclined symmetry axis is
not axisymmetric and then new features appear in the
intensity and polarization maps.
This is illustrated in Fig.~\ref{FigRingi60} by
the PSF$_{\rm G}$ convolved maps for the scattered intensity $I_d$ and the
polarization parameters $Q$, $U$, $Q_\phi$, $U_\phi$, and $P$ for 
RingI60 models with an inclination of $i=60^\circ$ and
$r_0=12.5~$mas, 25.2~mas and 50.4~mas and including
the intrinsic maps for the last case. The central star is assumed
to be unpolarized, or $Q=Q_d$, $U=U_d$ and $P=P_d$, and the stellar
intensity is not included in the $I_d(x,y)$ intensity maps.
All RingI60 models have the same intrinsic
polarization $\Sigma Q'_\phi$.

The ring front side is much brighter because
of the adopted forward scattering parameter $g=0.6$ for the dust
and this is clearly visible for the $I'(x,y)$ and $I(x,y)$ maps for
$r_0=50.4~$mas. This changes for the less resolved systems into a
front-side intensity arc for $r_0=25.2$~mas or $r_0=D_{\rm PSF}$
and a slightly elongated spot offset towards the front side
for $r_0=12.6~$mas ($0.5\,D_{\rm PSF}$).

The inclined models show a
left-right symmetry for the intensity $I_d(x,y)=I_d(-x,y)$, for
Stokes $Q$, the azimuthal polarization $Q_\phi$, and the
polarized flux $P$. The Stokes $U$-parameter has
a left-right antisymmetry $U_d(x,y)=-U_d(-x,y)$, as well as
the $U_\phi$-signal introduced by convolution effects.

The intrinsic Stokes parameters $Q'(x,y)$ and
$U'(x,y)$ show positive and negative regions which can be characterized
by the quadrant polarization parameters
$Q'_{000}$, $Q'_{090}$, $Q'_{180}$, and $Q'_{270}$ for Stokes $Q'$ and $U'_{045}$,
$U'_{135}$, $U'_{225}$, and $U'_{315}$ for Stokes $U'$ as indicated in
Fig.~\ref{FigRingi60}. They are useful for the characterization
of the azimuthal distribution of the polarization based on the natural
Stokes patterns produced by circumstellar scattering \citep{Schmid21}.
Quadrant parameters have been calculated for simple models
of debris disks \citep{Schmid21} and of transition disks \citep{Ma22}.

The intrinsic RingI60 models have
strong positive $Q_{090}$ and $Q_{270}$ quadrants centered
on the major axis of the projected disk, because of the high fractional
polarization produced by scatterings with $\theta\approx 90^\circ$. The front
side quadrant $Q_{180}$ shows for well resolved systems a clear
negative component, which is however less dominant than in 
intensity, because the fractional polarization of 
forward scattering is lower than for $90^\circ$ scattering.

The negative $Q_{000}$ and $Q_{180}$ components
disappear for not well resolved disks because the PSF smearing
of the two strong positive components $Q_{090}$ and $Q_{270}$
cancel the signal of the negative Stokes $Q$ quadrants. 
For the convolved RingI60 model with $r_0=12.6$~mas there remain
only two positive $Q_d$-spots but their relative position still
indicates the orientation of the projected major axis.
The brighter disk front side produces
in the Stokes $U_d$ map a left side dominated by the positive $U_{135}$
signal and a right side by the negative $U_{225}$ component. This 
feature is still visible for barely resolved disks and this
indicates the location of the disk front-side.

The intrinsic disk polarization of our models
is everywhere azimuthal and therefore the map $Q'_\phi(x,y)$ is equal
$P'(x,y)$, while $U'_\phi(x,y)=0$ according to
Eq.~\ref{EqPQUQphiUphi}. The convolved maps $Q_\phi(x,y)$
for $r_0=50.4$~mas represents well the intrinsic
map appart from the smearing, but for less well
resolved disks $Q_\phi(x,y)$ starts to display negative values
above and below the center and for
$r_0=12.6$~mas a strong, central quadrant pattern for
$Q_\phi(x,y)$ is visible as expected
for an unresolved, central source with a positive Stokes
$\Sigma Q$ signal and $\Sigma U=0$.

The convolution of the non-axisymmetric scattering polarization 
produces a $U_\phi$ signal and we call this effect the convolution
cross-talks for the azimuthal polarization\footnote{This convolution
cross-talk is different from instrumental polarization cross-talks which
are introduced by optical components \citep[e.g.,][]{Tinbergen07}.}
$Q_\phi\rightarrow U_\phi$.
This effect increases with
decreasing spatial resolution and therefore the $U_\phi(x,y)$ signal
becomes stronger for less well resolved disks. For an unresolved disk
the quadrant patterns for the azimuthal polarization
$Q_\phi(x,y)=-Q(x,y)\cos(2\phi_{xy})$ and
$U_\phi(x,y)=Q(x,y)\sin(2\phi_{xy})$ are equally strong
but rotated by $45^\circ$. For the polarized flux $P$
the convolved signal for large rings
is roughly equal to the azimuthal polarization
$P(x,y)\approx Q_\phi(x,y)$, and for barely resolved rings it evolves towards
$P(x,y) \approx Q(x,y)$ (Fig.~\ref{FigRingi60}).

\begin{figure}
\includegraphics[trim=2cm 13cm 3.8cm 3.0cm, width=8.8cm]{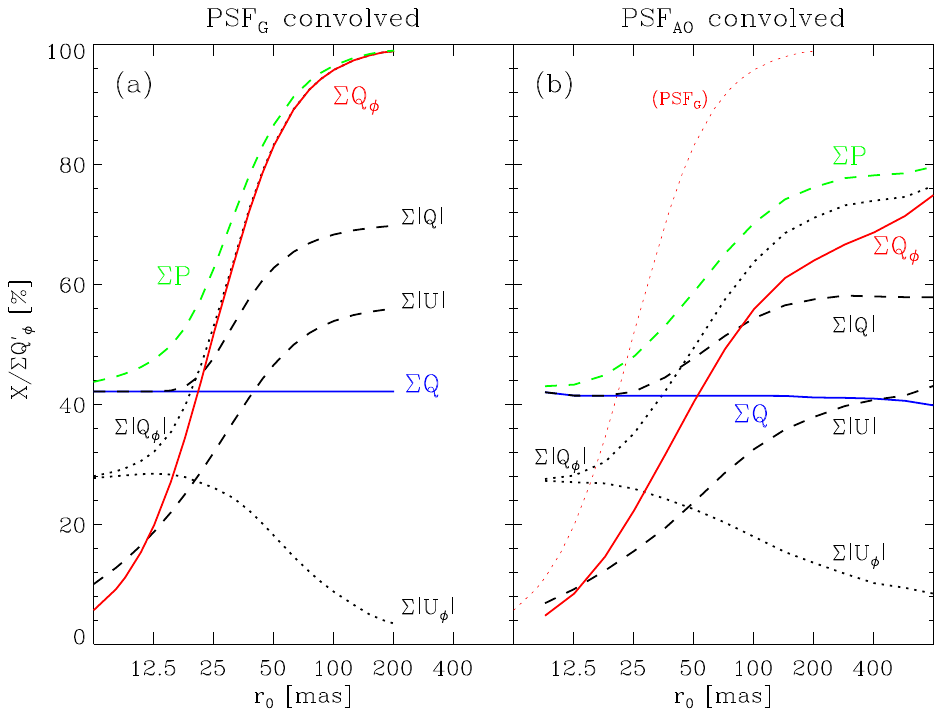}
\caption{Integrated polarization parameters for the RingI60 model
  as function of the ring radius $r_0$ after convolution with PSF$_{\rm G}$
  (a) and PSF$_{\rm AO}$ (b). 
  All parameters are normalized to the intrinsic value $\Sigma Q'_\phi=100~\%$.
  In panel (b) the $\Sigma Q_\phi$ curve from panel (a)
  is repeated as dotted line.}
\label{DiagAperg6i60Ring}
\end{figure}

\subsubsection{Convolution and integrated polarization parameters}

The convolution with a normalized PSF does not change the
  integrated Stokes polarization or the integrated intensity.
Thus, there is for the RingI60 models
$\Sigma {Q}=\Sigma Q' = 0.421~\Sigma Q'_\phi$ and $\Sigma U=\Sigma U'=0$,
independent of the disk size or the spatial resolution.} Contrary to this, the
integrated polarization parameters $\Sigma Q_\phi$, $\Sigma U_\phi$, and
$\Sigma P$ and the sums of absolute values
$\Sigma |Q|$ and $\Sigma |U|$, $\Sigma |Q_\phi|$ and $\Sigma |U_\phi|$
depend on the resolution and the PSF convolution. The dependencies are
plotted for the RingI60 models as function of ring radius $r_0$
convolved with PSF$_{\rm G}$ and PSF$_{\rm AO}$
in Fig.~\ref{DiagAperg6i60Ring}, and Table~\ref{TabRingi60} lists
numerical values.

The red curves in Fig.~\ref{DiagAperg6i60Ring} show
$\Sigma Q_\phi$, which is large for
well resolved disks and approaches zero for small, unresolved disks
$r_0\rightarrow 0$ when the $Q_\phi$ map shows a perfect
positive-negative quadrant pattern with no net $\Sigma Q_\phi$
polarization. For the intrinsic disk there is
$\Sigma P'=\Sigma Q'_\phi=\Sigma |Q'_\phi|$
and $\Sigma |U'_\phi|=0$ and these relations are still approximately
valid for well resolved disks convolved with PSF$_{\rm G}$ 
because the convolution effects are small (Fig.~\ref{DiagAperg6i60Ring}(a)).
The convolution with PSF$_{\rm AO}$ introduces even for
very extended disks strong smearing, because of the extended PSF halo
(Fig.~\ref{DiagAperg6i60Ring}(b)). Therefore, $\Sigma P$ is
for well resolved disks larger than $\Sigma Q_\phi$
by about 5~\% for the model with $r_0\approx 800$~mas and
about 18~\% for $r_0=201.6$~mas (Table~\ref{TabRingi60}). 

For an unresolved scattering region
$r_0\ll D_{\rm PSF}$ there remains only an unresolved
polarized source with $\Sigma P$. Because of the alignment of the RingI60
models with the $(x,y)$ disk coordinates, there is 
$\Sigma Q=\Sigma |Q|=\Sigma P$ and
$\Sigma U=\Sigma |U|=0$. The azimuthal
polarization is then zero $\Sigma Q_\phi\approx 0$ and the integrated
absolute values for the azimuthal polarization are
$\Sigma |Q_\phi|=\Sigma |U_\phi| = (2/\pi)\,|\Sigma P|$ for
the quadrant pattern of an unresolved source
\citep{Schmid21}.

In the models a net signal in azimuthal polarization
$\Sigma Q_\phi>0$ or an integrated absolute signal $\Sigma |Q_\phi|$
larger than $\Sigma |U_\phi|$ indicates the presence of at
least a marginally resolved circumstellar polarization signal
(Fig.~\ref{DiagAperg6i60Ring}). In observational data,
$\Sigma Q_\phi$ signals can also be produced
by noise effects and one needs to derive for a given 
data set the limits for a significant detection of a resolved
circumstellar $\Sigma Q_\phi$-signal. However, one can expect for
random noise sources introduced for example by the atmospheric
turbulence, photon and read-out noise, that they produce also random
positive and negative $Q_\phi$ and $U_\phi$ signals, and different
systematic effects like polarization offsets (see Sect.~\ref{Sect.Calib})
or alignment errors produce no or only small positive or negative
$\Sigma Q_\phi$ and $\Sigma U_\phi$ signals. Therefore, the
two criteria $\Sigma Q_\phi>0$ and $\Sigma |Q_\phi| > \Sigma |U_\phi|$
are reliable indicators for the presence of resolved
circumstellar polarization. Contrary to this,
the $\Sigma P$ signal is systematically
enhanced by random noise and also polarization offsets change
typically $\Sigma P$ substantially.

\begin{figure}
\begin{centering}  
  \includegraphics[trim=0.0cm 0.5cm 2.0cm 0.5cm, width=8.8cm]{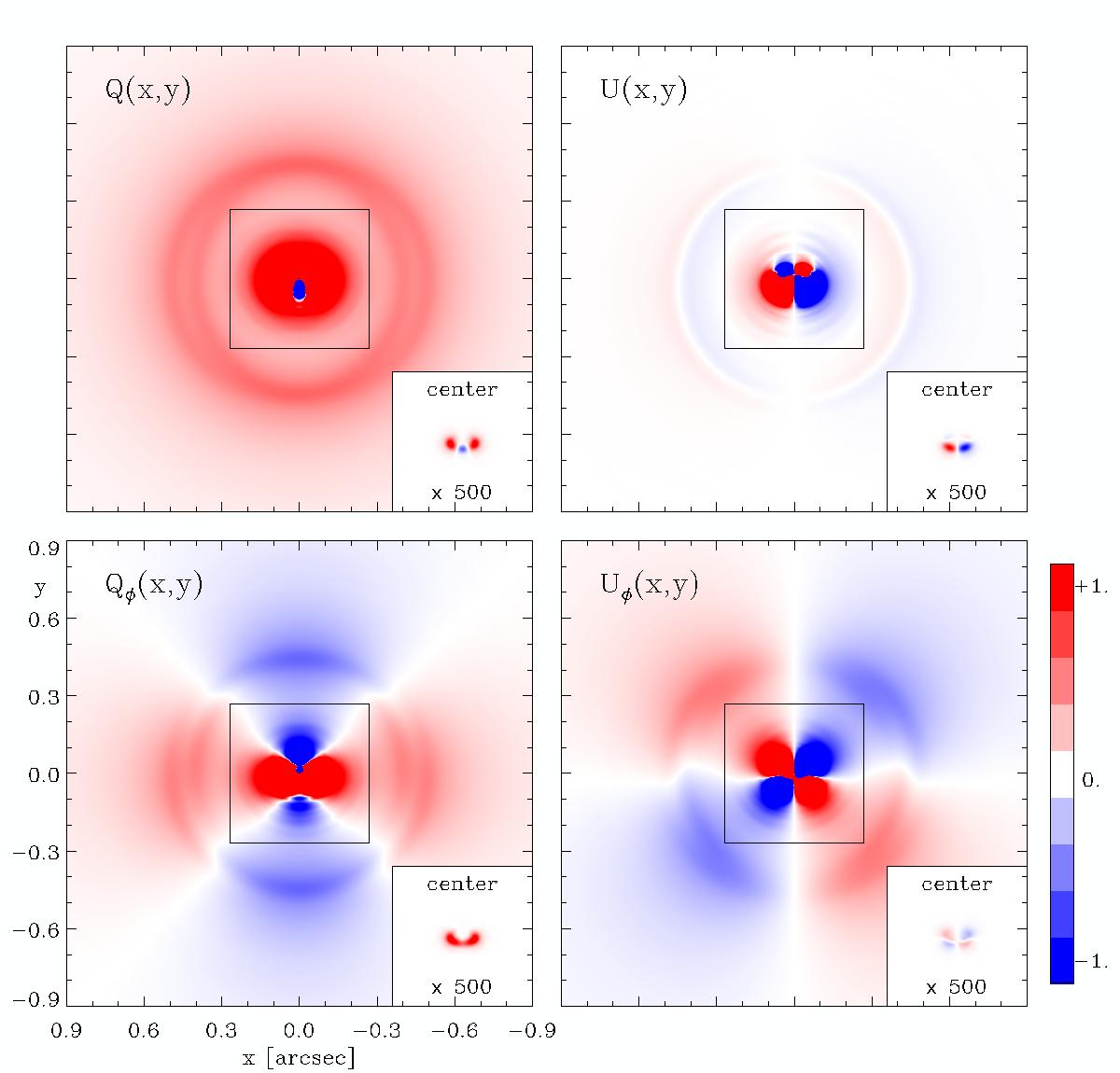}
\end{centering}  
\caption{Large scale halo signals for the Stokes parameters
  $Q$, $U$, and the azimuthal polarization 
  $Q_\phi(x,y)$, $U_\phi(x,y)$ for the RingI60 model
  with $r_0=50.4~$mas convolved with PSF$_{\rm AO}$. The inset
  on the lower right in each panel shows the disk ring in
  the center with a $500\times$ reduced color scale.}
\label{FigRingI60AOLRG}
\end{figure}

\subsubsection{Halo signals produced by an extended PSF$_{\rm AO}$}
\label{Sect.Halosignal}
The extended halo in the PSF$_{\rm AO}$ of an adaptive optics system
smears substantially the net Stokes polarization $\Sigma Q,\,\Sigma U$
of a system over a large area producing an extended $P(x,y)$ halo
of linear polarization for $r\gg r_0$.
This halo effect is illustrated in Fig.~\ref{FigRingI60AOLRG} for
the PSF$_{\rm AO}$ convolved RingI60 model with $r_0=50.4$~mas,
where the smeared $\Sigma Q$ signal produces a polarized speckle ring
ghost around $r\approx 0.4''$ and a halo. The
Stokes $U$ signal in the halo is much weaker because there is no net
$U$-signal for this model, and only a much weaker artefact pattern
of the kind described in Sect.~\ref{Sect.artifacts} is visible.
The $Q$-halo produces extended $Q_\phi(x,y)$, $U_\phi(x,y)$ quadrant patterns 
including the relatively strong speckle ring around $r\approx 0.4''$,
which shows the small scale imprint of the polarization
from the bright ring.

The surface brightness (SB) of the halo is low and it can be 
difficult to recognize it in real data because of observational noise.
Nonetheless, for the RingI60 model shown in Fig.~\ref{FigRingI60AOLRG}
the $Q$-halo integrated in an annulus from
$r=0.2$\arcsec to 1.5\arcsec is almost 30~\% of the
system integrated $\Sigma Q$-signal (Table~\ref{TabRingi60}).
Therefore, it is important to use large integration apertures for
the determination of $\Sigma Q$ and $\Sigma U$ for data convolved
by an extended PSF.

The $Q_\phi$ signal in the halo integrated
from $r=0.2''$ to 1.5$''$ contributes less than 0.2~\% to
the total $\Sigma Q_\phi$ (Tab.~\ref {TabRingi60}), because
this signal has a positive-negative $Q_\phi, U_\phi$
quadrant patterns with zero net signal. Therefore, the
measurement of $\Sigma Q_\phi$ for a compact circumstellar
scattering region should be restricted to
a circular aperture which excludes the outer halo regions
containing no net
$Q_\phi$ signal but possibly substantial observational noise. 

\subsubsection{Convolution cross-talk $Q_\phi \rightarrow U_\phi$ or
  intrinsic $U_\phi$ signal}
\label{UphiIntr}
The models considered in this work use a simpified description of
the dust scattering which does no produce an intrinsic $U'_\phi(x,y)$
polarization. However, it was pointed out in \cite{Canovas15} that
multiple-scattering by dust in optically thick protoplanetary disks
with a non-axisymmetric scattering geometry can
produce intrinsic $U'_\phi$ signals representing non-azimuthal polarization
components. Their simulations give $U'_\phi(x,y)$ signals of up
to about $\pm 5$~\% of the azimuthal polarization $Q'_\phi(x,y)$ and this
$U'$-polarization can be useful to constrain
the dust scattering properties and the disk geometry.
This effect was also recognized in the disk models presented
by \citet[][Fig.~6]{Ma22}.

\begin{figure}
\includegraphics[trim=2cm 13cm 3.0cm 4.5cm, width=8.8cm]{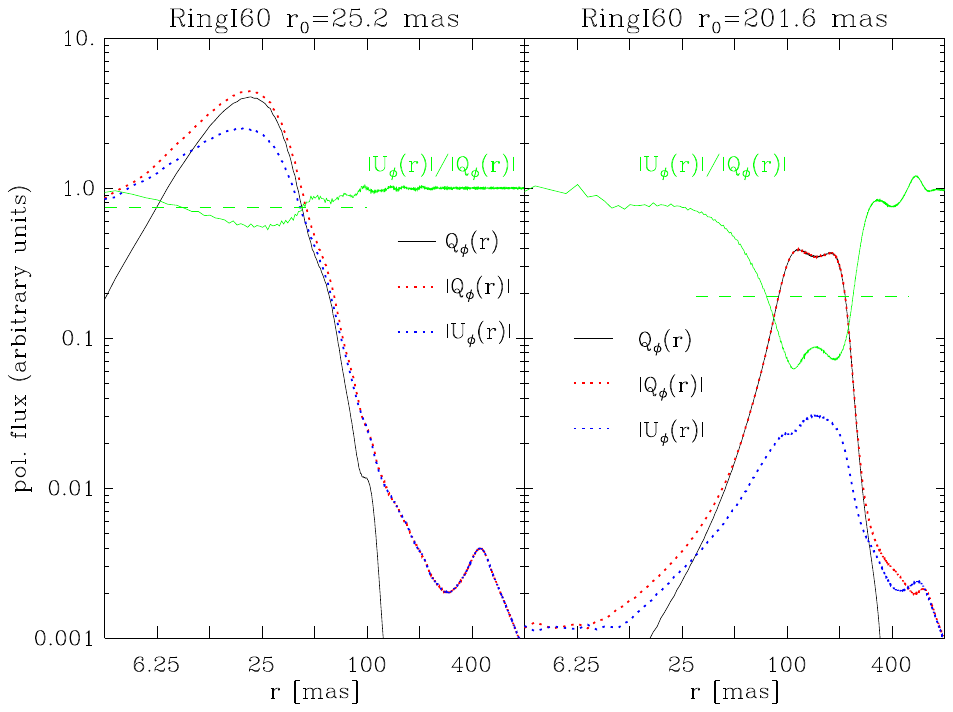}
\caption{Azimuthally averaged profiles for $Q_\phi(r)$, $|Q_\phi(r)|$ and
  $|U_\phi(r)|$ for the RingI60 models with $r_0=25.2$~mas and 201.6~mas
  convolved with PSF$_{\rm AO}$. The green curve for the ratio
  $|U_\phi(r)|/|Q_\phi(r)|$ provides a rough measure for the convolution
  cross talk and the dashed line
  indicates the system integrated value $\Sigma|U_\phi|/\Sigma|Q_\phi|$
  from Table~\ref{TabRingi60}}.
\label{FigRprofRingI60}
\end{figure}

As discussed above, also the PSF convolution can produce very substantial
$U_\phi(x,y)$ signals for models with zero intrinsic
$U_\phi'(x,y)$. This fact was already mentioned by
\cite{Canovas15} but they did not quantify this effect and their
modelling used only a Gaussian PSF for the convolution.

The RingI60 simulations can be used to quantify the 
$Q_\phi\rightarrow U_\phi$ convolution signal for different disk
sizes $r_0$. We use as simple metric for this effect the ratio
$\Sigma|U_\phi|/\Sigma|Q_\phi|$ given in Table~\ref{TabRingi60} (see
also the $\Sigma|U_\phi|$ and $\Sigma|Q_\phi|$ curves in
Fig.~\ref{DiagAperg6i60Ring}). For an unresolved disk
the convolution
gives as extreme limit a ratio  $\Sigma|U_\phi|/\Sigma|Q_\phi|=1$.
The ratio is 0.22 and still high for the PSF$_{\rm G}$
convolved disk with $r_0=50.4$~mas
plotted in Fig.~\ref{FigRingi60}. A larger disk of about $r_0=201.6$~mas
is requird to reach a low ratio of 0.03 so that an intrinsic $U_\phi$-signal
could be detectable. 

The situation is worse for disk models convolved
with PSF$_{\rm AO}$ producing substantially more $Q_\phi\rightarrow U_\phi$
cross-talk and the ratio $\Sigma|U_\phi|/\Sigma|Q_\phi|$ is 0.19 for
$r_0=201.6$~mas and still larger than 0.1 for $r_0=806.4$~mas
The strong smearing of the Stokes $Q$-signal produces in the
halo a ratio $\Sigma|U_\phi|/\Sigma|Q_\phi|\approx 1$.
For a more detailed analysis the radial dependence of the ratio
$|U_\phi(r)|/|Q_\phi(r)|$ should be considered, and
such profiles are shown in Fig.~\ref{FigRprofRingI60} for
PSF$_{\rm AO}$ convolved RingI60 models with
$r_0=25.2$~mas and $201.6$~mas. For the compact disk the
$|Q_\phi(r)|$ signal is only at the separation of the
ring $r\approx 12-35~$~mas substantially larger than 
$|U_\phi(r)|$, but nowhere more than a factor of 2.
For the unresolved polarization near the center and for the smeared
halo signal the ratio is $|U_\phi(r)|/|Q_\phi(r)|\approx 1$.
For the larger disk, the $Q_\phi(r)$ signal dominates the $U_\phi(r)$ cross-talk
signal by about a factor of 20 at the location of the
inclined ring $r\approx 100-200~$~mas.

This indicates that
an intrinsic $U'_\phi$ polarization at the level of $0.05\,Q'_\phi$
is only detectable for large disks
$r_0\gapprox 200~$mas in high quality data for currently
available AO systems, otherwise the convolution cross talks dominate.
Helpful is, that
the expected geometric structure of the observable $U_\phi(x,y)$
signal produced by multiple scattering  
\citep[see][]{Canovas15,Ma22} is different from the $Q_\phi\rightarrow U_\phi$
convolution artefacts, which can even be constrained strongly from the
observed $Q_\phi(x,y)$ signal. In any case, the detection 
of the presence of intrinsic $U'_\phi(x,y)$ polarization requires
high quality data of an extended scattering region and 
a very careful assessment of the convolution cross talk effects.

\paragraph{$U_\phi$ signal and $Q_\phi$ uncertainty.}
In many studies the $U_\phi$ signal is used as a proxy for the
observational uncertainty for the measured $Q_\phi$ signal. This
is a reasonable approach for cases with small $Q_\phi\rightarrow U_\phi$
cross talk, like axisymmetric or close to axisymmetric scattering geometries,
and very extended systems like RingI60 models with $r\gapprox 200$~mas.
The spurious signals introduces by speckle noise, read-out and photon 
noise, and small scale instrumental artefacts in the $U_\phi$ map are
then larger than the convolution cross-talk and therefore also
representative for observational uncertainties in the $Q_\phi$ map.

However, for non-axisymmetric compact scattering regions the
$U_\phi$ signal consist mainly of the well defined
systematic convolution cross-talk signal with high ratios 
$|U_\phi(r)|/|Q_\phi(r)|\gapprox 0.5$ like the RingI60 $r_0=25.2$~mas model
in Fig.~\ref{FigRprofRingI60}. Despite this the azimuthal
polarization signal $Q_\phi$
can still be highly significant, because for high quality data
the observational noise in $Q_\phi$ is much smaller than the systematic
$Q_\phi\rightarrow U_\phi$ cross talk. In such cases another metric
than the $U_\phi$ signal must
be used for the assessment of the observational uncertainties,
like the dispersion of the measured results for different data sets
or a detailed analysis of speckle and pixel to pixel noise in the data.

Nonetheless, a low $|U_\phi|$ signal can be used to identify high
quality data within a series of measurements taken under variable
observing conditions. Because the systematic
cross talk $Q_\phi\rightarrow U_\phi$ is anti-correlated with the quality of
the observational PSF one can select the polarization cycles
with low $|U_\phi|$ and this may provide a higher quality $Q_\phi$
map for a target.

\begin{figure}
\includegraphics[trim=2.5cm 13.0cm 6.0cm 5.7cm, width=8.8cm]{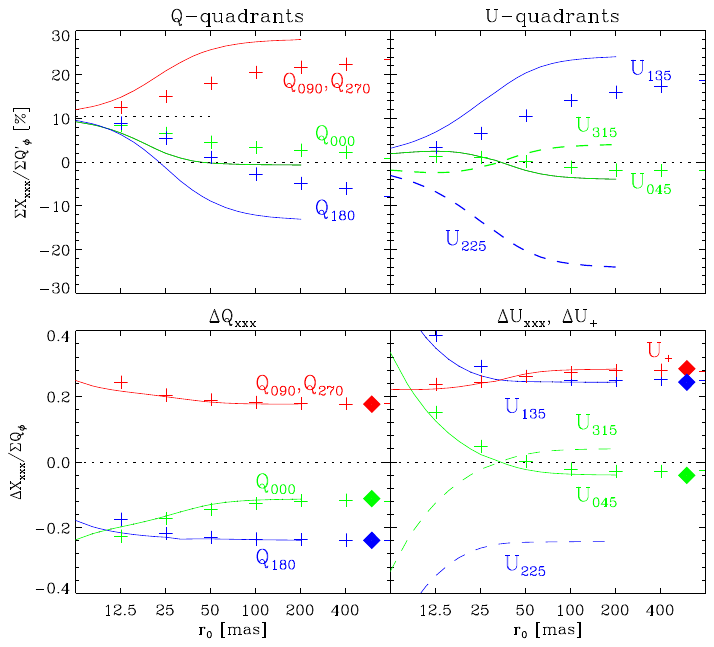}
\caption{Quadrant polarization parameters $\Sigma X_{xxx}$ normalized to the
  intrinsic azimuthal polarization $\Sigma Q'_{\phi}$ (upper panels)
  and differential quadrant parameters $\Delta X_{xxx}$ normalized to
  the convolved $\Sigma Q_\phi$ (lower panels) for Ringi60 models as
  function of radius $r_0$. Full lines give the values for
  PSF$_{\rm G}$ convolved, and crosses for PSF$_{\rm AO}$ convolved models
  while the filled diamonds are the intrinsic values.
  For clarity not all $\Delta U_{xxx}$ values are given.}
\label{DiagQuadg6i60}
\end{figure}

\subsubsection{Convolution and quadrant polarization parameters}
\label{Sect.QuadConv}
The convolution can change for not well resolved RingI60 models strongly
the azimuthal distribution of the polarization signal $Q_\phi(\phi)$
and $U_\phi(\phi)$.
This can be quantified with the quadrant polarization parameters
which provide a simple formalism for the description of polarimetric
convolution effects.
We use the symbol
$X_{xxx}$ for all quadrant parameters, and $Q_{xxx}$ or $U_{xxx}$
for the four Stokes $Q$ or Stokes $U$ quadrants, respectively. The
quadrant values are related to the integrated
Stokes parameters according to
\begin{eqnarray}
\Sigma Q =\Sigma Q_{000} + \Sigma Q_{090} + \Sigma Q_{180} + \Sigma Q_{270}\,,
  \quad {\rm and} \label{Eq.Sum4Quad}\\
  \Sigma U =\Sigma U_{045} + \Sigma U_{135} + \Sigma U_{225} + \Sigma U_{315}\,.
  \quad\phantom{\rm and} 
\end{eqnarray}
For the mirror-symmetric RingI60 models, there is also
$\Sigma Q_{000}=\Sigma Q_{270}$, $\Sigma U_{045}=-\Sigma U_{315}$ and
$\Sigma U_{135}=-\Sigma U_{225}$ \citep{Schmid21}.

The convolution changes the flux in the different Stokes quadrants
because of smearing and mutual cancellation as can be seen in
Fig.~\ref{FigRingi60}. This degradation is also illustrated for
the normalized quandrant parameters $\Sigma X_{xxx}/\Sigma Q'_\phi$
for the PSF$_{\rm G}$ and PSF$_{\rm AO}$ convolved RingI60 models
as function of $r_0$ in the upper panels of Fig.~\ref{DiagQuadg6i60}
(see also Table~\ref{TabRingi60}).
For smaller and less resolved disks all $Q$-quadrants approach the same value
$\Sigma Q_{xxx}=\Sigma Q_d/4$ as expected
for an unresolved system with an integrated polarization $\Sigma Q_d$.
For the Stokes $U$ quadrants the effects are equivalent on
both sides of the $y$-axis, and strong smearing turns the sign of the weaker
back-side quadrants values $U_{045}$ and $U_{135}$ to the sign of the
strong front side quadrants. This produces for compact disks
positive signals for $U_{045}$ and $U_{135}$ ``on the left side'' of
the star, and negative signals for $U_{225}$ and $U_{315}$ ``on the right''
side in (Fig.~\ref{FigRingi60}) for the RingI60 model with
$r_0=12.6$~mas.

\paragraph{Differential quadrant parameters.}
Despite the smearing and polarimetric cancellation the information
about the relative intrinsic strengths of the quadrant parameters
can still be recovered to some degree 
as long as the scattering region is partially resolved.
This is possible, because the mutual compensation of 
the polarization signal changes opposite sign quadrants roughly by similar
amounts (Fig.~\ref{DiagQuadg6i60} upper panel), and in step with
the degradation of the total azimuthal polarization $\Sigma Q_\phi$ shown
in Fig.~\ref{DiagAperg6i60Ring} for the same models.

Therefore, good values to constrain the intrinsic
$\Sigma Q'_{xxx}$ fluxes are the relative differential values
\begin{equation}
  \frac{\Delta Q_{xxx}}{\Sigma Q_\phi}
     =\frac{\Sigma Q_{xxx}-(\Sigma Q/4)}{\Sigma Q_\phi},
\label{Eq.Qdiff}  
\end{equation}
which quantify how much the individual Stokes $Q$ quadrants contribute
to the total Stokes $Q$ signal (Eq.~\ref{Eq.Sum4Quad}), or how
much more or less than the average contribution $Q/4$. According to
Fig.~\ref{DiagQuadg6i60} (lower left and Table~\ref{Tab.diffQuad})
the $\Delta Q_{xxx}/\Sigma Q_\phi$ values
are quite independent of the resolution for
disks with $r_0\geq 50~$mas and one can still recognize for a
RingI60 model disk with $r_0\approx 25~$mas that the front
side deviates more from the average than the back side. They are
also practically the same for a convolution with
PSF$_{\rm G}$ and PSF$_{\rm AO}$ and in very good agreement
with the intrinsic values
\begin{equation}
  \frac{\Delta Q_{xxx}}{\Sigma Q_\phi}\approx
             \frac{\Delta Q'_{xxx}}{\Sigma Q'_\phi}.
\end{equation}  

The equivalent quantities for the Stokes $U$ quadrants are just
ratios $\Sigma U_{xxx}/\Sigma Q_\phi$ because the
average value $\Sigma U/4$ is zero. Figure~\ref{DiagQuadg6i60}
(lower right) shows, that the relative Stokes
$U$ quadrant ratios deviate for small disks with $r_0\lapprox 25~$mas
substantially from a constant. This is caused by the morphology of the Stokes
$U$ map for the RingI60 model,
which has one dominant positive $U_{135}$ and one dominant
negative $U_{225}$ quadrant. Smearing and cancellation affect
the weak quadrants $\Sigma U_{045}$ and $\Sigma U_{315}$ stronger
than the reference value $\Sigma Q_\phi$, while the effects
between the strong components $U_{135}$ and $U_{225}$ are smaller than
for $\Sigma Q_\phi$, because the separation between the strong components
is relatively large.

A useful alternative are the relative differential values between the
$U$ quadrants on the positive or negative $x$-axis side
\begin{equation}
  \frac{\Delta U_+}{\Sigma Q_\phi}
  =\frac{\Sigma U_{135}-\Sigma U_{045}}{\Sigma Q_\phi}
  \quad{\rm and} \quad
  \frac{\Delta U_-}{\Sigma Q_\phi}
  = \frac{\Sigma U_{225}-\Sigma U_{315}}{\Sigma Q_\phi}\,.
\label{Eq.Udiff}  
\end{equation}  
There is $\Delta U_+=-\Delta U_{-}$ because of the
symmetry of the RingI60 models. The ratio
$\Delta U_+/\Sigma Q_\phi$ is almost independent of the disk size
$r_0$ and the used PSF, according to the red line and points in
Fig.~\ref{DiagQuadg6i60}
(lower right) and the values in Table~\ref{Tab.diffQuad},
very similar to the parameters $\Delta Q_{xxx}/\Sigma Q_\phi$. 
This demonstrates that differential quadrant parameters are useful
to push the characterization of
$Q_\phi(\phi)$ and $U_\phi(\phi)$ for circumstellar scattering regions
towards smaller separations.

\begin{figure}
  \includegraphics[trim=0.1cm 0.1cm 0.3cm 0.5cm, width=8.5cm]{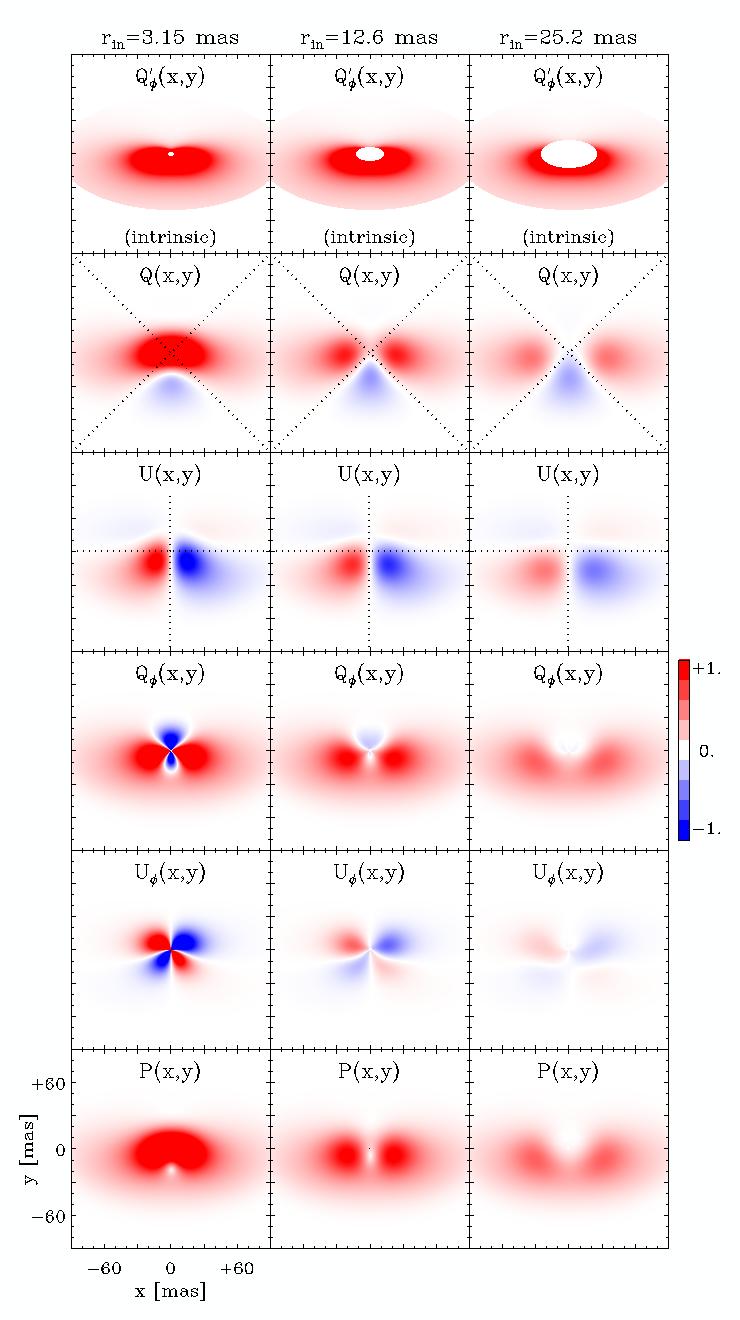}
  \caption{Intrinsic maps $Q'_\phi$ and PSF$_{\rm G}$ convolved maps for
    $Q$, $U$, $Q_\phi$, $U_\phi$, and $P$ for DiskI60$\alpha$-2 models
    with different cavity sizes $r_{\rm in}$. The radius $r_{\rm out}=100.8$~mas and the
    intrinsic surface brightness (SB) profile $Q'_\phi(r)\propto r^{-2}$ are the
    same for the three models.}
 \label{FigDiski60}
\end{figure}

\subsection{Convolution of inclined extended disks.}

Extended disks with unresolved or partially unresolved central
regions are frequently observed \citep[e.g.,][]{Garufi22} and
it is of interest to investigate regions close to the star $r<D_{\rm PSF}$ 
because they correspond to the zone of terrestrial
planet formation $\lapprox 3$~AU for systems in nearby star forming regions.
In particular, the polarization of the unresolved part of the
disk at $r\lapprox 0.5\,D_{\rm PSF}$ ($<12.6$~mas) can be compared with
the polarization of the resolved region $r\gapprox D_{\rm PSF}$ to constrain
the presence or absence of significant changes in the scattering
properties between the two regions.

Polarization maps for the inclined and extended DiskI60$\alpha$-2 models
are plotted in Fig.~\ref{FigDiski60} with radii for the inner cavities of
$r_{\rm in}=0.125\,D_{\rm PSF}$, $0.5\,D_{\rm PSF}$, and $D_{\rm PSF}$,
or $3.15$~mas, 12.6~mas, and 25.2~mas, respectively.
The inner disk rim is very bright for these models, and 
therefore the convolved disk maps look similar to the images in
Fig.~\ref{FigRingi60} for small disk rings. The model with
$r_{\rm in}=D_{\rm PSF}$ in Fig.~\ref{FigDiski60} shows a resolved
$Q_\phi$ disk image with small cross-talk residuals in $U_\phi$,
while the disk with a very small cavitiy of $r_{\rm in}=0.125\,D_{\rm PSF}$ has
strong quadrant patterns for $Q_\phi$ and $U_\phi$ as expected
from the net $Q$-signal of a bright, unresolved central disk region.

\begin{figure}
\includegraphics[trim=2cm 13cm 2.5cm 4.5cm, width=8.8cm]{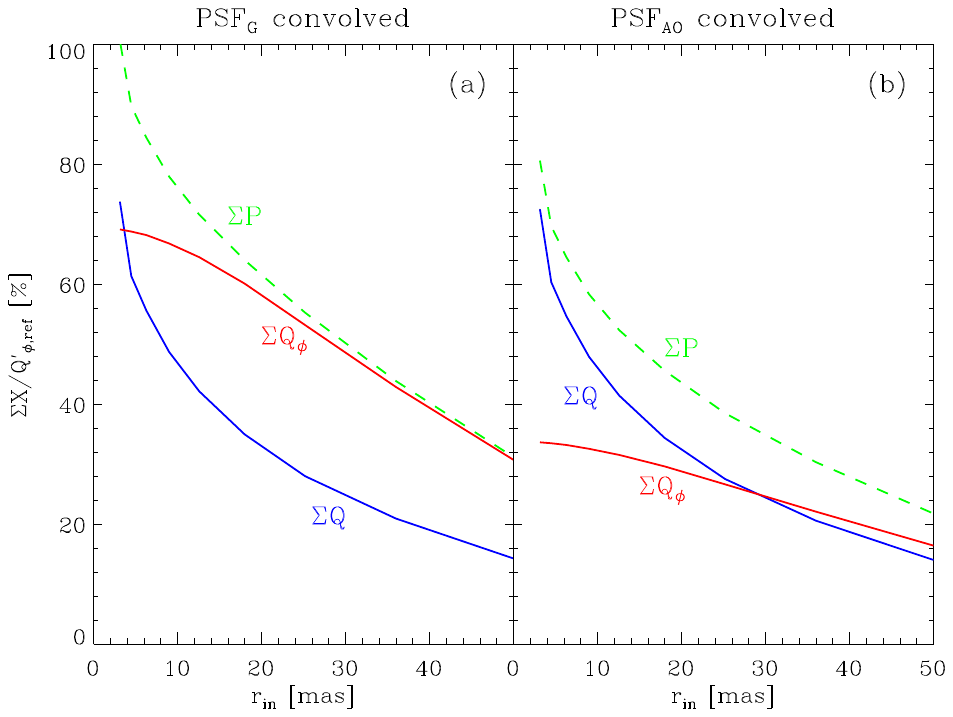}
\caption{Integrated polarization parameters $\Sigma Q_\phi$, $\Sigma P$,
  $\Sigma Q_d$ for the inclined DiskI60$\alpha$-2 model as function
  of the radius of the inner cavity $r_{\rm in}$ and for PSF$_{\rm G}$
  and PSF$_{\rm AO}$ convolution. All values are normalized to the
  intrinsic value $\Sigma Q'_\phi(r_{\rm in}=0.5\,D_{\rm PSF})$
  (or $r_{\rm in}=12.6$~mas).  
  } 
\label{DiagAperg6i60Disks}
\end{figure}

The integrated polarization parameters for the DiskI60$\alpha$-2
models depend strongly on the inner disk radius $r_{\rm in}$
according to Fig.~\ref{DiagAperg6i60Disks} or the numerical values given in  
Table~\ref{TabDiski60}. The parameters are all normalized to
the intrinsic azimuthal polarization
$Q'_{\phi,{\rm ref}}=\Sigma Q'_\phi$ for
the disk with $r_{\rm in}=0.5\,D_{\rm PSF}$.
The intrinsic polarization parameters
$\Sigma Q'_\phi=\Sigma P'$
and $\Sigma Q=\Sigma Q'= 0.421\,\Sigma Q'_\phi$ are much larger for disks
with small inner radii, for example by a factor 1.7
for $r_{\rm in}=0.125\,D_{\rm PSF}$ when compared
to $r_{\rm in}=0.5\,D_{\rm PSF}$. The convolution does not change the
integrated Stokes polarization $\Sigma Q=\Sigma Q'$ but only
redistributes spatially the signal $Q'(x,y)\rightarrow Q(x,y)$
and therefore the $\Sigma Q$ curves are identical in the two panels
of Fig.~\ref{DiagAperg6i60Disks} for the models convolved with
PSF$_{\rm G}$ and PSF$_{\rm AO}$. 

Contrary to this, the PSF convolution reduces or even cancels the
strong $\Sigma Q'_\phi$-signal of the central region and the effect is
more important for PSF$_{\rm AO}$ than for PSF$_{\rm G}$. Therefore, 
$\Sigma Q_\phi$ reaches for a given convolution PSF a limiting
value for models with a cavity smaller than
$r_{\rm in}<0.5\,D_{\rm PSF}$, despite the fact that the intrinsic
$\Sigma Q'_\phi$ and $\Sigma Q'$ increase steeply for
$r_{\rm in}\rightarrow 0$ for the DiskI60$\alpha$-2 model.

The central quadrant patterns in the convolved $Q_\phi$ and $U_\phi$ maps
have a strength proportional to the net $Q$ signal from the
unresolved inner disk region. The central Stokes signal
$\Sigma Q_c\approx \Sigma Q(r\lapprox 0.5\,D_{\rm PSF})$
and $\Sigma U_c$ can be used to constrain the
amount of polarization $P_c$ and the averaged polarization position
angle ($\theta_c=0.5\cdot {\rm arctan2}(\Sigma U_c,\Sigma Q_c)$)
for the unresolved part.
Differences between $\theta_c$ and the averaged polarization position
angle of the resolved signal $\theta_d$ for
$\Sigma Q(r\gapprox 0.5\,D_{\rm PSF})$ can point
to a structural change between the unresolved and
the resolved part of the scattering region, or
to a contribution
from interstellar or instrumental polarization as discussed in the
next section.

The intrinsic profile
$Q'_\phi(r)$ for the DiskI60$\alpha$-2 models is steep and
therefore the ratio of convolved parameters
$\Sigma Q/\Sigma Q_\phi$ 
changes strongly with the radius of the inner
cavity $r_{\rm in}$ (Fig.~\ref{DiagAperg6i60Ring}).
According to
Table~\ref{TabDiski60} the
dependence is smaller for the DiskI60$\alpha$0 and DiskI60$\alpha$-1
models. Nonetheless, $\Sigma Q/\Sigma Q_\phi$ could be a good parameter
to derive from observational data the radius $r_{\rm in}$ of an
unresolved central cavity or other disk properties at $r<D_{\rm PSF}$,
in particular when using constraints about disk inclination
and SB profile from the resolved part of the
disk at $r\gapprox D_{\rm PSF}$.

\begin{figure}
  \includegraphics[trim=0.5cm 0.1cm 0.1cm 0.5cm, width=9cm]{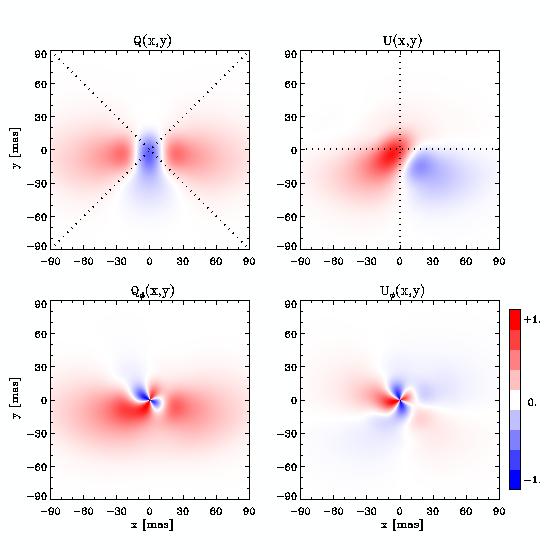}
  \caption{Polarization maps $Q$, $U$, $Q_\phi$, $U_\phi$ for 
    model DiskI60$\alpha$-2 ($r_{\rm in}=0.5\,D_{\rm PSF}$ or 12.6~mas) and a
    polarized central star with $\Sigma P'_s=0.2\cdot\Sigma Q'_\phi$
    and $\theta'_s=67.5^\circ$ convolved with PSF$_{\rm AO}$.}
 \label{FigDiski60Starpol}
\end{figure}

\subsection{The contribution of a polarized central star}
\label{Sect.Polstar}
The simulations presented up to now assume that the central star
is unpolarized ($Q'_s=0$ and $U'_s=0$ in Eqs.~\ref{QIntr},\ref{UIntr}),
and therefore it does not affect the polarization
signal of the circumstellar scattering region.
However, often the stars with resolved circumstellar dust
scattering regions have also unresolved components, as
inferred for example from the thermal emission of hot dust. This
dust can produce for the central, unresolved point-like
source an intrinsic polarization, if scattering occurs in a
non-symmetric structure.
The central star can also be
polarized by uncorrected contributions from interstellar or instrumental
polarization. As the star is typically much brighter than
the resolved circumstellar scattering already a small fractional
polarization of the order $p_s=\Sigma Q_s/\Sigma I_s\approx 0.001$
can have a strong impact on the observable polarization, and the
following Sect.~\ref{Sect.Calib} will address the question on how to
correct for this.

In this subsection we explore the impact of a polarized
central source on the imaging polarimetry of a circumstellar scattering region.
For this, we have to distinguish between the stellar
$Q_s,U_s$ and circumstellar
$Q_d,U_d$ polarization components. We consider a central point
source with a fractional polarization
$p_s=(q_s^2+u_s^2)^{1/2}$ and position angle
$\theta_s=0.5\cdot {\rm atan2}(u_s,q_s)$ defined in disk coordinates
$(x,y)$. The convolved intensity distribution
of the central source is $I_s(x,y)= \Sigma I_s \cdot {\rm PSF}(x,y)$
and the corresponding Stokes parameter maps are $Q_s(x,y) = q_s\,I_s(x,y)$
and $U_s(x,y) = u_s\,I_s(x,y)$. 
The integrated azimuthal polarization signals of the star are zero
$\Sigma Q_{\phi,s}=\Sigma U_{\phi,s}=0$,
but the corresponding convolved signal maps show the quadrant patterns
\begin{eqnarray}
Q_{\phi,s}(x,y)&=& -\, p_s\, I_s(x,y) \, \cos (2(\phi_{xy}-\theta_s)) \\
U_{\phi,s}(x,y)&=& +\, p_s\, I_s(x,y) \, \sin (2(\phi_{xy}-\theta_s)) \,.
\end{eqnarray}  
The strengths of these $Q_{\phi,s}$ and $U_{\phi,s}$ quadrant patterns
are independent of $\theta_s$ but their orientation is defined by
$\theta_s$.
The impact of the polarized intensity of the star
$\Sigma P_s=p_s\, \Sigma I_s$ on the disk polarization map in
convolved data depends of course on the relative strength between 
$\Sigma P_s$ and $\Sigma Q_{\phi,d}$.

Figure~\ref{FigDiski60Starpol} shows as example a disk plus
star system with an intrinsic ratio of
of $\Sigma P'_s/\Sigma Q'_{\phi,d}=0.2$ for a system with disk
intensity $\Sigma I'_d=0.05\, \Sigma I'_s$,
disk polarization $\Sigma Q'_{\phi,d}=0.1\,\Sigma I'_d$, and a
stellar polarization of $\Sigma P'_s=0.001\,\Sigma I'_s$. The
scattering region is simulated with the inclined disk model
DiskI60$\alpha$-2 with $r_{\rm in}=0.5\,D_{\rm PSF}$. The star polarization has
an orientation $\theta_s=67.5^\circ$ ($q_s=-0.71 p_s$, $u_s=+0.71 p_s$)
and the whole system is convolved with PSF$_{\rm AO}$. This
reduces the azimuthal polarization $Q_{\phi,d}$ of the disk
and it results a ratio of 
$\Sigma P_s/\Sigma Q_{\phi,d}=0.635$ for the convolved model.

The polarization of the star has a strong impact on the
polarization maps, despite the fact that $\Sigma P_s$ is weaker than
the circumstellar polarization $\Sigma Q_{\phi,d}$. This is apparent
in the maps in Fig.~\ref{FigDiski60Starpol}, in
particular when compared to the disk maps of Fig.~\ref{FigDiski60} without
stellar polarization. The star produces in the center of the
Stokes $Q(x,y)$ map a strong negative signal, and in $U(x,y)$ the
antisymmetric appearance for the inner disk region is distorted.
For the azimuthal polarization strong quadrant patterns for
$Q_\phi(x,y)$ and $U_\phi(x,y)$ are visible around the position of the central
star. 

\begin{figure}
\includegraphics[trim=2cm 13cm 6.5cm 7.5cm, width=8.8cm]{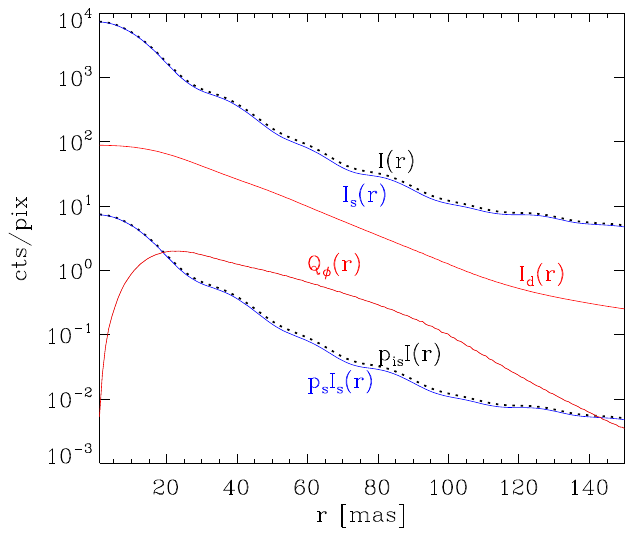}
\caption{Azimuthally averaged intensity profiles for the disk 
  $I_d(r)$, the star $I_s(r)$, and the total $I(r)$, for the disk
  polarization $Q_\phi(r)$, and the intrinsic stellar
  polarization $p_sI_s(r)$ or interstellar polarization $p_{\rm is}I(r)$
  introduced by a fractional offset of $p=0.1$~\%. The same disk model,
  polarization offset $p_sI_s(r)$, and PSF$_{\rm AO}$ convolution as
  in Fig.~\ref{FigDiski60Starpol} is used, while the
  interstellar offset for $p_{\rm is}I(r)$ is applied to the
  total intensity.} 
\label{FigDiskPoffsetProf}
\end{figure}

The azimuthally averaged radial profiles for $p_sI_s(r)$
and $Q_{\phi,d}(r)$ in Fig.~\ref{FigDiskPoffsetProf} show that
the stellar polarization dominates strongly for
$r<20$~mas, where the stellar intensity $I_s(r)$ is much higher than the
disk intensity $I_d(r)$, while the disk polarization is the main
polarization component in the range $r\approx 30$ to 120~mas.

There are three important properties for the interpretation of
the polarization signal: (i) the integrated Stokes parameters do not
depend on the PSF convolution and therefore the disk and
star signal are just adding up 
\begin{eqnarray}
\Sigma Q = \Sigma P_s \cos(2\theta_s) + \Sigma Q_d\,, \\
\Sigma U = \Sigma P_s \sin(2\theta_s) + \Sigma U_d\,; 
\end{eqnarray}  
(ii) $\Sigma Q_\phi$ and $\Sigma U_\phi$ do not depend on the
polarization signal of an unresolved central object,
because the introduced quadrant patterns add no net signal
$\Sigma Q_{\phi,{\rm s}}=0$ and $\Sigma U_{\phi,{\rm s}}=0$; 
(iii) the orientation of the central $Q_\phi$ and $U_\phi$ quadrant patterns
are defined by $\theta_s$ of the central source.

If the polarization of the central source is aligned with the
resolved disk $\theta_s=\theta_d$, then we expect a left right
symmetry $Q_\phi(x,y)=Q_\phi(-x,y)$ for the overall Stokes $Q_\phi$ map and
an antisymmetry $U_\phi(x,y)=-U_\phi(-x,y)$ for Stokes $U_\phi$ like
for the partially resolved disks shown in Fig.~\ref{FigDiski60}. Deviations
from these properties are indicative for a more complicated polarization
structure with a stellar polarization component not aligned with the
disk polarization.

\paragraph{Interstellar and instrumental polarization.}
The impact of polarization offsets from interstellar $p_{\rm is}I(x,y)$
or instrumental $p_{\rm inst}I(x,y)$ polarization are proportional
to the total intensity, while an offset from an
intrinsically polarized star $p_sI_s(x,y)$ is proportional to
the PSF of the star. However, for many typical cases one can approximate
$p_s\,I_s(x,y) \approx p_s\,I(x,y)$ because the
difference between stellar and total intensity, which is equivalent
to the scattered intensity from the dust $I_d(x,y)$,
is very small for circumstellar scattering regions.

This is is supported by the azimuthal profiles in
Fig.~\ref{FigDiskPoffsetProf} for the disk plus star
model described above (Fig.~\ref{FigDiski60Starpol}).
In this example the disk polarization in the range
$r\approx 30$ to 120~mas is much larger than
the difference between the polarization offsets of $p=0.001$
for the total intensity and the stellar intensity
$Q_\phi(r)\gg(p(I(r)-I_s(r))$. Therefore
it makes for this example practically no difference for the
disk polarization signal $Q_\phi(x,y)$
whether one uses $p\,I(x,y)$ or $p\,I_s(x,y)$ for the
polarization offset correction. 

This approximation is also valid for faint circumstellar
scattering regions, if the fractional scattering polarization is
at the same level $\Sigma Q_\phi\approx 0.1\, \Sigma I_d$ as in the example
shown above, because the difference for the two cases of polarization
offsets scales still with $\Sigma I_d$ according to
$p\,(I(r)-I_s(r))=p\,I_d(r)$. The approximation $I\approx I_s$ is
less good for a bright scattering region $I_d$, in particular
if the polarization offsets is large $p\gapprox 0.01$ and the scattering
region weakly polarized $\Sigma Q_\phi/\Sigma I_d\lapprox 0.1$, and
combination of these cases.

We treat in this study an intrinsic stellar polarization like
an interstellar polarization offset.
Interstellar and instrumental polarization offsets
can be corrected using $p_{\rm is}\, I(x,y)$ or $p_{\rm inst}\, I(x,y)$
with the advantage, that the system intensity $I(x,y)$ can be
derived from unsaturated observations of the target.
In the best case $I(x,y)$ is
obtained simultaneously with the polarization signal $Q(x,y)$ and $U(x,y)$
so that atmospheric variations can be taken accurately into account
\citep[e.g.,][]{Tschudi21}.
Using the system intensity $I(x,y)$ also as approximation for the
correction of an intrinsic stellar polarization
$p_s\, I(x,y)\approx p_s\,I_s(x,y)$ overestimates slightly the
offset, but avoids the difficult procedure of splitting 
$I(x,y)$ into a stellar PSF component $I_s(x,y)$ and a disk
component $I_d(x,y)$. Using $I_s(x,y)$ instead of $I(x,y)$
would provide in many cases only a marginal
improvement for the offset correction when considering other PSF
calibration issues in high contrast imaging polarimetry.

\section{Polarimetric calibration and zp-correction}
\label{Sect.Calib}

\subsection{The impact of a fractional polarization offset}
\label{Sect.Poloffset}
Already a small fractional polarization offset of
the order $p \approx 0.1~\%$ from an intrinsic polarization
of the central star, from interstellar polarization, or from
instrumental polarization can have a strong impact on the
observed polarization maps of circumstellar scattering regions
(Fig.~\ref{FigDiski60Starpol}).
Unfortunately, the different polarization effects are often not 
well known and after a calibration there can remain 
non-negligible polarization residuals $p_{\rm res}$ with an arbitrary
position angle $\theta_{\rm res}$, or residual fractional Stokes values $q_{\rm res}$ and $u_{\rm res}$
\begin{eqnarray}
  Q_{\rm obs}(\alpha,\delta)\approx Q(\alpha,\delta)
            + q_{\rm res}\,I(\alpha,\delta) \label{EqPolCompQ}\,, \\
  U_{\rm obs}(\alpha,\delta) \approx U(\alpha,\delta)
  + u_{\rm res} \, I(\alpha,\delta)\,.\label{EqPolCompU}
\end{eqnarray}
This can also be expressed by integrated parameters
\begin{equation}
  \Sigma Q_{\rm obs} = \Sigma Q+q_{\rm res}\Sigma I \label{EqSumPolCompQ}
\end{equation}
and equivalent for Stokes $U$.
The residual offset $p_{\rm res}\,\Sigma I$ can strongly disturb or
even dominate the weak polarization signal $Q_d,U_d$ of the circumstellar
scattering region. The standard method to improve the situation
is a polarimetric zp-correction, which cancels the integrated Stokes signals
\begin{eqnarray}
  \Sigma Q^{\rm z}
  = \Sigma Q_{\rm obs} - \langle q_{\rm obs}\rangle \, \Sigma I = 0 \,,
  \label{Eq.SumNormObs}\\
         \Sigma U^{\rm z}
          = \Sigma U_{\rm obs} - \langle u_{\rm obs} \rangle \, \Sigma I = 0 \,,
\end{eqnarray}
in a certain apperture $\Sigma$ using the fractional Stokes parameters
$\langle q_{\rm obs} \rangle = \Sigma Q_{\rm obs}/\Sigma I$ and
$\langle u_{\rm obs} \rangle = \Sigma U_{\rm obs}/\Sigma I$
\citep[e.g.,][]{Quanz11,Avenhaus14}.
This is a powerful method to find the circumstellar polarization
component in data dominated by residual interstellar, instrumental,
or stellar polarization offsets.

The zp-correction is also very useful for data, where a
circumstellar polarization is detected, but where the signal
might be affected by an unknown offset $q_{\rm res}$. Because
$\langle q_{\rm obs} \rangle = \Sigma Q/\Sigma I+q_{\rm res}$ according to
Eq.~\ref{EqSumPolCompQ} and using Eq.~\ref{Eq.SumNormObs} gives
\begin{equation}
\Sigma Q^{\rm z} = \Sigma Q - \langle q \rangle\, \Sigma I =0 \label{Eq.SumNorm}
\end{equation}  
with $\langle q \rangle=\Sigma Q/\Sigma I$ and equivalent for
$\Sigma U^{\rm z}$. This means, that the zp-corrected signals
of a Stokes map
with a fractional polarization offset, e.g. because of calibration
uncertainties, is equal to the zp-corrected signal of the
map without offsets
like for perfectly calibrated data. Therefore, the zp-correction
provides data with a well defined offset correction, which can
be re-calibrated later
$Q^{\rm z}(\alpha,\delta)\rightarrow Q(\alpha,\delta)$ in a second step
once the intrinsic offset value $\langle q \rangle$ is known from
more accurate polarimetry or constraints from scattering models.
For example, for a given object the same zp-corrected Stokes signals
$\Sigma Q^{\rm z}, \Sigma U^{\rm z}$ should result for data affected by different
instrumental polarization offsets $p_{\rm res}$. The Stokes maps
$Q^{\rm z}(\alpha,\delta),\,U^{\rm z}(\alpha,\delta)$ still depend on
the observational PSF. 

The zp-correction has practically no impact on the integrated
azimuthal polarization $\Sigma Q_\phi$, and there is
\begin{equation}
\Sigma Q_\phi^{\rm z} \approx \Sigma Q_\phi\,.
\label{Eq.QphiStability}
\end{equation}  
because a fractional Stokes polarization offset introduces
for the $Q_\phi(\alpha,\delta)$ and $U_\phi(\alpha,\delta)$ maps
positive-negative quadrant patterns with a zero net
signal. Therefore, $\Sigma Q_\phi$ is a very robust
quantity, which is hardly affected by polarization offsets
introduced by calibration uncertainties. However, it is
important to note that
the zp-correction changes the signal distribution in the
Stokes maps $Q^{\rm z}(\alpha,\delta)$, and
$U^{\rm z}(\alpha,\delta)$ and this can produce
a strong bias for the measured azimuthal
distribution of the $Q^{\rm z}_\phi(\phi)$ polarization
which must be considered for the interpretation of zp-corrected data.

\subsection{Cases without zp-correction bias}
\label{Sect.Nobias}
Polarimetric calibration uncertainties can be corrected with a
zp-correction
without introducing a bias, if the central source is intrinsically
unpolarized ($\Sigma Q'_s\approx 0,\,\Sigma U'_s\approx 0$ or
$\Sigma Q_d=\Sigma Q,\,\Sigma U_d=\Sigma U$) and can be used as
zero polarization reference. The zp-correction value
$\langle q_1 \rangle = \Sigma_1Q/\Sigma_1 I$ derived for a small stellar
aperture $\Sigma_1$ accounts then for the
polarization offsets $q_{\rm res}$ introduced by
the interstellar and instrumental polarization 
according to
\begin{equation}
  Q^{\rm z1}(\alpha,\delta)
  = Q(\alpha,\delta)+q_{\rm res}I(\alpha,\delta)
          -\langle q_1\rangle\, I(\alpha,\delta) \approx Q(\alpha,\delta)
\label{Eq.StarCal}     
\end{equation}
and equivalent for Stokes $U^{\rm z1}$.
This approximation assumes that the scattering region does not
contribute to the polarization signal in $\Sigma_1$.

The central star cannot be used as calibration source for
coronagraphic observations. However, if the object has an axisymmetric
scattering geometry with zero or close to zero net polarization
$\Sigma_2 Q\approx 0$ and $\Sigma_2 U\approx 0$,
measured in an annular aperture $\Sigma_2$,
then this signal can be used as zero polarization
reference. Good examples for such axisymmetric systems are
circumstellar disks seen pole-on, like for TW Hya,
HD 169142 or RX J1604 \citep{Rapson15,vanBoekel17,Poteet18,Tschudi21,Ma23}.

\subsection{Bias introduced by the polarimetric zp-correction}
\label{Sect.Bias}

The zp-correction is usually applied to the
$Q(\alpha,\delta)$ and the $U(\alpha,\delta)$ frames in the sky
coordinate system. We investigate the zp-correction effects
in the $(x,y)$ coordinate system of inclined disks, because the
impact of polarization offsets does not depend on the orientation
of the selected coordinate system.
This simplifies the comparison of
$Q^{\rm z}(x,y),\,U^{\rm z}(x,y)$
with the signal of the convolved models $Q(x,y),\,U(x,y)$ representing
perfectly calibrated data.
The corrected signal depends on the
selected correction region $\Sigma_{\rm zp}$ and different
cases are considered including the zp-correction of coronagraphic
data, or systems with partly unresolved disks. 

\begin{figure}
  \includegraphics[trim=0.5cm 0.1cm 0.1cm 0.1cm, width=9cm]{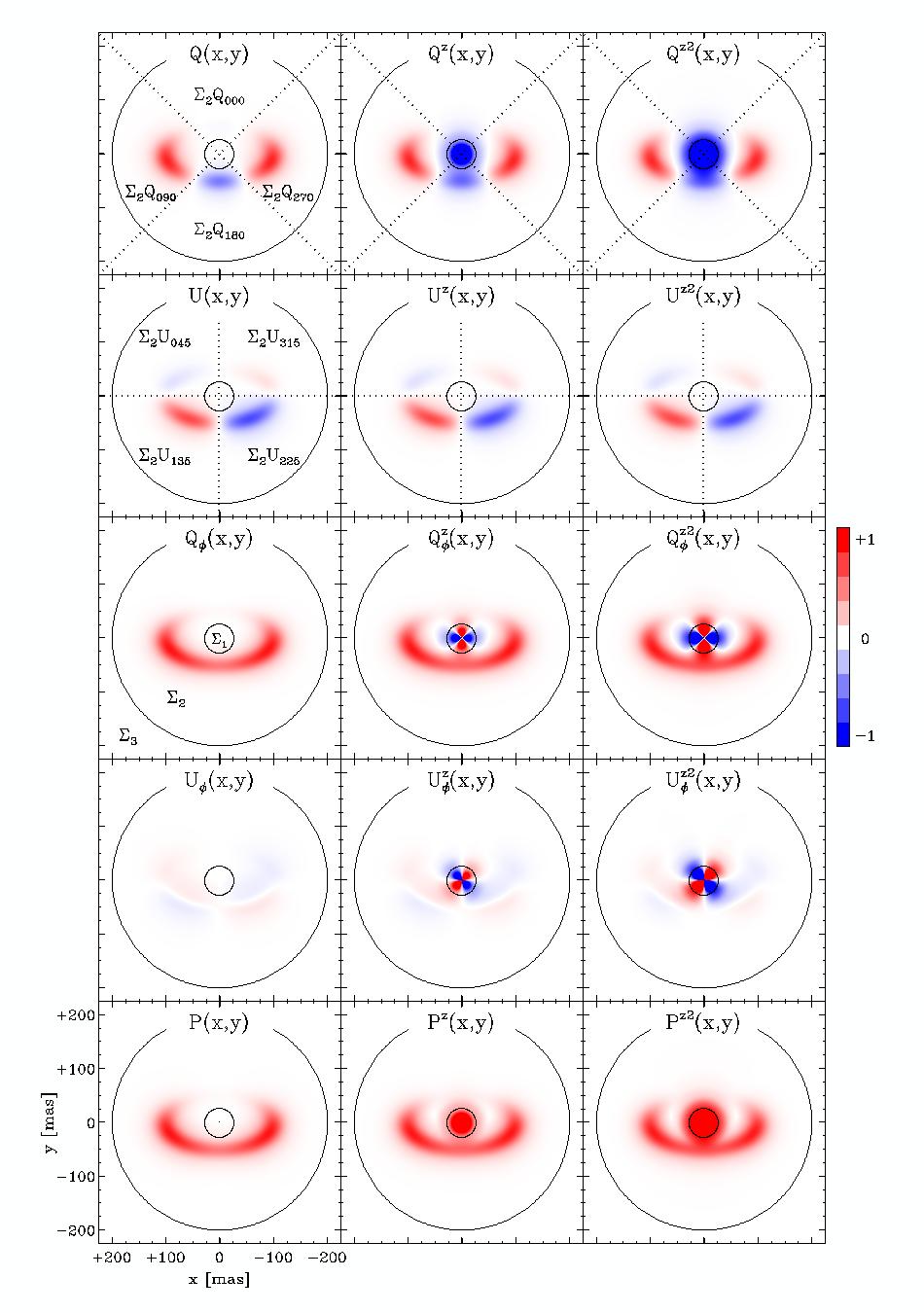}
  \caption{Polarization maps for the RingI60 model
    with $r_0=100.8~$mas convolved with PSF$_{\rm AO}$. The columns
    shows from left to right the convolved maps $X(x,y)$, the maps
    $X^{\rm z}(x,y)$ for a system zp-correction, and $X^{\rm z2}(x,y)$
    for disk zp-correction. The circles describe the integration
    regions $\Sigma_1$, $\Sigma_2$, and $\Sigma_3$ as indicated in the
    $Q_\phi(x,y)$ panel and quadrant regions $\Sigma_2 X_{xxx}$ are
    identified in the $Q(x,y)$ and $U(x,y)$ panels.}
 \label{Fig.ConvNorm}
\end{figure}

\subsubsection{ZP-correction for an inclined disk ring model}
\label{Sect.NormPolRing}
We consider the inclined model RingI60 with $r_0=100.8$~mas
in $(x,y)$-coordinates convolved with the extended PSF$_{\rm AO}$,
apply a zp-correction, and compare the corrected Stokes signal
$Q^{\rm z}(x,y)$ with the corresponding disk signal $Q(x,y)$ without
offsets. An overview on the used polarization parameters
for zp-corrected models is given in Table~\ref{ParXYTabzp}.  

The zp-correction offset derived from a large aperture
$\Sigma_{\rm zp} = \Sigma$ with a radius of $r= 1.5''$ 
yields for the corrected signal integrated in the same aperture 
$\Sigma Q^{\rm z} = \Sigma Q - \langle q\rangle \,\Sigma I = 0$
according to Eq.~\ref{Eq.SumNorm}. This procedure compensates
the positive disk polarization $\Sigma Q$ by a negative offset
$-\langle q \rangle\,\Sigma I$. 
The spatial distribution of the Stokes $Q$ signal differs between the
zp-corrected and the initial signals by
\begin{equation}
\Delta Q(x,y)=Q^{\rm z}(x,y) - Q(x,y) = - \langle q \rangle \,I(x,y)\,.
\label{Eq.QNorm}
\end{equation}
For Stokes $U$ there is $U^{\rm z}(x,y)=U(x,y)$ for the RingI60 models 
because $\Sigma U=0$ and therefore $\langle u\rangle=0$. Thus, a
Stokes $Q$ zp-correction has no impact on the Stokes $U$ signal and
contrariwise. Therefore the zp-correction effects described for
Stokes $Q$ can be generalized to an offset with $Q$ and $U$ components.

The zp-correction effects are illustrated in Fig.~\ref{Fig.ConvNorm}
with RingI60 maps of the polarization parameters
$X=\{Q,\,U,\,Q_\phi,\,U_\phi,\,P\}$ for the convolved model $X(x,y)$,
the $\Sigma_{\rm zp}=\Sigma$ corrected model $X^{\rm z}(x,y)$, and
the $\Sigma_2$ corrected case $X^{\rm z2}(x,y)$ to be discussed
later (Sect.~\ref{Sect.NormCoro}).
For Stokes $Q$ a strong negative signal is introduced in the
$Q^{\rm z}(x,y)$ map at the position of the star, and because of the
extended PSF$_{\rm AO}$, there are also negative contributions further
out. Therefore, the corrected polarization $Q^{\rm z}$ is also lower 
than $Q$ at the location of the disk ring,
while there is no difference between $U^{\rm z}(x,y)$ and 
$U(x,y)$. The zp-corrected azimuthal polarizations $Q^{\rm z}_\phi(x,y)$
and $U^{\rm z}_\phi(x,y)$ show the additional quadrant patterns for
a convolved central point source 
with a negative Stokes $Q$ polarization. For the polarized intensity
$P$ the offset $-\langle q \rangle\,\Sigma I$ adds predominantly
a positive component to the $P^{\rm z}(x,y)$ map.

The impact of the zp-correction in the Stokes polarization maps depends
on the separation from the center and therefore we consider besides
the total system integration region $\Sigma_{\rm int}=\Sigma$ also
three radial integration regions $\Sigma_1$, $\Sigma_2$,
and $\Sigma_3$ indicated in Fig.~\ref{Fig.ConvNorm} in the $Q_\phi(x,y)$ panel.
They represent the star, the disk and the halo regions, respectively, and there
is $\Sigma=\Sigma_1+\Sigma_2+\Sigma_3$.

In the following a detailed description for the system corrected map
$Q^{\rm z}(x,y)$ in Fig.~\ref{Fig.ConvNorm}
is given. The negative signal $\Sigma_1Q^{\rm z}$ in the stellar aperture
$\Sigma_1$ with $r < 0.027''$ introduced by the
zp-correction can be estimated using
$\Sigma_1Q^{\rm z}=\Sigma_1Q-\langle q\rangle\,\Sigma_1I$,
which is Eq.~\ref{Eq.SumNorm} applied to the integration region $\Sigma_1$.
The disk contribution is very small $\Sigma_1 Q\approx 0$ and 
the PSF$_{\rm AO}$ convolved intensity $\Sigma_1 I$ is about
40~\% of the total intensity $\Sigma I$
(see Table~\ref{Tab.ConvNorm} for accurate values)
and we obtain 
\begin{equation}
  \Sigma_1 Q^{\rm z} \approx - \langle q \rangle\, \Sigma_1 I
  \approx -0.4\,\Sigma Q\,
\label{Eq.Sum1Q}  
\end{equation}  

In the halo region $\Sigma_3$ ($0.2''<r<1.5''$)
there is roughly everywhere the same fractional polarization
$Q(x,y)/I(x,y)\approx \Sigma_3 Q/\Sigma_3I\approx\langle q \rangle$
for the convolved model,
because far from the center the PSF$_{\rm AO}$ smearing of the
intrinsic Stokes signal
$Q'(x,y)$ of a compact disk is very similar
to the smearing of the total intensity $I'(x,y)$ dominated by
the central star.
Therefore, the halo signal is practically
cancelled by the zp-correction and Eq.~\ref{Eq.SumNorm} for
$\Sigma_3$ can be approximated by 
\begin{equation}
  \Sigma_3 Q^{\rm z}\approx 0 \,.
\label{Eq.Sum3Q}
\end{equation} 

Of importance for the analysis of the disk polarization is the annular
aperture $\Sigma_2$ with $0.027''<r<0.2''$ covering the
disk and there is (using Eq.~\ref{Eq.SumNorm} for $\Sigma_2$)
\begin{equation}
\Sigma_2 Q^{\rm z} = \Sigma_2 Q-\langle q \rangle\, \Sigma_2 I \,.
\label{Eq.Sum2Q}
\end{equation}
Because $\Sigma Q^{\rm z}=0$ for the entire system $r<1.5''$, and
$\Sigma_3 Q^{\rm z}\approx 0$ for the halo, the polarization
for the zp-corrected disk region $\Sigma_2 Q^{\rm z}$ is equal to
the the negative signal of the central star or
\begin{equation}
  \Sigma_2 Q^{\rm z}\approx - \Sigma_1 Q^{\rm z}\approx
              \langle q \rangle\, \Sigma_1 I\,,
\end{equation}  
where the second relation is equal to Eq.~\ref{Eq.Sum1Q}.
The  Stokes $Q$ polarization is after zp-correction
$\Sigma_2 Q^{\rm z}\approx +0.44~\Sigma Q$,
which is significantly lower than for 
the non-corrected disk $\Sigma_2 Q\approx +0.71~\Sigma Q$
(Table~\ref{Tab.ConvNorm}). Thus, the zp-correction introduces
for the PSF$_{\rm AO}$ convolved RingI60 model a strong bias for Stokes $Q$, 
  changing also the angular distribution of the azimuthal 
polarization $Q^{\rm z}_\phi(\phi)$.

\begin{figure}
  \includegraphics[trim=2.2cm 12.8cm 6.5cm 3.0cm, width=8.8cm]{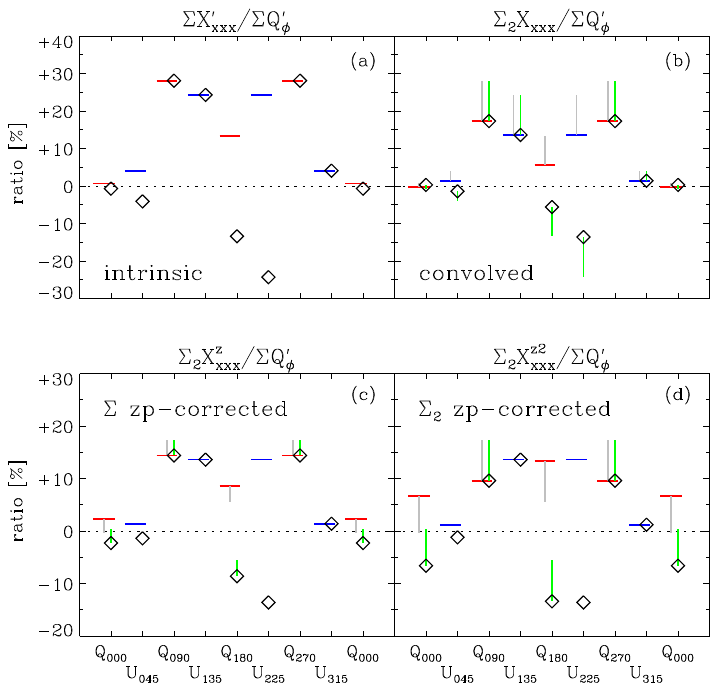}
  \caption{Relative quadrant polarization parameters $\Sigma_2X_{xxx}$
    ($\diamondsuit$) and their azimuthal values $\Sigma_2X_{xxx}|_\phi$
    (horizontal bars) for the disk integration region $\Sigma_2$ for
    RingI60 with $r_0=100.8$~mas: (a)
    instrinsic model, (b) convolved model, (c) 
    $\Sigma$ corrected, and (d) $\Sigma_2$ corrected models.
    The green and grey lines in (b) indicate the change
    of $\Sigma_2X_{xxx}$ and $\Sigma_2X_{xxx}|_\phi$, respectively, with respect
    to panel (a), and in
    the panels (c) and (d) with respect to panel (b). All vallues are
    normalized to $\Sigma Q'_\phi$.}
 \label{Fig.QuadNorm}
\end{figure}

\subsubsection{ZP-correction and quadrant polarization parameters}
\label{Sect.NormQuad}

The intrinsic angular distribution of the azimuthal polarization
$Q'_\phi(\phi)$ is changed in polarimetric observations, first by
the instrumental convolution as described by the
quadrant parameters $\Sigma X_{xxx}$ in Sect.~\ref{Sect.QuadConv},
and, if applied, also by the polarimetric zp-correction. 
A Stokes $Q$ zp-correction adds a signal
$-\langle q \rangle\,I(x,y)$, and because we adopted only axisymmetric
PSFs and approximate $I(x,y)\approx I_s(x,y)$, this introduces
for all Stokes quadrants $\Sigma_{\rm int} Q_{xxx}$ practically the same offset.
These simple dependencies of the quadrant parameters is very
useful to describe the convolution and zp-correction effects 
of the observed azimuthal polarization signal $Q_\phi(\phi)$, which
contains important information about the scattering geometry
and the dust scattering properties. 

The impact of the convolution and zp-correction are illustrated
in Fig.~\ref{Fig.QuadNorm} (upper panels) for the RingI60
($r_0=100.8$~mas) model with the quadrant polarization parameters
$\Sigma_2 X_{xxx}$ (diamonds) 
measured inside the integration region $\Sigma_2$ extending from
$r=27$~mas to $200~$mas as defined in Fig.~\ref{Fig.ConvNorm}
(panels $Q(x,y)$ and $U(x,y)$). The quadrant values $X_{xxx}$
represent different azimuthal parts $\phi_{xxx}$ of the polarizaton
signal $Q_\phi(\phi)$, but
one needs also to consider whether the sign of $X_{xxx}$ corresponds to
a positive or negative contributions to the azimuthal polarization $Q_\phi$.
Therefore, we define azimuthal quadrant values $\Sigma_i X_{xxx}|_\phi$, which
account for the sign of the Stokes polarization with respect
to $Q_\phi(\phi_{xxx})$ as follows:
\begin{itemize}
\vspace{-0.2cm}
\item{} for the Stokes $Q$ quadrants
  $\Sigma_iQ_{000}|_\phi=-\Sigma_i Q_{000}$, $\Sigma_i Q_{090}|_\phi=+\Sigma_i Q_{090}$, $\Sigma_i Q_{180}|_\phi=-\Sigma_i Q_{180}$,
  and $\Sigma_i Q_{270}|_\phi=+\Sigma_i Q_{270}$,
\smallskip  
\item{} and for the Stokes $U$ quadrants
  $\Sigma_i U_{045}|_\phi=-\Sigma_i U_{045}$, $\Sigma_i Q_{135}|_\phi=+\Sigma_i U_{135}$, $\Sigma_i U_{225}|_\phi=-\Sigma_iU_{225}$,
  and $\Sigma_i U_{315}|_\phi=+\Sigma_i U_{315}$. 
\end{itemize}
\vspace{-0.2cm}
The $\Sigma_i X_{xxx}|_\phi$ values
are plotted in Fig.~\ref{Fig.QuadNorm} as horizontal bars. For the
intrinsic disk model they are just equal to the absolute value of the
quadrant values $\Sigma X'_{xxx}|_\phi=\Sigma |X'_{xxx}|$, but this
is not always the case for convolved and corrected models.
From the change of the $\Sigma_i X_{xxx}|_\phi$ values for the
different cases one can estimate the change of the
azimuthal distribution of $Q_\phi(\phi)$ for angles $\phi_{xxx}$
representative for a particular quadrant.

The convolution changes the relative strengths of the quadrants
as already described in Sect.~\ref{Sect.QuadConv} for the total system
integration region $\Sigma$. In this section, we concentrate
on the quadrant signals in the disk integration region $\Sigma_2$
of the model RingI60 $(r_0=100.8~$mas)
and in Fig.~\ref{Fig.QuadNorm} (panel b) the differences between
the intrinsic and the convolved quadrant values $\Sigma_2X_{xxx}$
(diamonds) are illustrated by the vertical green lines and for
the azimuthal quadrant values $\Sigma_2X_{xxx}|_\phi$ (bars)
by grey lines.
The mutual cancellation and the smearing of signal into the
halo region $\Sigma_3$ reduces the signal in $\Sigma_2$ for the positive
quadrants and enhances it for the negative quadrants and this
corresponds to a substantial reduction of the absolute quadrant values
$|\Sigma_2 X_{xxx}|$ and
of the corresponding azimuthal values $\Sigma_2 X_{xxx}|_\phi$ in step
with reduction for $\Sigma Q_\phi$ (see also Table~\ref{Tab.ConvNorm}) .

The zp-correction compensates the disk polarization
$\Sigma_2 Q$ by adding a negative signal $-\Sigma_2 Q $
with a distribution like $I(x,y)$
and this reduces the polarization in $\Sigma_2$ by
$\Delta_2 Q= \Sigma_2 Q^{\rm z} - \Sigma_2 Q=-\langle q \rangle \, \Sigma_2 I$ 
according to Eq.~\ref{Eq.Sum2Q}. For the RingI60 model with
$r_0=100.8$~mas the effect is equal to $\Delta_2 Q/\Sigma Q'_\phi = -11.2~\%$
(Table~\ref{Tab.ConvNorm}). All the Stokes $Q$ 
quadrant values are changed by roughly the same amount
\begin{equation}
\Sigma_2 Q^{\rm z}_{xxx}\approx \Sigma_2 Q_{xxx}+(\Delta_2 Q)/4\,,
\label{Eq.QuadNorm}  
\end{equation}
with $(\Delta_2 Q/4)/\Sigma Q'_\phi\approx -2.8~\%$.
The Stokes $U$ quadrants are not changed, because $\Sigma_2 U=0$
for the used model. Also the integrated azimuthal polarization
$\Sigma_2 Q_\phi$ is practically not changed by the zp-correction 
offset.

Because of the correspondence between $Q_\phi(\phi)$ for different
$\phi$-wedges and the azimuthal quadrant parameters,
the negative $\Sigma_2 Q/4$ contribution
is equivalent to a positive contribution for $\Sigma_2 Q_{000}|_\phi$
and $\Sigma_2 Q_{180}|_\phi$, and a negative contribution
for $\Sigma_2 Q_{090}|_\phi$ and $\Sigma_2 Q_{270}|_\phi$ as illustrated
by the grey vertical lines in panel (c) of Fig.~\ref{Fig.QuadNorm}.
This zp-correction enhances
the relative signal of the disk front side by more than 50~\% or from
$\Sigma_2 Q_{180}|_\phi/\Sigma Q'_\phi=5.6~\%$ to
$\Sigma_2 Q^{\rm z}_{180}|_\phi/\Sigma Q'_\phi=8.6~\%$ between
the ``non-corrected'' and the {\rm zp-corrected} disk maps
(Table~\ref{Tab.ConvNorm}). If one compares the ratio
$\Sigma_2 Q_{180}|_\phi/\Sigma_2 Q_{090}|_\phi$ between
the signal on the front side with respect to the signal
in the left or right quadrants, which are reduced by the zp-correction 
then the initial ratio is boosted from
$\Sigma_2 Q_{180}|_\phi/\Sigma_2 Q_{090}|_\phi=0.32$ to
$\Sigma_2 Q^{\rm z}_{180}|_\phi/\Sigma^{\rm z}_2 Q_{090}|_\phi=0.60$, or almost
a factor of two. This example illustrates that the
azimuthal distribution of the
polarization signal $Q_\phi(\phi)$ can be very significantly changed
by a polarimetric zp-correction.

\subsubsection{ZP-correction for coronagraphic observations}
\label{Sect.NormCoro}

Annular appertures must be used for the zp-correction of
coronagraphic observations or data with detector saturation
at the position of the bright central star. Thus, one needs to
consider other zp-correction regions than $\Sigma_{\rm zp}=\Sigma$ for the
whole system. The resulting
zp-corrected maps, integrated Stokes parameters, or quadrant parameters
are identified with a superscript like 
$Q^{\rm z2}$, $Q^{\rm z3}$ or $Q^{\rm z2+3}$ for zp-corrections
applied to the annuli
$\Sigma_{\rm zp}=\Sigma_2$, $\Sigma_3$ or $\Sigma_{2+3}$ representing
the disk region, the halo region, or the disk plus halo region, respectively.
Table~\ref{Tab.ConvNorm} gives an overview on how
the polarization parameters for different integration regions
$\Sigma_i$ depend on the used correction region $\Sigma_{\rm zp}$ for the
model RingI60 with $r_0=100.8~$mas. Important are the resulting
polarization values for the annular aperture $\Sigma_2$ which
includes the disk ring.

The zp-correction applied to a given annulus $\Sigma_{\rm zp}=\Sigma_i$ sets
the corrected Stokes value in this region $\Sigma_i Q^{{\rm z}i}$ to zero
according to
$\Sigma_i Q^{{\rm z}i} = \Sigma_i Q - \langle q_{{\rm z}i} \rangle \, \Sigma_i I=0$,
which is a generalization of Eq.~\ref{Eq.SumNorm}.
Selecting different zp-correction regions introduces different
offsets $\langle q_{{\rm z}i} \rangle $ factors and for the
RingI60 models the $\langle q_{{\rm z}i} \rangle$ values behave like
\begin{equation}
  \langle q_2 \rangle > \langle q_{2+3} \rangle > 
  \langle q_3 \rangle 
  \quad {\rm and}\quad
    \langle q_3 \rangle \approx \langle q \rangle \,.
\end{equation}

\paragraph{ZP-correction for the disk.} The largest
offset $\langle q_2 \rangle$ results for a zp-correction
based on the disk region with a strong positive
$\Sigma_2 Q$ polarization. This sets 
$\Sigma_2Q^{{\rm z}2}$ polarization to zero, or the
positive quadrants ($\Sigma_2Q^{{\rm z}2}_{090}+\Sigma_2Q^{{\rm z}2}_{270}$)
equal to the negative quadrants
$-(\Sigma_2Q^{{\rm z}2}_{000}+\Sigma_2Q^{{\rm z}2}_{180})$
and produces apparently a very strong $\Sigma_2Q^{{\rm z}2}_{180}|_\phi$ signal
for the disk front side as shown in panel (d) of
Fig.~\ref{Fig.QuadNorm} (Table~\ref{Tab.ConvNorm}).
This is only a zp-correction effect and should not be interpreted
as real azimuthal distribution of the $Q_\phi(\phi)$ signal, 
as would be expected for a disk with highly forward scattering dust.

Moreover, the derived correction offset
$\langle q_2 \rangle \Sigma_2I=\Sigma_2 Q$
depends on the quality of the PSF. It is difficult to account
for this, particularly for extended
disks without sharp structures as in the RingI60 models.
Thus, whenever possible one should
avoid a zp-correction based on the disk region. This
can be problematic for coronagraphic observation of faint
disks with large outer radii $r_0$, where the halo region $r>r_0$
has not enough signal for a well defined offset correction.

A zp-correction based on the whole
region outside the coronagraphic mask $\Sigma_{\rm zp}=\Sigma_{2+3}$
reduces the correction offset
(Table~\ref{Tab.ConvNorm}). In the halo
the smeared stellar intensity dilutes more efficiently the
smeared disk Stokes signal, and there is 
$\langle q_{2+3}\rangle < \langle q_2 \rangle$. Thus, the
zp-correction offset is reduced, but it depends still
significantly on the PSF profile.

\paragraph{ZP-correction for the halo.}
For the halo there is $\langle q_3\rangle \approx \langle q \rangle$
because the smearing of a $\Sigma Q$ signal of a compact disk model
like RingI60 with $r_0=100.8$~mas is very similar to the
smearing of the intensity, which is dominated by the star. Therefore
the halo zp-corrected disk polarization is practically identical to the
system zp-corrected signal (Table~\ref{Tab.ConvNorm}), or 
\begin{equation}
  \Sigma_2 Q^{\rm z3}
  \approx \Sigma_2 Q^{\rm z}\,.
\end{equation}
The offset defined by the system zp-correction
$-\langle q \rangle \Sigma I$
does not depend on the PSF profile, and this is also the case for the halo
corrected models. This also applies, at least approximately,
for coronagraphic
observations and they should be corrected based on the
halo region $\Sigma_{\rm zp}=\Sigma_3$ outside the disk because of the
well defined and relatively small bias offset.

For coronagraphic observations or data with detector saturation
at the position of the star there is the fundamental issue that the
PSF profile and the total intensity of the system $\Sigma I$
cannot be derived from the same data. Flux and PSF calibrations are
therefore required for quantitative measurements to allow for
a correction of the convolution effects and to relate the
measured polarization signal to
the intensity of the central star, which is typically a
good flux reference.

\begin{figure}
  \includegraphics[trim=1.5cm 1.5cm 1.0cm 1.0cm, width=8.8cm]{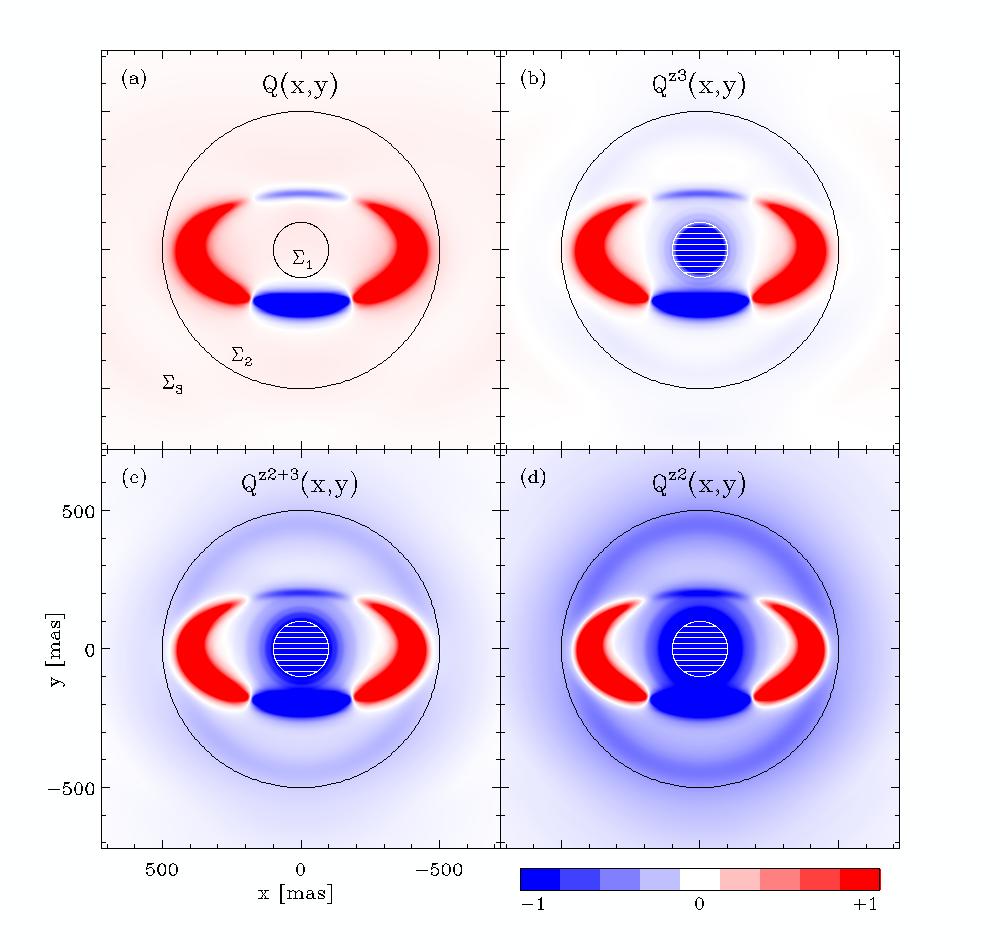}
  \caption{Effects of the zp-correction for Stokes $Q$ for
    the PSF$_{\rm AO}$ convolved RingI60 model with large $r_0=403.2~$mas for
    observations with a coronagraphic mask as indicated by the hatched
    central area. Panel (a) show the convolved signal
    $Q(x,y)$, (b) the map after a correction based on the halo
    $Q^{\rm z3}$, (c) on the disk plus halo $Q^{\rm z2+3}$, and (d) on the disk
    $Q^{\rm z2}$. The peak $Q$ signal is in all panels between 22 to 23 units
    with respect to the indicated color scale.}
 \label{Fig.ConvNormLarge}
\end{figure}

\paragraph{Coronagraphic observations of a large ring.}
A RingI60 model with $r_0=403.2~$~mas convolved with PSF$_{\rm AO}$
provides a good example for the effects of a zp-correction
for coronagraphic observations of an extended disk taken with a currently
available instrument. Figure~\ref{Fig.ConvNormLarge} shows
the Stokes $Q(x,y)$ map, and the zp-corrected maps $Q^{\rm z3}(x,y)$,
$Q^{\rm z2+3}(x,y)$, and $Q^{\rm z2}(x,y)$ while corresponding
numerical values are given in Tab.~\ref{Tab.ConvNorm400}.
The $\Sigma_1$ region
with the central star is defined by $r<100~$mas and assumed to
be covered by a coronagraphic mask. The region $\Sigma_2$ with
the circumstellar disk is described by the annulus
$0.1''<r<0.5''$ and $\Sigma_3$ for the halo by $0.5''<r<1.5''$.
A very narrow color scale was selected to illustrate the weak
extended signal.

The intrinsic disk has a positive net signal $\Sigma Q$ and
this produces in the convolved map $Q(x,y)$ an extended halo
with a faint, positive $Q$ signal (Fig.~\ref{Fig.ConvNormLarge}(a)).
A zp-correction applied to the coronagraphic
data $\Sigma_{\rm zp}=\Sigma_{2+3}$ cancels 
the net positive polarization in the $\Sigma_{2+3}$ region
and this turns the positive background
in $Q(x,y)$ into a negative background
in the corrected map for $Q^{\rm z2+3}$ (Fig.~\ref{Fig.ConvNormLarge}(c)).
Moreover, the subtraction $-\langle q_{2+3} \rangle\,I(x,y)$
produces a quite significant ring of negative $Q$ polarization
at the position of the PSF speckle ring typical for AO systems.
This correction artifact can disturb significantly
the analysis despite the fact that the disk signal is very well resolved.
The bias effects is smaller, if the halo region is used for the
zp-correction $\Sigma_{\rm zp}=\Sigma_3$
and larger, if only the disk region is used $\Sigma_{\rm zp}=\Sigma_2$
(see Fig.~\ref{Fig.ConvNormLarge}(b,d) or Tab.~\ref{Tab.ConvNorm400}).

Determining the zp-offset $\langle q_3 \rangle$ from
the halo is therefore also for coronagraphic observations of
extended disks, like the example in Fig.~\ref{Fig.ConvNormLarge},
a good approach for a well defined
$\langle q_3 \rangle \approx \, \langle q \rangle$ zero point
correction. For this, one should select a region which represents
well the average fractional polarization of the system, like the
region outside the AO speckle ring. This feature should be avoided,
because it overrepresents the stellar contribution. 

\paragraph{Speckle ring as stellar polarization signal.}
The AO speckle ring in high contrast coronagraphic data
  can be strong and could be used to measure or estimate the
 polarization of the star $\Sigma Q_s/\Sigma I_s$ and $\Sigma U_s/\Sigma I_s$,
The requirement is, that stellar light in the speckle ring can be separated
well from the circumstellar scattering signal.

This could be the case for a small scattering region located clearly
between coronagraphic mask and the stellar speckle ring in particular
if the disk polarization has a small net Stokes $Q$ and
$U$ signal so that the smeared halo of the disk is almost
unpolarized. The polarization of the stellar speckle ring can
probably also be measured quite well a circumstellar regions similar to the
coronagraphic model in Fig.~\ref{Fig.ConvNormLarge}.
because the speckle halo at $\Delta y\pm 450$~mas above and below the
star is well separated from the disk signal. 

The derived fractional polarization from the stellar speckles can then
be used as approximation for $\Sigma Q_s/\Sigma I_s$ and
$\Sigma U_s/\Sigma I_s$ and be used for a polarimetric zp-correction.
If the star in unpolarized, then one would get well calibrated
coronagraphic polarization data $Q(x,y)$ and $U(x,y)$ like
for the case described in Sect.~\ref{Sect.Nobias}.
Applying specific procedures for a given data set must probably be
considered to achieve the best results, but this is beyond
the scope of this paper.

\subsection{Partially resolved circumstellar scattering regions}
\begin{figure}
  \includegraphics[trim=0.3cm 0.5cm 0.1cm 1.5cm, width=9cm]{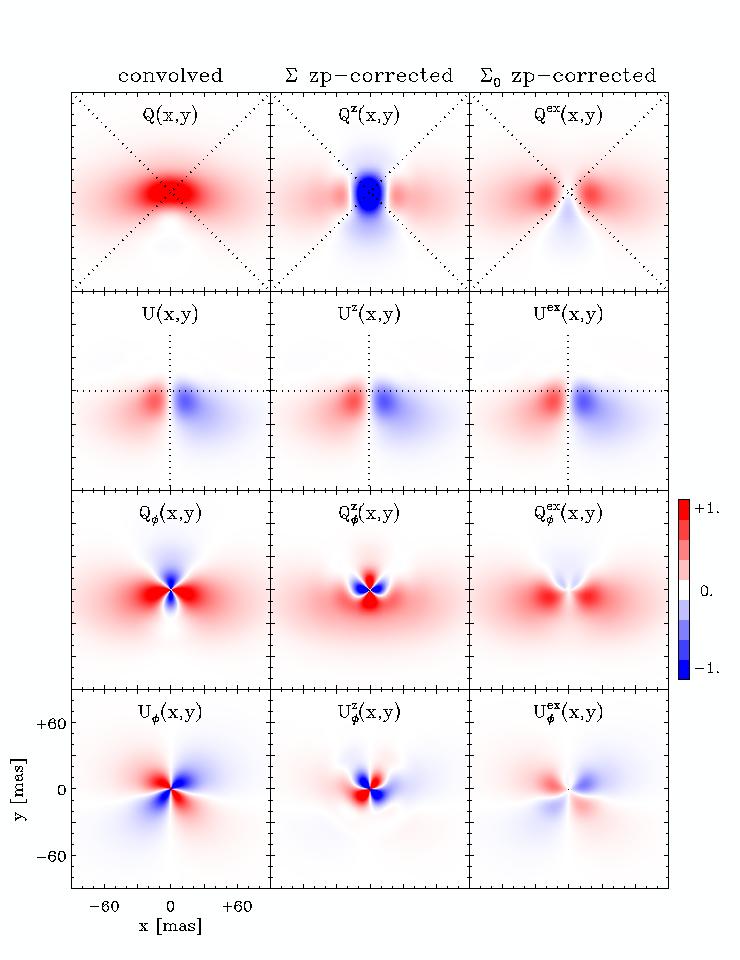}
  \caption{ZP-correction effects for DiskI60$\alpha$-2 with
    small $r_{\rm in}=0.125 D_{\rm PSF}$ (3.15~mas) and strong
    unresolved scattering polarization. The panel columns give from
    left to right polarization
    maps for the PSF$_{\rm AO}$ convolved disk $X(x,y)$, system
    corrected maps $X^{\rm z}(x,y)$, and 
    center corrected maps for the extended polarization region $X^{\rm ex}(x,y)$.
 }
 \label{Fig.NormDisk}
\end{figure}

Many circumstellar disks and shells are too small to be fully
separated from the star. The PSF$_{\rm AO}$ convolved
DiskI60$\alpha$-2 model with a small inner cavity $r_{\rm in}=3.15$~mas
represents such a situation. The circumstellar scattering in
this model produces a substantial amount of unresolved Stokes $Q_d$
signal in the center (Fig.~\ref{Fig.NormDisk}, left panels).
The zp-correction for the system (middle panels) sets the integrated
$\Sigma Q^{\rm z}$ signal to zero and produces therefore a strong $-Q$ signal in the
center and correspondingly strong central quadrant patterns for
the $Q_\phi$ and $U_\phi$ map with opposite sign when compared to
the convolved model.
In both cases the strong central $Q$ signal disturbs substantially
the relatively faint signal of the spatially resolved part
of the disk, and it is difficult to
separate the two polarization components.

\subsubsection{ZP-correction for the center.}

The polarization offset introduced by the central source
can be removed with a star peak or center zp-correction, where
the offsets
$\langle q_0 \rangle=\Sigma_0 Q/\Sigma_0 I$ and
$\langle u_0 \rangle=\Sigma_0 Q/\Sigma_0 I$
are derived from a few pixels centered on the intensity
peak. This sets the polarization signal at the center to zero
\begin{equation}
Q^{\rm z0}(0,0) = Q(0,0) - \langle q_0 \rangle\, I(0,0) = 0\,, 
\end{equation}  
and equivalent for Stokes $U^{\rm z0}(0,0)$ and removes in
Fig.~\ref{Fig.NormDisk} the strong central Stokes $Q$-component
and quadrant patterns in the $Q_\phi$ and $U_\phi$ maps.
The center correction
leaves an extended polarization signal
from circumstellar scattering called hereafter $Q^{\rm ex}$, $U^{\rm ex}$ according to
\begin{eqnarray}
Q^{\rm z0}(\alpha,\delta) = Q(\alpha,\delta) - \langle q_0 \rangle\, I(\alpha,\delta) = Q^{\rm ex}(\alpha,\delta)\,,
  \label{Eq.QNormc} \,\\
U^{\rm z0}(\alpha,\delta) = U(\alpha,\delta) - \langle u_0 \rangle\, I(\alpha,\delta) = U^{\rm ex}(\alpha,\delta) \,.
  \label{Eq.UNormc}
\end{eqnarray}

The polarization in the PSF peak
$\langle q_0 \rangle\, I(0,0)$ and $\langle u_0 \rangle\, I(0,0)$
can include contributions from the following components:
(i) offsets introduced by (residual)
interstellar and instrumental polarization $(q_{\rm is}+q_{\rm inst})\,I$,
$(u_{\rm is}+u_{\rm inst})\,I$, and
(ii) the intrinsic central polarization $Q_c=q_sI_s+q_{d,c}I_s$ and
$U_c=u_sI_s+u_{d,c}I_s$ from the star and the unresolved part of
the scattering region.

The offset from the center correction is proportional
to $I(\alpha,\delta)$ and corrects therefore exactly the
signal introduced by the interstellar and
instrumental polarization. In addition also the contribution
from the intrinsic stellar polarization $Q_c,U_c$ are corrected,
but only approximately because the intensity distribution of the
convolved central signal is
$\propto {\rm PSF}(\alpha,\delta)$ and differs
slightly from $I(\alpha,\delta)$ used for the correction
offset.
This introduces a small overcorrection at the level of
$(q_s+q_{d,c})\,I^{\rm ext}(\alpha,\delta)/I_s(\alpha,\delta)$
and similar for Stokes $U$, where $I^{\rm ext}$ corresponds to
the resolved part of the scattering intensity.
There is typically $(q_s+q_{d,c}) I^{\rm ext}(r) \ll Q_\phi(r)$ based
on simular arguments as discussed in Sect.~\ref{Sect.Polstar},
and therefore we can assume that the center correction
accounts for many cases well for all contributions to the
polarization of the central source.

The center corrected maps $X^{\rm ex}(x,y)$ for the DiskI60$\alpha$-2 model
with small cavity $r_{\rm in}=0.125~D_{\rm PSF}$ in Fig.~\ref{Fig.NormDisk}
look very similar to the non-corrected DiskI60$\alpha$-2 model with a
larger inner cavity with $r_0=0.5~D_{\rm PSF}$ plotted in
Fig.~\ref{FigDiski60}. Thus, the effect of the center zp-correction
is very similar to the removal of the polarization signal
from the non-resolved inner disk region $r<0.5~D_{\rm PSF}$
(Table~\ref{Tab.CentNorm}).

\begin{figure}
\includegraphics[trim=1.8cm 13.0cm 1.8cm 3.0cm, width=8.8cm]{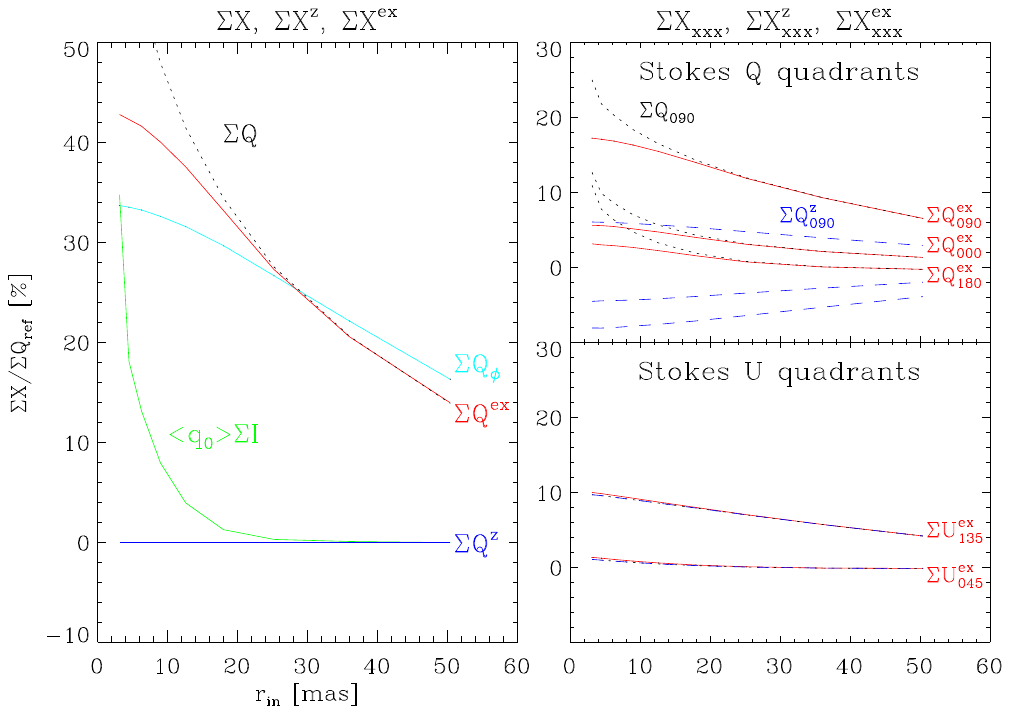}
\caption{Dependence of the polarization $\Sigma X/Q_{\phi,{\rm ref}}$ and
  quadrant parameters $\Sigma X_{xxx}/Q_{\phi,{\rm ref}}$ for 
  DiskI60$\alpha$-2 as function of $r_{\rm in}$ for the
  PSF$_{\rm AO}$ convolved, system corrected $X^{\rm z}$, and center
  corrected $X^{\rm ex}$ model parameters.
  All disks have the same outer boundary $r_{\rm out}=100.8$~mas.
  For Stokes $Q$ quadrants there is
  $\Sigma U_{xxx}=\Sigma U^{\rm z}_{xxx}=\Sigma U^{\rm ex}_{xxx}$.} 
\label{DiagDiskI60Norm}
\end{figure}

The zp-correction for the center
splits the integrated polarization into a central unresolved
component and an extended resolved component
\begin{equation}
\Sigma Q=\Sigma Q_0+\Sigma Q^{\rm ex}\quad {\rm and} \quad 
\Sigma U=\Sigma U_0+\Sigma U^{\rm ex} \,.  \label{Eq.QUsplit}
\end{equation}
For simulations with a given intrinsic model but using
different PSFs for the convolution the total Stokes signals
$\Sigma Q$ and $\Sigma U$ are conserved but
the splitting between the $\Sigma Q_0$ and $\Sigma Q^{\rm ex}$ or the
$\Sigma U_0$ and $\Sigma U^{\rm ex}$ components depends on the PSF profile.
For a more extended PSF a larger fraction of the polarization signal
will contribute to the unresolved central source and a smaller fraction to the
extended (resolved) component.

The impact of the center correction for the $\Sigma Q$ splitting
is illustrated in the left panel of Fig.~\ref{DiagDiskI60Norm}
for DiskI60$\alpha$-2 models with different
$r_{\rm in}$. For large central cavities
$r_{\rm in}>D_{\rm PSF}$ ($> 25~$mas) there is practically no difference
between $\Sigma Q^{\rm ex}$ and $\Sigma Q$ because of the lack of
a central component.
For small cavitity $r_{\rm in}< D_{\rm PSF}$ the unresolved central signal 
$\Sigma Q_0\approx\langle q_0 \rangle \,\Sigma I$ plotted by the
green curve increases strongly for $r_{\rm in}\rightarrow 0$,
while the extended polarization 
$\Sigma Q^{\rm ex}$ approaches a limiting value.

Equation~\ref{Eq.QUsplit} for Stokes $Q$ can also be written as
$\Sigma Q^{\rm ex}/\Sigma I = \Sigma Q/\Sigma I - \Sigma Q_0/\Sigma I
= \langle q \rangle - \langle q_0 \rangle \, \Sigma I_0/\Sigma I$
and the same applies for Stokes $U$. In observational data, one
can often assume that the central object
dominates the total intensity of the system, and use the rough
approximation $\Sigma I_0/\Sigma I=1-\epsilon\approx 1$. For such
cases one can then derive the approximative Stokes flux
$Q^{\rm ex}$ for the resolved
scattering region from the total intensity and the difference
between the fractional polarization measured for the whole system and the
center according to 
\begin{equation}
  \Sigma Q^{\rm ex} \approx ( \langle q_{\rm obs}
  \rangle - \langle q_{0,{\rm obs}} \rangle )\, \Sigma I\,,
\end{equation}
and equivalent for the $\Sigma U^{\rm ex}$.
The two fractional polarization values
$\langle q_{\rm obs} \rangle $ and $\langle q_{0,{\rm obs}} \rangle$ depend
equally or practically equally on the various polarization
offsets listed above. Therefore the differential value is
barely affected by polarimetric offset, even if their nature is unclear.
This could provides for many cases useful Stokes $\Sigma Q^{\rm ex}$ and
$\Sigma U^{\rm ex}$ parameters for the spatially resolved
circumstellar scattering polarization.

The resulting values for the Stokes parameters $\Sigma Q^{\rm ex}$
and $\Sigma U^{\rm ex}$ describe like $\Sigma Q_\phi$
properties of the resolved part of the circumstellar scattering
and all three parameters depend on the spatial resolution of
the observations.
Simulations of the convolution can be used to obtain from
the ratios of convolved or observed polarization values
$\Sigma Q^{\rm ex}/\Sigma Q_\phi$
and $\Sigma U^{\rm ex}/\Sigma Q_\phi$
stronger constraints on the intrinsic polarization
$Q'(x,y)$, $U'(x,y)$, and $Q_\phi'(x,y)$.

\subsubsection{Quadrant parameters for the extended polarization}

The maps for the extended polarization $Q^{\rm ex}$ and $U^{\rm ex}$
are useful for the characterization of the azimuthal polarization
$Q_\phi(\phi)$. This is illustrated for the DiskI60$\alpha$-2 models
in the right panels of Fig.~\ref{DiagDiskI60Norm}, which show
$\Sigma X_{xxx}/Q_{\phi,{\rm ref}}$ for the convolved models, and
also for the system corrected  
$\Sigma X^{\rm z}_{xxx}/Q_{\phi,{\rm ref}}$, 
and center corrected  
$\Sigma X^{\rm ex}_{xxx}/Q_{\phi,{\rm ref}}$ models 
plotted by black dotted, blue dashed and solid red lines respectively
(see also Tab.~\ref{Tab.CentNorm}).

The behaviour of the $Q$ quadrants is very similar
to the integrated Stokes parameter $\Sigma Q$, with
$\Sigma Q^{\rm ex}_{xxx}=\Sigma Q_{xxx}$ for $r_{\rm in}>D_{\rm PSF}$,
with a strong increase of $\Sigma Q_{xxx}$ for $r_{\rm in}<D_{\rm PSF}$,
and much lower values for the system corrected
quadrants $\Sigma Q^{\rm z}_{xxx}$. The Stokes $U$ quadrant values are
for all three cases the same because they are not affected
by Stokes $Q$ zp-corrections.

The differences between the lines for $Q$ quadrants with
$xxx=090,\,000$ and $180$ are the same for the three cases, and this
is equivalent to the differential quadrant values (Eq.~\ref{Eq.QuadNorm}),
which practically do not depend on a polarization offset. Thus, the
differential quadrant values and the
corresponding information on the azimuthal distribution of
$Q_\phi(\phi)$ is preserved by the zp-correction for the central peak.

\begin{figure}
\includegraphics[trim=2.8cm 15.3cm 4.0cm 5.2cm, width=8.8cm]{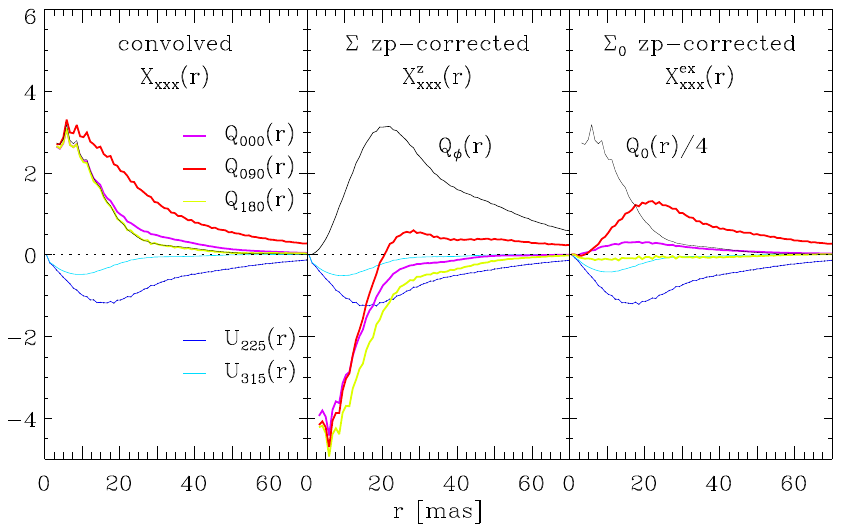}
  \caption{Radial profiles for the polarization signal in
    the Stokes quadrants for the DiskI60$\alpha$-2 model with
    $r_{\rm in}=0.125~D_{\rm PSF}$ (3.15~mas).
    The panels give $X_{xxx}(r)$ for the convolved,
    $X_{xxx}^{n}(r)$ for the system corrected, and
    $X_{xxx}^{\rm ex}(r)$ for the center corrected models.
    For all three cases the Stokes $U_{xxx}(r)$ profiles are
    practically the same.} 
\label{NormQuadProf}
\end{figure}

\paragraph{Central zero.}
A most important advantage of the center corrected maps
$Q^{\rm ex}(x,y)$ is the central zero (Fig.~\ref{Fig.NormDisk})
with facilitates
for observational data very significantly the splitting of
the $Q^{\rm ex}$-signal into quadrant values.
This is illustrated in Fig.~\ref{NormQuadProf} with radial profiles
for the polarization signal 
$X_{xxx}(r)$ obtained by averaging azimuthally the polarization signal
within the $90^\circ$ wedges of the individual quadrants. The profiles
show for small $r$ some sampling noise because of the used finite
pixel size of $0.9~{\rm mas}\times 0.9~{\rm mas}$ for the model maps.

The strong central $Q_c$ signal, which is positive for the convolved
map and negative for the map with a system zp-correction, dominates
at small $r$ strongly the $Q_{xxx}(r)$ and $Q^{\rm z}_{xxx}(r)$ profiles.
In these cases the
determination of quadrant values $\Sigma Q_{xxx}$ and $\Sigma Q^{\rm z}_{xxx}$
depend critically on the splitting of the strong
central signal.
This can introduce significant uncertainties for AO data with a
variable and not exactly axisymmetric PSF core \citep[e.g.,][]{Cantalloube19}.
The splitting of the Stokes signal into quadrant values
is less critical for the center corrected $Q^{\rm ex}$ map
with a signal close to zero in the center, where the borderlines between the
quadrants intersect. Therefore, the center zp-correction is
be very helpful for the derivation of the azimuthal distribution of
the Stokes parameters $Q(\phi)$ and $U(\phi)$ using quadrant parameters
and for the azimuthal parameters
$Q_\phi(\phi)$ and $U_\phi(\phi)$ for barely resolved observational data.
This is also supported by recent results for observations of
the circumstellar dust around post-AGB stars \citep{Andrych24,Andrych25}. 

\section{Summary and discussion}
High resolution imaging polarimetry is a very attractive
differential technique for the detection of a circumstellar
scattering regions because the produced 
polarization signal can be separated from the strong
radiation of the central star \citep[e.g.,][]{Kuhn01,Perrin04,Hinkley09}.
However, imaging polarimetry is affected by observational
convolution effects and by calibration offsets which must be taken
into account for a quantitative analysis
\citep[e.g.,][]{Schmid06,Monnier19,Hunziker21,Tschudi21,Ma24b}.
This work investigates these effects systematically and provides 
guide lines for the derivation of quantitative
polarimetric results from observational
data. Basic effects are treated with simple model simulations
using axisymmetric PSFs, considering only axisymmetric or mirror
symmetric scattering geometries aligned with the Stokes $Q$ polarization
direction, and circular or annular
apertures for the determination of zp-correction offsets
and integrated flux parameters.

The convolved azimuthal polarization for intrinsically
axisymmetric models is also axisymmetric $Q_\phi(r)$ and depends
on the PSF profile. The mirror-symmetric models RingI60 and
DiskI60 for inclined disks have an azimuthal dependence
for $Q'_\phi(\phi)$ and they produce a net Stokes
$\Sigma Q$ signal. The PSF convolution and polarization offsets both
change the azimuthal distribution of the signal,
which is described by Stokes quadrant polarization parameters
$\Sigma X_{xxx}$. Because only faint disks
are considered $\Sigma I_d\lapprox 0.05\, \Sigma I_s$ a polarimetric
offset for Stokes $Q$ is close to axisymmetric $q\,I(x,y)\approx q\,I(r)$
and this changes the four quadrants $\Sigma Q_{xxx}$ 
practically equally, while
the Stokes $U$ signal is not changed by a Stokes $Q$ offset.
Observational data are more complex than the presented simple
simulations and instrumental effects introduce additional noise,
but many basic results of this study are still applicable or
approximately valid for the interpretation of real data.

\paragraph{Presence of circumstellar polarization.}
In all models a net positive azimuthal
polarization $\Sigma Q_\phi>0$ indicates 
the presence of resolved circumstellar scattering polarization.
Measurements of $\Sigma Q_\phi$ are ideal for the detection of faint,
extended circumstellar scattering regions, because the signal is
not biased by pixel to pixel noise and can
be summed up for large image areas \citep{Schmid06}. For observational
data one needs to define $\Sigma Q_\phi$ detection limits considering
spurious signals from PSF variations or systematic instrumental noise
{\citep[e.g.,][]{Cantalloube19,Tschudi24}. 
According to the presented noise free simulations 
a measurable $\Sigma Q_\phi$-signal can be obtained for very compact
scattering regions down to a separation of about $r \approx D_{\rm PSF}$
and this has also been achieved for observational data
\citep[e.g.,][]{Avenhaus17,Schmid18,Andrych25}
In addition the $\Sigma Q_\phi$ value is not significantly changed by
a polarization offset $p\,I(r)$ because this adds
for $Q_\phi(x,y)$ only a positive-negative quadrant pattern with 
practically no net $\Sigma Q_\phi$ signal.

\paragraph{Degradation of the $Q_\phi$ signal.}
The PSF convolution degrades the $\Sigma Q_\phi$ signal because of
the smearing and polarimetric cancellation \citep{Schmid06}.
The effect is particularly strong for small separations from
the star and there results practically no net 
$\Sigma Q_\phi$ signal for a scattering regions with separation
$r\lapprox 0.5~D_{\rm PSF}$ from the star.
For resolved but compact scattering regions $0.5~{\rm D}_{\rm PSF}<r <5~D_{\rm PSF}$
the degradation depends strongly on the PSF structure and is
about a factor of a few for the PSF$_{\rm AO}$ used in the simulations.
For scattering regions at larger separation $r\gapprox 10~D_{\rm PSF}$
the $\Sigma Q_\phi$ degradation is at a level of about $\approx 10~\% - 30~\%$
according to Table~\ref{Tab.PoleOnRing} and the tables in the appendix. For
AO observations with high PSF Strehl ratio of about $S\approx 0.8$ 
or a space instrument with $S\approx 1$, the degradation would be closer
or comparable to the model results using the Gaussian PSF$_{\rm G}$.

A measurement of the azimuthal polarization $\Sigma Q_\phi/\Sigma I$ should 
always include an assessment of the convolution effects based on
the observational PSF and provide an estimate or even a derivation
of the intrinsic circumstellar polarization $\Sigma Q'_\phi/\Sigma I'$.
This allows for individual objects comparisons
between results from different observations and enables investigations about
temporal variations and the wavelength dependence of the polarized
reflectivity \citep[e.g.,][]{Ma24a,Ma24b,Andrych25}.

The PSF$_{\rm AO}$ convolution of inclined disks introduces
a cross-talk signal $Q_\phi\rightarrow U_\phi$, significant differences
between the azimuthal polarization $Q_\phi$ and the polarized flux $P$,
extended Stokes $Q$, $U$ polarization halos, and 
central Stokes signals $Q$, $U$ for unresolved inner scattering regions.
These extended and central Stokes components produce in the
azimuthal polarization maps $Q_\phi(x,y)$ and $U_\phi(x,y)$ quadrant
patterns with practically zero net $Q_\phi,U_\phi$ signal,
if measured in axisymmetric apertures
as illustrated in Figs.~\ref{FigRingi60} and \ref{FigRingI60AOLRG}.

\paragraph{Stokes polarization parameters.}
The integrated Stokes parameters $\Sigma Q$ and $\Sigma U$ 
do not depend on the PSF convolution but the smearing leads to
a mutual averaging of the polarization between the positive and
negative Stokes $\Sigma Q_{xxx}$ or $\Sigma U_{xxx}$ quadrants towards
the mean value $\Sigma Q/4$ or $\Sigma U/4$, respectively. This changes also
the azimuthal distribution $Q_\phi(\phi)$ of the circumstellar polarization
in step with the degradation of the integrated
$\Sigma Q_\phi$ signal.
Differential quadrant values expressed relative to
$\Sigma Q_\phi$, like $(\Sigma Q_{xxx}-\Sigma Q/4)/\Sigma Q_\phi$,
are much less changed by the PSF convolution, and therefore
they provide for convolved data still strong constraints on the
intrinsic azimuthal distribution of $Q'_\phi(\phi)$.
For complex scattering geometries it is more difficult to
account for the mutual averaging effect between the polarization quadrants,
but the basic principle is still valid and can be considered
for the interpretation of the data.

An unresolved central scattering region can produces a central 
signal $\Sigma Q_c, \Sigma U_c$ with a spatial distribution
like the PSF.
This can be compared with the spatially resolved polarization
$Q(x,y)$ and $U(x,y)$ at $r\gapprox D_{\rm PSF}$
providing potentially strong constraints on the presence and the radius
of a central cavity caused by dust sublimation for accretion disks
or dust condensation for shells around mass losing stars.

\begin{figure*}
\sidecaption  
\includegraphics[trim=0.1cm 1.5cm 1.0cm 0.10cm, width=12cm]{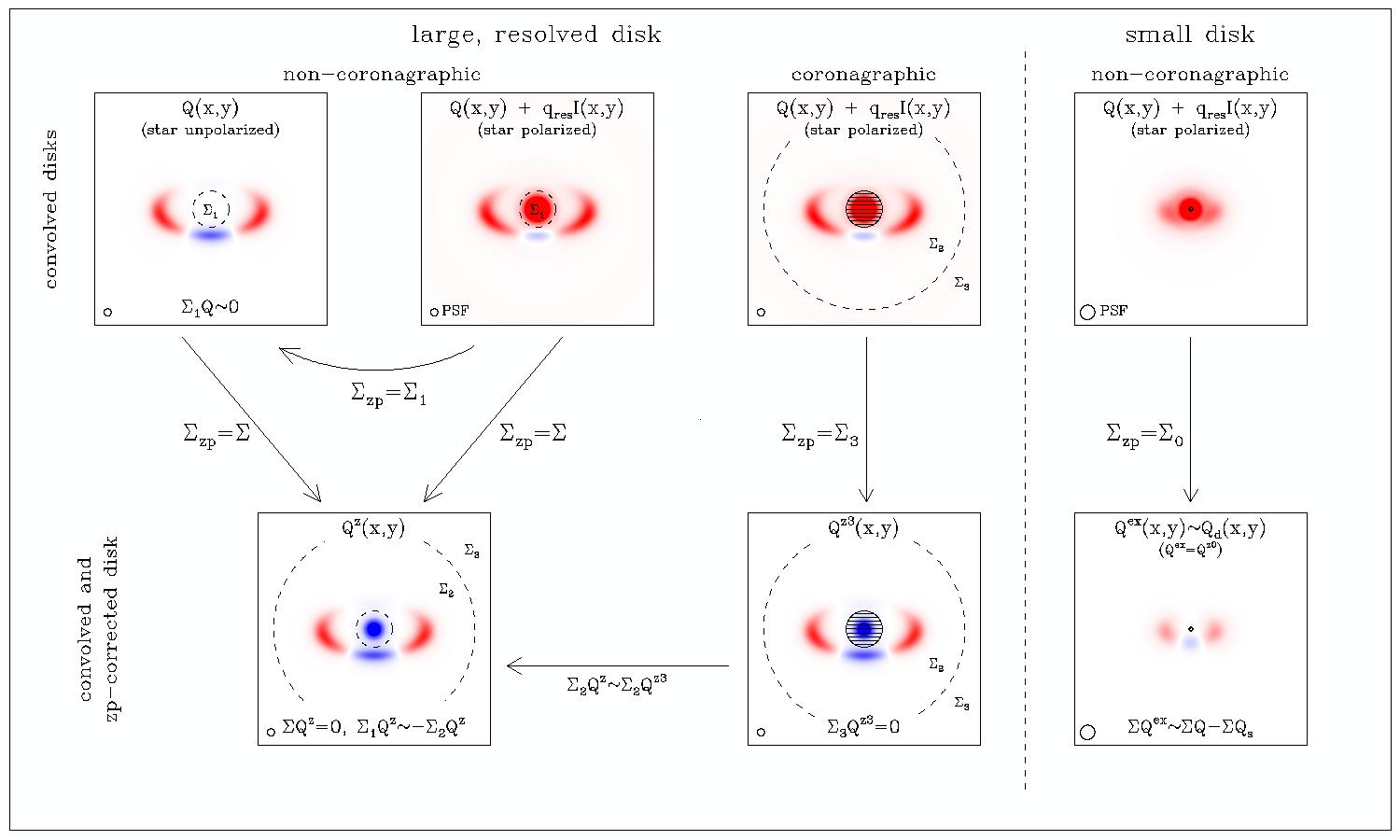}
\caption{Sketch for the polarimetric zp-corrections procedures for 
  Stokes $Q$. The system correction $\Sigma_{\rm zp}=\Sigma$ gives
  the same $Q^{\rm z}$ for the first two cases, while a star correction
  $\Sigma_{\rm zp}=\Sigma_1$ can recover the convolved $Q^{\rm z1}=Q$ case
  for data with a polarization offset. A halo correction
  $\Sigma_{\rm zp}=\Sigma_3$ is
  a good approach for coronagraphic data ($Q^{\rm z3}\approx Q^{\rm z}$ outside of
  the coronagraphic disk). A center correction
  $\Sigma_{\rm zp}=\Sigma_0$ provides the resolved circumstellar
  signal $Q^{\rm ex}$ for compact regions.}
\label{SummarySketch}
\end{figure*}

\paragraph{Polarization offsets and zp-correction.}
Measurements of Stokes parameters $\Sigma Q$, $\Sigma U$
and quadrant values $\Sigma Q_{xxx}$, $\Sigma U_{xxx}$ are often affected by
residual polarimetric offsets $p_{\rm res}\,I(x,y)$ introduced by not
so well known amounts of intrinsic stellar polarization,
or by interstellar and instrumental polarization.
Already, an offset of $p_{\rm res}\gapprox 0.001$ can introduce strong
bias effects or even mask the circumstellar polarization signal.

A polarimetric zero point or zp-correction accounts for
a fractional Stokes polarization
offset $\langle q \rangle I(x,y)$ and $\langle u \rangle I(x,y)$,
and this is for many observations a key step to reveal
better the circumstellar polarization signal.

A correction offset $-\langle q \rangle\,I(x,y)$ has
  for all four Stokes $Q$ quadrants the same impact
 with practically no change for the differential quadrant values
  $\Sigma_2 Q_{xxx}-\Sigma_2 Q/4$. The
same would applies for Stokes $U$ offset, and the
corrected Stokes values $\Sigma Q^{\rm z}$, $\Sigma U^{\rm z}$
and quadrant $\Sigma Q^{\rm z}_{xxx}$, $\Sigma U^{\rm z}_{xxx}$ 
can be recalibrated, if accurate determination
of the intrinsic $\Sigma Q'$ and $\Sigma U'$ are available.

A zp-correction does not change
significantly the integrated azimuthal polarization $\Sigma Q_\phi$,
but it introduces a bias effect for the azimuthal signal distribution
$Q_\phi(\phi)\neq Q_\phi^{\rm z}(\phi)$.
A negative offset $-\langle q \rangle\,I(x,y)$ reduces for the
RingI60 and DiskI60 models the positive $Q_{090}$ and $Q_{270}$
quadrants and makes the negative quadrants $Q_{000}$ and $Q_{180}$
even more negative. For $Q_\phi(\phi)$, this is equivalent to a
weakening around $\phi\approx 90^\circ$ and $\approx 270^\circ$,
while the signals around $\phi\approx 0^\circ$ and $\approx 180^\circ$
are enhanced. The zp-corrected RingI60 model shows for the
quadrant ratio $\Sigma Q^{\rm z}_{180}/\Sigma Q^{\rm z}_{090}$ a value
which is more than 50~\% enhance when compared to the model without
zp-correction (Fig.~\ref{Fig.QuadNorm}, Table~\ref{Tab.ConvNorm}),
equivalent to a much enhanced front side brightness.
This effect must be taken into account for the
derivation of the polarized scattering phase functions for the
dust in the disk.

A positive Stokes $U$ offset would enhance the $U_{135}$ and $U_{315}$
components and make the negative quadrants $U_{045}$ and $U_{225}$ even
more negative as shown in Fig.~\ref{FigDiski60Starpol} and
in principle similar to the case of Stokes $Q$.
However, because the $U$ direction is not aligned with the symmetry
of the adopted model geometries an $U$ offset distorts the
antisymmetric or symmetric appearance of the $U$, $U_\phi$ and $Q_\phi$ maps.

Important for the zp-correction 
are well defined correction region $\Sigma_{\rm zp}$
and different useful cases are illustrated
in the sketch in Fig.~\ref{SummarySketch}. The reference case is
a circumstellar signal without intrinsic stellar
polarization and observational polarization offset. A system
normalization $\Sigma_{\rm zp}=\Sigma$ sets the net polarization
to zero $\Sigma Q^{\rm z}=0$, and introduces for the RingI60 model
a negative polarization for the which compensates the
circumstellar signal $\Sigma_1 Q^{\rm z}=-\Sigma_2 Q^{\rm z}$.

Practically the same $Q^{\rm z}$ maps are obtained
for the same circumstellar model, but ``affected''
by an offset from interstellar, instrumental or intrinsic
stellar polarization. Therefore, the zp-correction is
very useful to mitigate offsets and calibration
uncertainties, because one can search for an intrinsic
$Q'(x,y),U'(x,y)$ model, which matches after convolution and a
zp-correction the
observed and zp-corrected data despite some undefined offsets.

If other data provide accurate values
$(\Sigma Q',\Sigma U')=(\Sigma Q,\Sigma U)$,
then one can correct the zp-corrected data 
and derive $Q(x,y),\,U(x,y)$ maps without offsets.
Alternatively, if the central star can be used as zero polarization
standard, then a zp-correction based on the star
$\Sigma_{\rm zp}=\Sigma_1$ provides directly the intrinsic 
Stokes signal 
$\Sigma Q'_d=\Sigma Q^{\rm z1}$ and $\Sigma U'_d=\Sigma U^{\rm z1}$.

A good solution for coronagraphic observations or data with 
saturated star is the use of
the halo region $\Sigma_{\rm zp}=\Sigma_3$
for the zp-correction, because the fractional polarization of
the smeared signal in the
halo approximates well the value for the whole system
$\langle q_3 \rangle \approx \langle q \rangle $. This correction
provides then a good match for the reference case
$Q^{\rm z3}(x,y)\approx Q^{\rm z}(x,y)$ outside of the coronagraphic mask
(Fig.~\ref{SummarySketch}).
This procedure requires enough signal and the
absence of significant systematic errors in the halo
for the determination of $\langle q_3 \rangle$ , what may be
critical for some observations. In addition, one
needs also a separate PSF calibration to estimate 
the smearing degradation of the derived $Q_\phi$ signal.

For partially resolved scattering regions a
zp-correction based on a few brightest
pixels of the PSF peak $\Sigma_{\rm zp}=\Sigma_0$ can be very favourable.
This approach splits the signal into two components:
(i) an unresolved polarization
for the central source $Q_0$, $U_0$ with a spatial distribution
like the PSF and zero azimuthal polarization
$\Sigma Q_{\phi,{\rm 0}}\approx 0$ 
and (ii) an extended polarization maps
$Q^{\rm ex}(x,y)$, $U^{\rm ex}(x,y)$ and $Q_\phi$ with a central zero,
and quadrant values $\Sigma Q^{\rm ex}_{xxx},\,\Sigma U^{\rm ex}_{xxx}$
constraining the azimuthal distribution of $Q_\phi(\phi)$ down to a
separation of $r\approx D_{\rm PSF}$. The
center correction accounts for offsets introduced by interstellar and
instrumental polarization, and for the contributions from the
intrinsic polarization of the unresolved, central source. The
resulting maps represent the resolved part of the circumstellar
polarization signal $Q^{\rm ex}(x,y)\approx Q_d(x,y)$
and $U^{\rm ex}(x,y)\approx U_d(x,y)$.}

\paragraph{Conclusions.} The presented model simulations
explore systematically the PSF convolution effects and the impact of
polarization offsets for imaging polarimetry of circumstellar
scattering regions. The used simple models are ideal for a
description of the basic principles which also apply for more
complex systems and should be considered for the
planning of new observations and
for the interpretation and analysis of observational data.
Important is, that observational data are taken which can be corrected for
the instrumental PSF convolution using PSF calibrations,
and that polarimetric offsets are corrected with a well defined
zp-correction procedure. In particular, for
coronagraphic high contrast observations a
good PSF calibration and polarimetric measurements for
the central bright object with non-coronagraphic exposures
could provide higher quality results.

Future instruments may target more demanding
objects and effects introduced by non-axisymmetric features in
the PSF convolution and field dependent polarimetric offsets
could affect significantly the data. Analysis procedures considering
only axisymmetric PSFs and simple fractional polarization
offset may not be sufficient.
However, an important first
step are investigations of the achievable limits using the
procedures discussed in this work for available data,
which have barely been started despite the existence of many high
quality observations in data archives \citep[e.g.,][]{Benisty23}. This will provide
useful insights on the next level of challenges to be considered
for future improvements in polarimetric imaging.

\begin{acknowledgements}
We thank the referee, Frans Snik, for very thoughtful and supportive
comments, which improved the manuscript.
JM thanks the Swiss National Science Foundation for financial support
under grant number P500PT\_222298.   
\end{acknowledgements}

\bibliographystyle{aa} 
\bibliography{aa56371-25} 

\appendix

\section{Polarization parameters}
\label{SectPar}

Stokes polarization parameters $Q,\,U$ and
azimuthal polarization parameters $Q_\phi,\, U_\phi$ are used in this work
for the description of the model results.
Convolution and polarization offset are simple mathematical
operations for the Stokes parameters, but quite complex for the
azimuthal parameters. Therefore, a comprehensive quantitative description
of the circumstellar scattering polarization requires an
extensive framework of parameters,
which is outlined in Table~\ref{ParXYTab} for the convolved models and
in Table~\ref{ParXYTabzp} for zp-corrected models in the
system aligned $x,y$-coordinates.

A few remarks on the entries in the tables. The integration regions
$\Sigma_{\rm int}$ and the zp-calibration regions $\Sigma_{\rm zp}$ are
defined on a model by model basis, with the aim to quantify the
polarization signal for the whole system $\Sigma$, or only the
central source $\Sigma_1$, the disk $\Sigma_2$, and the halo $\Sigma_3$,
and combination of these subregions like $\Sigma_{1+2}$.

It is also important to distinguish between $\Sigma P$ and
${\cal{P}}(\Sigma)$. $\Sigma P$ is the sum of the degree of linear
polarization measured for each pixel in the resolved map in the
integration region $\Sigma$ (Eq.~\ref{EqPQUQphiUphi}).
This does not considering the position angle $\theta$ of the
polarization. In addition, the polarization $\Sigma P$ depends
in a complex way on the spatial resolution, on instrumental
noise, and on polarization offsets. 

Therefore it is useful to characterize the ``aperture'' polarization
${\cal{P}}(\Sigma_i)$ for a given integration region $\Sigma_i$
using the sums of the Stokes parameters according to
\begin{equation}
{\cal{P}}(\Sigma_i) = \bigl((\Sigma_i Q)^2 + (\Sigma_i U)^2\bigr)^{1/2}
\label{EqAperP}  
\end{equation}
with the corresponding position angle 
\begin{equation}
\theta(\Sigma_i)=0.5\,{\rm atan2}(\Sigma_i U,\Sigma_i Q)\,.
\label{EqApertheta}  
\end{equation}
The fractional polarization can then be defined by
$p_i={\cal{P}}(\Sigma_i)/\Sigma_i I$ or for the Stokes components
by $q_i=\Sigma_i Q/\Sigma_iI$ and $u_i=\Sigma_i U/\Sigma_iI$, from
which one can also derive a position angle $\theta_i$. The polarization
of the star, or all the fractional polarization parameters
like $q_{is}$, $\langle q_i\rangle$, or $q_0$, are defined
with these ``aperture'' polarization parameters.

\begin{table}
  \caption{Polarization parameters for convolved and intrinsic models.} 
  \label{ParXYTab}
  \begin{tabular}{p{1.0cm}p{7.2cm}}
\noalign{\hrule\smallskip}    
{param.} & description \\
\noalign{\smallskip\hrule\smallskip}    
\noalign{parameters for the convolved maps}
\noalign{\smallskip}
$I(x,y)$           & intensity \\
\noalign{\smallskip}
$Q(x,y)$           & linear Stokes polarization $Q=I_0-I_{90}$    \\
$U(x,y)$           & linear Stokes polarization $U=I_{45}-I_{135}$  \\
\noalign{\smallskip}
$Q_\phi(x,y)$ $U_\phi(x,y)$
                   & azimuthal parameters for the linear polarization \par 
                     ~~~~~with respect to $x_0,y_0$
                     (Eqs.~\ref{EqQphiQU},\ref{EqUphiQU}).\\
\noalign{\smallskip}
$P(x,y)$           & intensity of the polarized flux
                     (Eq.~\ref{EqPQUQphiUphi}) \\
\noalign{\smallskip}
$X(x,y)$           & expression for different parameter maps, \par
                     ~~~~~ e.g., $X=\{I,Q,U,P,Q_\phi,U_\phi\}$ \\
$\Sigma_{\rm int} X$ & summed parameter in intergration
                     region $\Sigma_{\rm int}$ \\
\noalign{\smallskip}
\noalign{used centro-symmetric apertures $\Sigma_{\rm int}$\tablefootmark{a}}
\noalign{\smallskip}
$\Sigma$          & round aperture for whole system \\
$\Sigma_1$        & small, round aperture centered on the star \\
$\Sigma_2$        & annular aperture including the disk \\
$\Sigma_3$        & annular aperture for the halo \\
$\Sigma_i$        & expression for all apertures
                    $\Sigma_i=\{\Sigma,\Sigma_1,\Sigma_2,\Sigma_3\}$\tablefootmark{b} \\
\noalign{\smallskip}
\noalign{quadrant polarization parameters\tablefootmark{c}}
\noalign{\smallskip}
$\Sigma_i Q_{xxx}$   &
Stokes $Q$ quadrants $\Sigma_i Q_{000}$, $\Sigma_i Q_{090}$,
                    $\Sigma_i Q_{180}$, $\Sigma_i Q_{000}$\tablefootmark{d} \\
$\Sigma_i U_{xxx}$   &  Stokes $U$ quadrants $\Sigma_i U_{045}$, $\Sigma_i U_{135}$,
                    $\Sigma_i U_{225}$, $\Sigma_i U_{315}$ \\
$\Sigma_i X_{xxx}$   & expression for all quadrant parameters \\ 
$\Sigma_i X_{xxx}|_\phi$ & azimuthal quadrant parameters considering sign
                       for \par
                       ~~~~~~positive $Q_\phi$ contribution
                       (Sect.~\ref{Sect.NormQuad})\\ 
\noalign{\smallskip}
$\Delta Q_{xxx}$    & differential quadrant value for Stokes $Q$
                     (Eq.~\ref{Eq.Qdiff})\\
$\Delta U_{\pm}$     & differential quadrant values for Stokes $U$
                     (Eq.~\ref{Eq.Udiff})\\ 
\noalign{\smallskip}  
\noalign{azimuthal polarization in radial and azimuthal coordinates}
\noalign{\smallskip}
$Q_\phi(r,\phi)$   & used in the discussion for the $Q_\phi$ distribution in \par
                    ~~~~~radial apertures $\Sigma_i Q_\phi$
                         and the angular \par
                    ~~~~~distribution in quadrants $\Sigma_iX_{\rm xxx}|_\phi$\\
\noalign{\smallskip}
\noalign{components of intrinsic models}
\noalign{\smallskip}
$I'_s,I'_d$  & stellar and circumstellar (dust or disk) intensity \\
$I'$         & total intensity $I'(x,y)=I'_d(x,y)+I'_s(x_0,y_0)$ \\
$Q'_s,Q'_d$  & same for Stokes $Q'$ (Eq.~\ref{QIntr})\\
$U'_s,U'_d$  & same for Stokes $U'$ (Eq.~\ref{UIntr})\\
$Q'_\phi$    & circumstellar azimuthal polarization $Q'_\phi=Q'_{\phi,d}$ \\  
$U'_\phi$    & $U'_\phi=0$ in the adopted models \\
\noalign{\smallskip}  
${\cal{P}}'_d(\Sigma)$ $\theta'_d(\Sigma)$ & aperture polarization and
  position angle for disk \par (Eqs.~\ref{EqAperP} and \ref{EqApertheta}) \\
\noalign{\smallskip}    
${\cal{P}}'_s, \theta'_s$ & same for point like star (${\cal{P}}'_s=P'_s(x_0,y_0)$) \\  

\noalign{\smallskip\hrule}    
\end{tabular}
  \tablefoot{The polarization parameters are aligned with the
  $x,y$-coordinates of the scattering models.
  \tablefoottext{a}{see Fig.~\ref{Fig.NormDisk};}
  \tablefoottext{b}{there is $\Sigma=\Sigma_1+\Sigma_2+\Sigma_3$;}
  \tablefoottext{c}{see Figs.~\ref{FigRingi60}, \ref{Fig.NormDisk};}
  \tablefoottext{d}{there is $\Sigma_i Q=\Sigma_i Q_{000}+\Sigma_i Q_{090}
         +\Sigma_i Q_{180}+\Sigma_i Q_{000}$}.}
\end{table}

\begin{table}
  \caption{Polarization parameter for the zp-corrected models.}
  \label{ParXYTabzp}
  \begin{tabular}{p{1.0cm}p{7.2cm}}
\noalign{\hrule\smallskip}    
\noindent param. & description \\
\noalign{\smallskip\hrule\smallskip}    
$\Sigma_{\rm zp}$ & zero-point correction region $\Sigma_{\rm zp}=\Sigma_i$\\
\noalign{\smallskip}
\noalign{fractional polarization correction factor}
$\langle q_i \rangle$ & equal to $\Sigma_i Q/\Sigma_i I$ \\
$\langle u_i \rangle$ & equal to $\Sigma_i U/\Sigma_i I$ \\
\noalign{\smallskip}
\noalign{fractional Stokes polarization offsets}
$q_{\rm is},u_{\rm is}$ & for interstellar polarization
                  (Eqs.~\ref{Eq.qis} and \ref{Eq.uis})\\
$q_{\rm inst},u_{\rm inst}$ & for instrumental polarization \\
$q_s,u_s$           & for intrinsic stellar polarization \\
$q_{d,c},u_{d,c}$     & for intrinsic, unresolved circumstellar polarization \\
\noalign{\smallskip}
\noalign{zp-corrected maps}
\noalign{\smallskip}
$Q^{{\rm z}i}(x,y)$ & zp-corrected Stokes $Q$ map
                  for $\Sigma_{\rm zp}=\Sigma_i$) \\
$U^{{\rm z}i}(x,y)$ & same for zp-corrected Stokes $U$ map \\
\noalign{\smallskip}
$Q^{{\rm z}i}_\phi(x,y)$ $U^{{\rm z}i}_\phi(x,y)$
                    & $\Sigma_i$ zp-corrected azimuthal polarization
                        maps \par
                    ~~~~~ derived from $Q^{{\rm z}i}(x,y)$ and $U^{{\rm z}i}(x,y)$\tablefootmark{a} \\
\noalign{\smallskip}
\noalign{integrated zp-corrected parameters}                   
\noalign{\smallskip}
$\Sigma_{i'}X^{{\rm z}i}$ & $\Sigma_{\rm zp}=\Sigma_i$ corrected $X$-parameter
                        integrated in $\Sigma_{i'}$ \\
\noalign{\smallskip}
$\Sigma_{i'}X^{{\rm z}i}_{xxx}$ & same for quadrant parameters \\

\noalign{\bigskip}
\noalign{special quantities for center zp-correction}
\noalign{\smallskip}
$\Sigma_0$    & aperture for central pixels of the stellar PSF peak\\
\noalign{\smallskip}
$q_0,u_0$     & fractional polarization for central pixels  \\
$Q_0(x,y)$ $U_0(x,y)$ & Stokes flux $q_0\Sigma I(x,y)$ and \par
                   Stokes flux $u_0\Sigma I(x,y)$ \\
\noalign{\smallskip}
$Q^{\rm ex}(x,y)$ $U^{\rm ex}(x,y)$ & Stokes $Q$ and $U$ maps
                         for center corrected, spatially \par
                         ~~~~~resolved polarization\tablefootmark{b} \\
\noalign{\smallskip}
$Q_\phi^{\rm ex}(x,y)$ $U_\phi^{\rm ex}(x,y)$ & azimuthal polarization maps
for center corrected, \par
                   ~~~~~spatially resolved polarization \\

\noalign{\smallskip}
\noalign{integrated zp-corrected parameters}                   
\noalign{\smallskip}
$\Sigma Q^{\rm ex}$ $\Sigma U^{\rm ex}$ & integrated Stokes
                     polarization for center corrected, \par
                     ~~~~~resolved polarization signal \\
\noalign{\smallskip}
$\Sigma X^{\rm ex}_{xxx}$ & corresponding quadrant polarization values \\ 
\noalign{\smallskip}
$\Sigma Q_\phi^{\rm ex}$ $\Sigma U_\phi^{\rm ex}$ & integrated azimuthal
                     polarization for center \par
                     ~~~~~corrected, resolved polarization signal \\
\noalign{\smallskip}
$\Sigma Q,\,\Sigma U$ & total signal $\Sigma Q=\Sigma Q_0+\Sigma Q^{\rm ex}$
                   and $\Sigma U=\Sigma U_0+\Sigma U^{\rm ex}$ \\
\noalign{\smallskip\hrule}    
\end{tabular}
  \tablefoot{
   The polarization parameters are aligned with the
   $x,y$-coordinates of the scattering models.
    \tablefoottext{a}{see Table~\ref{Tab.ConvNorm} for all
      possible combinations;}
    \tablefoottext{b}{$Q^{\rm ex}$ and $U^{\rm ex}$ are equal to $Q^0$ and $U^0$}
    .}
\end{table}

\section{Faint PSF$_{\rm AO}$ artifacts}
\label{Sect.artifacts}

The extended halo of the PSF$_{\rm AO}$,
in particular the strong speckle ring at a separation
of about 0.45\arcsec, can introduce weak artefacts in the
convolved polarization maps.
For an axisymmetric, compact system with $r_0<100~$mas the
positive-negative Stokes $Q$ and $U$ quadrant patterns are not
resolved by the PSF halo and therefore there is no
net $\Sigma Q$ and $\Sigma U$ signal which can produce a
polarized halo ghost signal.
However, for
more extended scattering regions the quadrant patterns are also
resolved by the PSF halo and this can produce various kinds of
weak spurious features.

\begin{figure}
\includegraphics[trim=1.0cm 1.7cm 1cm 1.2cm, width=8.8cm]{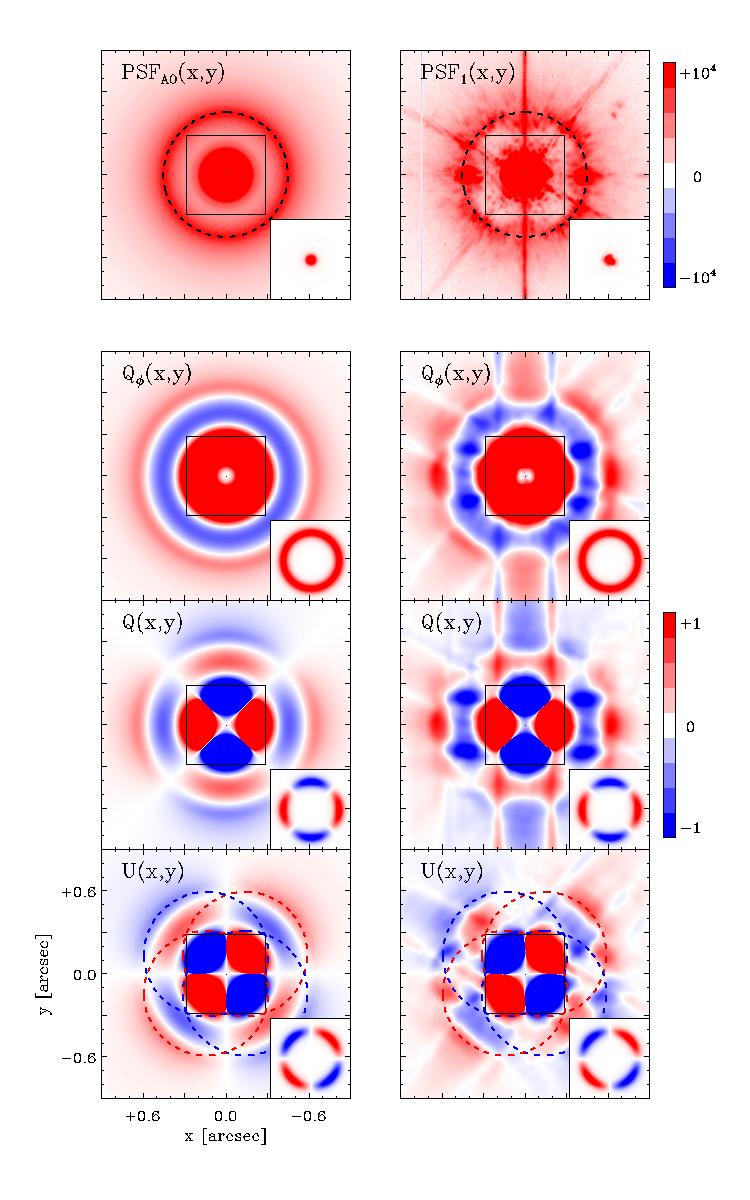}
\caption{Faint polarization articfacts in the 
  maps $Q_\phi(x,y)$, $Q(x,y)$, $U(x,y)$ of the Ring0 model
  $r_0=201.6$~mas produced by the axisymmetric
  PSF$_{\rm AO}(x,y)$ (left column), and a 2D single frame speckle
  PSF$_1$ of an AO system (right column). Dashed circles are indicated
  for the PSFs and the Stokes $U$ maps to indicate the location of
  the halos produced by the speckle ring. The color scale
  is strongly saturated in the centre of the main maps and therefore
  this region is plotted as inset on the lower right
  with a 200~$\times$ wider color scale.}
\label{FigRing200mas}
\end{figure}

One example is illustrated in Fig.~\ref{FigRing200mas} for the
Ring0 model with $r_{\rm 0}=201.6~$mas convolved with PSF$_{\rm AO}$, but also 
with PSF$_1$ representing the full 2D speckle structure of a single
frame from SPHERE/ZIMPOL
\citep[see][for PSF details]{Schmid18}. If multiple polarimetric
measurements are taken in field rotation mode and then combined
in a derotated data set, then the patterns in PSF$_1$ will average towards
the case PSF$_{\rm AO}$. The panels in Fig.~\ref{FigRing200mas}
emphasize the weak outer structure in the convolved $Q_\phi(x,y)$
and $Q(x,y)$, $U(x,y)$ maps using a strongly saturated color scale.
The insets on the lower right show the central ring signals with a
200~$\times$ wider color scale.

The PSF$_{\rm AO}$ speckle ring, indicated in Fig.~\ref{FigRing200mas} 
for the PSFs by a dashed circle, produces in the Stokes maps
for the positive or negative Stokes quadrants of the circumstellar
polarization ring, a corresponding circular positive
and negative ghost on top of the disk signal. This is indicated
in the $U(x,y)$ maps with four dashed circles centered on the
$U$ ring quadrants. These ghosts add up to a weak
positive/negative tile pattern for
the convolution with PSF$_1$ with its quadratic
substructure, while the pattern is much smoother for the
PSF$_{\rm AO}$ convolved $U$ maps. The same ghosts, but rotated by $45^\circ$,
are present in the Stokes $Q$ maps. Therefore, there
results for the convolved $Q_\phi(x,y)$ map a weak, negative ring
(a radial polarization signal!) just outside the disk ring, and
a weak positive ring further out.
This is a special case
because the polarization is concentrated in a narrow ring
with a radius of about half the size of the PSF$_{\rm AO}$
speckle ring. The SB of the spurious negative $Q_\phi$-signal at
$r\approx 460$~mas is at the level of $-0.2~\%$ of the positive
$Q_\phi$ SB peak signal of the disk ring. The negative $Q_\phi$ signal introduces
also a small discrepancy between the integrated $\Sigma Q_\phi$ and
$\Sigma P$ values despite the strict axisymmetry.

Another type of circular ghost pattern appears for the large Ring0 model with
$r_0\approx 800$~mas, where the PSF$_{\rm AO}$ convolution
produces two weak, positive $Q_\phi$ ghost rings, one just outside
and one just inside of the bright disk ring. 
These features are also very faint and can often be neglected 
in typical polarimetric data taken with current AO systems.
However, one should be careful with the interpretation of
faint structures near bright scattering regions. The
protoplanetary disk around HD 169142 is a case with a narrow
bright ring, where this type of ghost had to be taken into
account in the measurements of \cite{Tschudi21}
for the determination of the faint signal from the disk region
located further out.

\section{Parameters for Disk0 models}

\begin{table}
  \caption{Results for the axisymmetric models Disk0$\alpha$0, Disk0$\alpha$-1,
    and Disk0$\alpha$-2.}
  \label{TabPoleOnDisk}
  \begin{tabular}{lcccccc}
\noalign{\hrule\smallskip}    
\multispan{2}{\hfil $r_{\rm in}$\hfil}  &
     $\hspace{-0.8mm}\Sigma Q'_\phi/Q'_{\phi,{\rm ref}}\hspace{-0.8mm}$ & 
     \multispan{2}{\hfil $\Sigma Q_\phi/Q'_{\phi,{\rm ref}}$ \hfil} &  
     \multispan{2}{\hspace{-2mm}$r({\rm max}(Q_\phi))~~r_h(Q_\phi)$
        \hspace{-0.2mm}} \\      
      &       & \hspace{-1mm}intrin.\hspace{-1mm} & PSF$_{\rm G}$ & PSF$_{\rm AO}$
              & \hspace{-0.4mm}PSF$_{\rm AO}$\hspace{-0.2mm} & \hspace{-0.3mm}PSF$_{\rm AO}$\hspace{-0.3mm} \\
     $[{\rm mas}]$  & \hspace{-8mm}$D_{\rm PSF}$\hspace{-8mm}
                           & [$\%$]  & [$\%$]  & [$\%$]
              & \hspace{-0.4mm}[$D_{\rm PSF}$]\hspace{-0.2mm} & \hspace{-0.3mm}[$D_{\rm PSF}$]\hspace{-0.3mm} \\ 
\noalign{\smallskip\hrule\smallskip}
\noalign{Disk0$\alpha$0}
\noalign{\smallskip}
0     & 0   & 101  & 91.6  & 51.1
                      & \hspace{-0.2mm} (2.70)\hspace{-0.2mm} & 0.84  \\
12.6  & .5 & $\hspace{-0.2mm}Q'_{\phi,{\rm ref}}\hspace{-0.2mm}$
& 91.3  & 51.0
                      & \hspace{-0.2mm} (2.70)\hspace{-0.2mm} & 0.88 \\
25.2   & 1   & 95.2  & 88.9  & 50.0
                      & \hspace{-0.2mm} (2.73)\hspace{-0.2mm} & 1.09 \\
50.4   & 2   & 76.2  & 73.0  & 42.4
                     & \hspace{-0.2mm} (2.98)\hspace{-0.2mm} & 1.86 \\
\noalign{\smallskip}
\noalign{Disk0$\alpha$-1}
\noalign{\smallskip}
0      & 0    & 114  & 86.4  & 46.0  & \hspace{-0.2mm} 1.05 &  0.43 \\
3.15   & \hspace{-0.2mm}.125\hspace{-0.2mm}
              & 111  & 86.4  & 46.0  & \hspace{-0.2mm} 1.05 &  0.43 \\
6.30   & .25  & 107  & 86.2  & 45.9  & \hspace{-0.2mm} 1.09 &  0.44 \\
12.6   & .5  & $\hspace{-0.2mm}Q'_{\phi,{\rm ref}}\hspace{-0.2mm}$
                       & 85.0  & 45.4  & \hspace{-0.2mm} 1.23 &  0.50 \\
25.2   & 1  & 85.7     & 78.0  & 42.5  & \hspace{-0.2mm} 1.70 &  0.86 \\
50.4   & 2  & 57.1     & 54.6  & 31.4  & \hspace{-0.2mm} 2.63 &  1.77 \\

\noalign{\smallskip}
\noalign{Disk0$\alpha$-2}
\noalign{\smallskip}
3.15   & \hspace{-0.2mm}.125\hspace{-0.2mm}
             & 171:
                      & 81.1  & 40.0  &  \hspace{-0.2mm}0.77 &  0.34 \\
6.30   & .25 & 133  & 79.5  & 39.3  &  \hspace{-0.2mm}0.80 &  0.35 \\
12.6   & .5  & $\hspace{-0.2mm}Q'_{\phi,{\rm ref}}\hspace{-0.2mm}$
                      & 74.1  & 37.1  &  \hspace{-0.2mm}0.91 &  0.41 \\
25.2   & 1  & 66.6    & 58.6  & 30.6  &  \hspace{-0.2mm}1.41 &  0.77 \\
50.4   & 2  & 33.3    & 31.7  & 18.1  &  \hspace{-0.2mm}2.48 &  1.71 \\
\noalign{\smallskip}
err    &    &  $\pm 0.3$ &  $\pm 0.2$ &  $\pm 0.2$ &  \hspace{-0.4mm}$\pm 0.01$ &\hspace{-0.2mm}$\pm 0.01$ \\
\noalign{\smallskip\hrule}
  \end{tabular}
  \tablefoot{Results for the intrinsic azimuthal polarization $\Sigma Q'_\phi$,
    and the PSF$_{\rm G}$ and PSF$_{\rm AO}$ convolved values $\Sigma Q_\phi$,
    are listed for models with different inner radii $r_{\rm in}$.
    The $\Sigma Q'_\phi$ and $\Sigma Q_\phi$
    values for a given model type Disk0$\alpha$0, Disk0$\alpha$-1,
    or Disk0$\alpha$-2 are normalized to the corresponding reference values
    $Q'_{\phi,{\rm ref}}=\Sigma Q'_\phi (r_{\rm in}=0.5 D_{\rm PSF})$.
    Also given are the
    characteristic radii $r({\rm max}(Q_\phi))$ and $r_h(Q_\phi)$ for
    PSF$_{\rm AO}$ convolved models. }
\end{table}  

Table~\ref{TabPoleOnDisk} gives numerical values for the simulation
of the axisymmetric, extended models Disk0$\alpha$0, Disk0$\alpha$-1,
and Disk0$\alpha$-2 described in Sect.~\ref{Sect.Exti0Models}.
For all models with the same $\alpha$-index the intrinsic
$\Sigma Q'_\phi$, and the PSF$_G$ and PSF$_{\rm AO}$ convolved $\Sigma Q_\phi$
values are given relative to the reference value
$Q'_{\phi,{\rm ref}}=\Sigma Q'_\phi (r_{\rm in}=0.5\,D_{\rm PSF})$.

The intrinsic $\Sigma Q'_\phi/Q'_{\phi,{\rm ref}}$ value for the
Disk0$\alpha$0 model with flat surface brightness (SB) plotted in
Fig.~\ref{ProfFlatPoleOn} changes not much between
$r_{\rm in}=0$ and $0.5\,D_{\rm PSF}$. For the convolved
signal there is almost no difference in $\Sigma Q_\phi/Q'_{\phi,{\rm ref}}$
for $r_{\rm in}\lapprox D_{\rm PSF}$ and one cannot distinguish easily
between different $r_{\rm in}$ cases. One can use the hole radius
$r_{\rm h}(Q_\phi)$ to distinguish between models with $r_{\rm in}=0.5\,D_{\rm PSF}$
and $1.0\,D_{\rm PSF}$, but this would require for observational
studies high quality data. 

For the Disk0$\alpha$-2 the intrinsic 
$\Sigma Q'_\phi/Q'_{\phi,{\rm ref}}$ signal for a small cavity
$r_{\rm in}=0.125~D_{\rm PSF}$
is much larger than the reference value, because the brightness
increases rapidly towards the center (see Fig.~\ref{Profalpham2PoleOn}).
For the convolved signal $\Sigma Q_\phi/Q'_{\phi,{\rm ref}}$ the contributions
from the central disk regions are
strongly suppressed by polarimetric cancellation, but
there is still a difference of about 20~\% to 30~\%  between the
model with $r_{\rm in}=0.5\,D_{\rm PSF}$ and $1.0\,D_{\rm PSF}$. Also the
radii $r({\rm max}(Q_\phi))$ and $r_{\rm h}(Q_\phi)$
are quite different between these two cases. For the Disk0$\alpha$-1 models
these relative differences are intermediate between Disk0$\alpha$0 and
Disk0$\alpha$-2.

\section{Parameters for inclined disks}

\begin{table*}
  \caption{Results for RingI60 models with different $r_0$.}
\label{TabRingi60}
\begin{tabular}{l ccccc ccccc ccccc cc}
\noalign{\hrule\smallskip}
$r_0$ & \multispan{7}{\hfil integrated parameters
                          $\Sigma X/\Sigma Q'_\phi$ \hfil}
             & \hspace{-3mm} cross talk \hspace{-3mm}
      & \multispan{5}{\hfil quadrant values
                          $\Sigma X_{xxx}/\Sigma Q'_\phi$ \hfil} & note \\
       & $\Sigma Q_\phi$ & $\Sigma Q$ & $\Sigma P$
                          & $\Sigma |Q_\phi|$ & $\Sigma |U_\phi|$
                                & $\Sigma |Q|$ & $\Sigma |U|$ 
       & $\Sigma |U_\phi|$ & $Q_{000}$ & $Q_{090}$ & $Q_{180}$ & $U_{045}$ & $U_{135}$
                                          &  \\
mas   & [\%] & [\%] & [\%] & [\%] & [\%] & [\%] & [\%] &
                            $\overline{\Sigma |Q_\phi|}$ &
             [\%] & [\%] & [\%] & [\%] & [\%] &  \\
\noalign{\smallskip\hrule\smallskip}
\noalign{intrinsic RingI60 model}
\noalign{\smallskip}
all   & 100 & 42.1 & 100 & 100 & 0 & 70.3 & 56.6 & 0
           & $-0.7$ & $+28.1$ & $-13.4$  & $-4.1$ & $+24.3$  \\   
\noalign{\smallskip}
\noalign{unresolved RingI60 model}
\noalign{\smallskip}
unres.   & 0 & 42.1 & 42.1 &
26.8\tablefootmark{a} & 26.8\tablefootmark{a} & 42.1 & 0 & 1
         & $+10.5$ & $+10.5$ & $+10.5$ & $0 $ & $0$ & \\
\noalign{\smallskip}
\noalign{PSF$_{\rm G}$ convolved}
\noalign{\smallskip}
3.15\tablefootmark{b}
      & 1.5 & 42.1 & 42.6 & 27.1 & 27.1 & 42.1 & 5.1: & 1.00  
       & $+10.0$ & $+11.1$ & $+10.4$  & $+1.1$ & $+1.5$ \\    
6.3   & 5.7  & 42.1 & 43.7 & 28.0 & 27.7 & 42.1 & 10.0  & 0.99 
        & $+9.2$ & $+11.9$ & $+9.5$  & $+1.9$ & $+3.1$ \\ 
12.6  & 19.2 & 42.1 & 47.4 & 31.9 & 28.5 & 42.1 & 18.7 & 0.89 
        & $+7.1$ & $+14.3$ & $+5.7$  & $+2.7$ & $+6.7$ &    \\
25.2  & 51.3 & 42.1 & 62.8 & 52.7 & 26.1 & 47.8 & 32.0 & 0.50
        & $+2.1$ & $+20.7$ & $-1.6$  & $+1.4$ & $+13.4$ &   \\
50.4  & 83.2 & 42.1 & 86.6 & 83.4 & 18.0 & 62.8 & 46.6 & 0.22
        & $-0.3$ & $+25.7$ & $-9.0$  & $-2.0$ & $+20.4$ &    \\
100.8 & 95.7 & 42.1 & 98.4 & 95.7 &  8.5 & 68.3 & 53.9 & 0.09
        & $-0.6$ & $+27.4$ & $-12.2$ & $-3.6$ & $+23.3$ &    \\
201.6 & 98.7 & 42.1 & 98.9 & 98.7 &  3.2 & 69.6 & 55.8 & 0.03 
        & $-0.7$ & $+28.0$ & $-13.3$ & $-3.9$ & $+24.0$ &    \\
\noalign{\smallskip}
\noalign{PSF$_{\rm AO}$ convolved}
\noalign{\smallskip}
6.3\tablefootmark{b}
      &  2.5 & 41.2 & 41.7 & 26.6 & 26.6 & 41.2 & 4.8 & 1.00 
        & $+9.5$ & $+11.3$ & $+10.0$ & $+1.0$ & $+1.7$ &    \\
12.6  &  8.4 & 41.4 & 43.2 & 28.1 & 27.0 & 41.4 & 9.2  & 0.96
        & $+8.4$ & $+12.4$ & $+8.9$  & $+1.3$ & $+3.2$ &    \\
25.2  & 22.3 & 41.4 & 48.0 & 35.1 & 25.9 & 42.1 & 15.5 & 0.74
        & $+6.5$ & $+14.9$ & $+5.5$  & $+1.1$ & $+6.5$ &   \\
\noalign{\smallskip}
50.4  & 40.4 & 41.4 & 58.7 & 49.5 & 22.4 & 47.7 & 23.7  & 0.45
        & $+4.5$ & $+18.0$ & $+1.0$  & $0.0$ & $+10.6$ &    \\  
~~disk & 40.3 & 29.4 & 46.7 & 41.8 & 14.8 & 35.7 & 22.9 & 0.35
         & $+1.5$ & $+15.0$ & $-1.9$ & $0.0$ & $+10.4$
                                                 &  $<0.2$\arcsec \\
~~halo &  0.1 & 12.0 & 12.0 &  7.7 & 7.7 & 12.0 & 0.9 & 1.00
         & $+3.0$ & $+3.0$ & $+3.0$  & $+0.1$ & $+0.1$ 
                                                 &  $>0.2$\arcsec \\
\noalign{\smallskip}
100.8 & 55.9 & 41.4 & 70.2 & 63.7 & 17.9 & 54.3 & 32.5  & 0.28
        & $+3.2$ & $+20.5$ & $-2.8$  & $-1.2$ & $+14.0$ &    \\
201.6 & 63.9 & 41.1 & 76.1 & 71.1 & 13.8 & 57.4 & 37.9  & 0.19 
        & $+2.7$ & $+21.6$ & $-4.9$  & $-1.9$ & $+15.9$ &    \\
403.2\tablefootmark{c}
     & 68.6 & 40.9 & 78.1 & 73.9 & 10.1 & 57.9 & 40.7  & 0.14
        & $+2.2$ & $+22.4$ & $-6.1$  & $-2.0$ & $+17.2$ &    \\
806.4\tablefootmark{c}
     & 74.9 & 39.8 & 79.6 & 76.4 & 8.4 & 57.8  & 43.1   & 0.11
        & $+0.9$ & $+23.4$ & $-7.9$  & $-2.0$ & $+18.7$ &    \\
\noalign{\smallskip\hrule}
\end{tabular}
\tablefoot{The table lists the integrated polarization
    $\Sigma X$ and quadrant polarization parameters $\Sigma X_{\rm xxx}$, and 
    the $\Sigma |U_\phi|/\Sigma |Q_\phi|$ convolution cross talk ratios for
    RingI60 models with different $r_0$ convolved with PSF$_{\rm G}$
    and PSF$_{\rm AO}$. All $\Sigma X$ and $\Sigma X_{xxx}$ values are normalized
   by the intrinsic polarization $\Sigma Q'_\phi=100~\%$.
   \tablefoottext{a}{Analytical result for an unresolved model
     convolved with an extended PSF,\,\,}
  \tablefoottext{b}{small models with uncertainties larger than
    two unit in the last digit for some parameters because
    of the limited numerical sampling for the innermost region,\,\,}
  \tablefoottext{c}{large disks, for which
    more than 2~\% of $\Sigma Q$ falls outside of the $r<1.5''$
    integration region.}}
\end{table*}  

\subsection{RingI60 models}
Results of the RingI60 simulations discussed in Sect.~\ref{Sect.RingI60}
are listed in Table~\ref{TabRingi60}.
All values for the integrated polarization parameters are normalized
$\Sigma X/\Sigma Q'_\phi$ to the intrinsic azimuthal polarization
with $X=\{Q_\phi,Q,P,|Q_\phi|,|U_\phi|,|Q|,|Q|\}$.

The same normalization is applied to the five quadrant polarization
values $\Sigma X_{xxx}/\Sigma Q'_\phi$ with
$X_{xxx}=\{Q_{000},Q_{090},Q_{180},U_{045},U_{135}\}$. Also given is the
ratio between $\Sigma |U_\phi|/\Sigma |Q_\phi|$ as measure for
the $Q_\phi\rightarrow U_\phi$ convolution cross-talk.
The values in Table~\ref{TabRingi60} were numerically integrated
for an aperture with $r=1.5''$.

The first line in Table~\ref{TabRingi60} gives the values
for the intrinsic model and the second line
for a convolved but spatially unresolved RingI60 model.
For $\Sigma |Q_\phi|$ and $\Sigma |U_\phi|$ of the
unresolved model the value $(2/\pi)\,\Sigma Q$ is given, which
follows from the analytic solution for the azimuthal integration of
$Q_\phi(\phi) = -Q \cos(2\phi)$ and $U_\phi = -Q \sin(2\phi)$
of a ``perfect'' quadrant pattern for a PSF convolved point source
with a polarization $Q$ 

The $\Sigma X/\Sigma Q'_\phi$-values for the PSF$_{\rm G}$
and PSF$_{\rm AO}$ convolved models with different $r_0$ correspond to the
diagrams in Fig.~\ref{DiagAperg6i60Ring}, while the curves for
the quadrant values
$\Sigma X_{xxx}/\Sigma Q'_\phi$ are shown in the upper panels of
Fig.~\ref{DiagQuadg6i60}.

For the PSF$_{\rm AO}$ convolved model with $r_0=50.4~$mas not
only values for an aperture with $r=1.5''$ are given,
but also the splitted values for the disk region using a
circular aperture with $r=0.2''$ and for the halo region using
an annular aperture from $r=0.2''$ to $1.5''$. These values are
given to support the discussion of the weak halo signal in
Sect.~\ref{Sect.Halosignal} and Fig.~\ref{FigRingI60AOLRG}. 

The numerical integration of the polarization fluxes
is affected for the smallest rings by the limited spatial sampling of
$0.9~{\rm mas} \times 0.9~{\rm mas}$ per pixel.
This produces errors larger than two units in the
last indicated digit for some tabulated values for the
PSF$_{\rm G}$ convolved model with $r_0=3.15$~mas and for
the PSF$_{\rm AO}$ convolved model with $r_0=6.3$~mas. 

Smearing for PSF$_{\rm AO}$ convolved models puts some flux
outside of the aperture radius $r=1.5''$. The strength of this
effect follows from convolved small disk models $r_0\leq 201.6$~mas
which give $\Sigma Q/\Sigma Q'_\phi\approx 41.4~\%$ instead of
the intrinsic value 42.1~\%.
For extended disks with $r_0>200$~mas additional disk polarization 
is ``lost'' outside the aperture because of the convolution.

\begin{table}
  \caption{Differential quadrant values $\Delta Q_{xxx}$ and $\Delta U_+$
    for RingI60 models.}
  \label{Tab.diffQuad}
  \begin{tabular}{lcccc}
\noalign{\hrule\smallskip}    
$r_0$  & \multispan{3}{\hfil $\Delta Q_{xxx}/\Sigma Q_\phi$ \hfil} &      
                                    $\Delta U_+/\Sigma Q_\phi$ \\
mas  & $Q_{000}$ & $Q_{090}$ & $Q_{180}$  &   \\ 
\noalign{\smallskip\hrule\smallskip}
\noalign{intrinsic RingI60 model}
\noalign{\smallskip}
all   & $-0.112$ & $+0.176$ & $-0.239$ & $+0.284$ \\
\noalign{\smallskip}
\noalign{convolved with Gaussian PSF$_{\rm G}$}
\noalign{\smallskip}
12.6  & $-0.178$ & $+0.197$ & $-0.251$ & $+0.208$     \\ 
25.2  & $-0.164$ & $+0.198$ & $-0.236$ & $+0.238$     \\ 
50.4  & $-0.130$ & $+0.182$ & $-0.235$ & $+0.269$        \\ 
100.8 & $-0.116$ & $+0.176$ & $-0.237$ & $+0.281$     \\ 
\noalign{\smallskip}
\noalign{convolved with extended PSF$_{\rm AO}$}
\noalign{\smallskip}

12.6  & $-0.232$ & $+0.244$ & $-0.173$ & $+0.226$      \\
25.2  & $-0.173$ & $+0.204$ & $-0.217$ & $+0.241$      \\
50.4  & $-0.145$ & $+0.189$ & $-0.231$ & $+0.262$     \\
100.8 & $-0.128$ & $+0.182$ & $-0.235$ & $+0.272$     \\

201.6 & $-0.119$ & $+0.177$ & $-0.237$ & $+0.279$      \\
403.2 & $-0.117$ & $+0.177$ & $-0.238$ & $+0.280$     \\
806.4 & $-0.121$ & $+0.180$ & $-0.238$ & $+0.276$     \\

\noalign{\smallskip\hrule}
\end{tabular}
\tablefoot{The differential quadrant values are normalized to
  $\Sigma Q_\phi$ of the corresponding RingI60 models characterized by
  $r_0$ and the used convolution PSF$_{\rm G}$ or PSF$_{\rm AO}$. The
  intrinsic values do not depend on $r_0$.}
\end{table}

The differential Stokes quadrant values
$\Delta Q_{000}$, $\Delta Q_{090}$, $\Delta Q_{180}$, and $\Delta U_+$
are listed in Table~\ref{Tab.diffQuad}. They are
expressed as ratio relative to $\Sigma Q_\phi$ for the RingI60 models
with different $r_0$ based on the values given in
Table~\ref{TabRingi60}. They are derived 
according to the Eqs.~\ref{Eq.Qdiff} and \ref{Eq.Udiff}
and are also plotted in the lower panels of Fig.~\ref{DiagQuadg6i60}.

\subsection{DiskI60 models}

\begin{table}
  \caption{Results for the DiskI60$\alpha$0, DiskI60$\alpha$-1, and
    DiskI60$\alpha$-2 models.}
\label{TabDiski60}  
\begin{tabular}{lccccccc}
\noalign{\hrule\smallskip}
\multispan{2}{\hfil $r_{\rm in}$\hfil}  & \multispan{2}{\hfil intrinsic \hfil}
       & \multispan{2}{\hfil PSF$_{\rm G}$ conv. \hfil}
                          & \multispan{2}{\hfil PSF$_{\rm AO}$ conv. \hfil} \\
    &  & $\Sigma Q'_\phi$ & $\Sigma Q'$ 
           & $\Sigma Q_\phi$ & $\Sigma P$
                   & $\Sigma Q_\phi$ & $\Sigma P$ \\
${\rm [mas]}$ & $D_{\rm PSF}$ & [\%]   & [\%]   & [\%]
                              & [\%]   & [\%]   & [\%]   \\
\noalign{\smallskip\hrule\smallskip}
\noalign{\noindent DiskI60$\alpha$0}
\noalign{\smallskip}
0     &  0  & 102  & 42.8  &  85.9  & 88.0 & 45.6  & 62.2  \\ 
12.6  & 0.5 & $Q'_{\phi,{\rm ref}}$   
                    & 42.1  &  85.8  & 87.5 & 45.5  & 61.7  \\
25.2  & 1    & 95.6 & 40.0  &  83.9  & 85.3 & 44.7  & 59.8  \\
50.4  & 2    & 76.2 & 32.1  &  70.2  & 71.1 & 38.4  & 50.1  \\
\noalign{\smallskip}
\noalign{\noindent DiskI60$\alpha$-1}
\noalign{\smallskip}
0     &  0    & 114  & 47.4  &  78.5 &  85.5  & 40.2 &  62.2 \\
3.15  & 0.125 & 111  & 46.9  &  78.5 &  84.8  & 40.2 &  61.5 \\
6.30  & 0.25  & 107  & 45.0  &  78.4 &  83.5  & 40.1 &  60.1 \\
12.6  & 0.5   & $Q'_{\phi,{\rm ref}}$
             & 42.1  &  77.5 &  81.0  & 39.7 &  57.6 \\ 
25.2  & 1    & 85.6 & 36.0  &  72.4 &  74.2  & 37.5 &  51.8 \\
50.4  & 2    & 57.2 & 24.1  &  52.3 &  53.0  & 28.3 &  37.2 \\
\noalign{\smallskip}
\noalign{\noindent DiskI60$\alpha$-2}
\noalign{\smallskip}
3.15  & 0.125 & 167  & 73.7  &  69.1  &  101   & 33.6  &  80.6 \\ 
6.30  & 0.25  & 135  & 55.6  &  68.2  &  84.4  & 33.2  &  64.5 \\ 
12.6  & 0.5   & $Q'_{\phi,{\rm ref}}$
              & 42.1  &  64.5  &  71.6  & 31.5  &  52.3 \\
25.2  & 1     & 66.6  & 28.0  &  53.2  &  55.2  & 26.6  &  38.5 \\
50.4  & 3     & 33.4  & 14.1  &  30.4  &  30.8  & 16.3  &  21.5 \\  
\noalign{\smallskip\hrule}
\end{tabular}
\tablefoot{The table lists integrated azimuthal polarization
  $\Sigma Q'_\phi/Q'_{\phi,{\rm ref}}$ for the intrinsic models,
  and $\Sigma Q_\phi/Q'_{\phi,{\rm ref}}$ and the polarization
  $\Sigma P/Q'_{\phi,{\rm ref}}$ for the models convolved with the
  Gaussian PSF$_{\rm G}$ and PSF$_{\rm AO}$. The reference values
  are defined by $Q'_{\phi,{\rm ref}}=\Sigma Q'_\phi=100~\%$ for the model
  with $r_{\rm in}=0.5\,D_{\rm PSF}$ (12.6~mas). The values for the models
  DiskI60$\alpha$-2 are plotted in Fig.~\ref{DiagAperg6i60Disks}. 
} 
\end{table}

Table~\ref{TabDiski60} gives integrated polarization parameters for
the extendend, $i=60^\circ$ inclinded disk models
DiskI60$\alpha$0, DiskI60$\alpha$-1, and DiskI60$\alpha$-2 for
different cavity sizes $r_{\rm in}$. The outer
disk radius is in all cases $r_{\rm out}=100.8~$mas. 
The model with $r_{\rm in}=0.5\,D_{\rm PSF}$ (12.6~mas) represents for
all models with the same $\alpha$-index the reference value
$Q'_{\phi,{\rm ref}}=\Sigma Q'_\phi(r_{\rm in}=0.5\,D_{\rm PSF})$ as in
Table~\ref{TabPoleOnDisk}.

The model parameters
$\alpha$, $r_{\rm in}$, and $r_{\rm out}$ for the inclined disks
corresponds to Table~\ref{TabPoleOnDisk} for pole-on models,
except for the disk inclination.
Therefore, the intrinsic polarization values
$\Sigma Q'_\phi/Q'_{\phi,{\rm ref}}$ are practically the same
apart for some numerical sampling errors for disks with bright ($\alpha=-2$)
and barely resolved ($r_{\rm in}< 0.5\,D_{\rm PSF}$) inner regions.  
For the selected disk inclination the ratio $\Sigma Q'/\Sigma Q'_\phi$
is for all intrinsic models 0.421. The Stokes value $\Sigma Q$ is not
changed by the convolution $\Sigma Q/\Sigma Q'=1$, but for the
PSF$_{\rm AO}$ convolved results this value is $\Sigma Q/\Sigma Q'=0.98$
because a small amount of the smeared $Q$-signal is located outside
the used aperture with $r=1.5''$. 

The values for $\Sigma Q_\phi$, $\Sigma Q$
($\approx \Sigma Q'=0.421\, \Sigma Q'_\phi$) and $\Sigma P$ are plotted
for convolved DiskI60$\alpha$-2 models in Fig.~\ref{DiagAperg6i60Disks}
as function of $r_{\rm in}$. Of interest is the ratio $\Sigma Q_\phi/\Sigma Q$, 
which depends on the cancellation of the $Q_\phi$ signal in particular
for bright, barely resolved inner disk regions. Table~\ref{TabDiski60}
gives the values for this dependence also for the models DiskI60$\alpha$0 and
DiskI60$\alpha$-1. 

The convolved polarized flux $\Sigma P$ differs for DiskI60 models
significantly from $\Sigma Q_\phi$, unlike for the pole-on models where
$\Sigma P \approx \Sigma Q_\phi$. The polarized flux $\Sigma P$ represents
a complex mix (Eq.~\ref{EqPQUQphiUphi}) of the azimuthal polarization
$Q_\phi$ and $U_\phi$ or of the Stokes $Q$ and $U$ components.

\section{ZP-corrected disk models}

\subsection{RingI60 model with $r_0=100.8$~mas}

\begin{table}
  \caption{Results for the RingI60 models with $r_0=100.8~$mas using
    different zp-corrections.}
\label{Tab.ConvNorm}  
\begin{tabular}{ll ccc ccc }
\noalign{\hrule\smallskip}
$\Sigma_{\rm zp}$ & $\Sigma_{\rm int}$ &
               \multispan{3}{\hfil $\Sigma_{i'} X/\Sigma Q'_\phi$ [\%] \hfil} &
               \multispan{3}{\hfil $\Sigma_{i'}Q_{xxx}/\Sigma Q'_\phi$ [\%] \hfil}  \\
$\Sigma_i$ & $\Sigma_{i'}$ & $Q_\phi$ & $Q$ & $P$ & $Q_{000}$ & $Q_{090}$ & $Q_{180}$ \\
\noalign{\smallskip\hrule\smallskip}
\noalign{\smallskip \noindent intrinsic parameters
  \smallskip}
      & $\Sigma$  &
    100. & $+42.1$ & 100. & $-0.7$ & $+28.1$ & $-13.4$ \\
\noalign{\smallskip \noindent PSF$_{\rm AO}$ convolved parameters \smallskip}
       & $\Sigma$  &  
    55.9 & $+41.4$ & 70.2 & $+3.2$ & $+20.5$ & $-2.8$ \\
\noalign{\smallskip}
      & $\Sigma_1$ & 
    0.2~ & $-$0.2 &  0.3  & $0.0$ & $0.0$ & $-0.2$  \\
      & $\Sigma_2$ & 
    55.2 & $+29.3$ & 57.5 & $+0.3$ & $+17.3$ & $-5.6$  \\
     & $\Sigma_3$ & 
    0.5 & $+12.2$ & 12.4 & $+3.0$  & $+3.2$  &  $+3.0$  \\
\noalign{\noindent $\Sigma_{\rm zp}$: whole system\smallskip}
$\Sigma$ & $\Sigma$ &
     55.8 & 0.0  &  82.6  & $-6.6$  & $+9.6$ & $-13.6$    \\
\noalign{\smallskip}
$\Sigma$ & $\Sigma_1$ &
      0.2 & $-18.4$ & 18.4 & $-4.2$ & $-4.9$ & $-5.1$   \\ 
$\Sigma$ & $\Sigma_2$ &
     55.2 & $+18.1$ & 61.7 & $-2.3$ & $+14.4$ & $-8.6$  \\ 
$\Sigma$ & $\Sigma_3$ &
     0.5  & $+0.3$  & 2.4 &  0.0      & $+0.2$  &  0.0  \\
\noalign{\smallskip \noindent $\Sigma_{\rm zp}$: central star \smallskip}
$\Sigma_1$ & $\Sigma$ &
    55.8 & $+41.8$ & 70.4 & $+3.4$ & $+20.6$ & $-2.6$   \\
\noalign{\smallskip}
$\Sigma_1$ & $\Sigma_1$ &
    0.2  & 0.0    & 0.4   & 0.0   &  0.0    & $-0.1$  \\
$\Sigma_1$ & $\Sigma_2$ &
    55.2 & $+29.7$ & 57.5 & $+0.3$ & $+17.3$ & $-5.5$ \\
$\Sigma_1$ & $\Sigma_3$ &
    0.5 & $+12.4$ & 12.5 & $+3.0$ & $+3.2$ & $+3.0$ \\
\noalign{\smallskip}
\noalign{\smallskip \noindent $\Sigma_{\rm zp}$: disk \smallskip}
$\Sigma_2$ & $\Sigma$ & 
   55.8 & $-66.3$ & 140. & $-22.3$ & $-7.7$ & $-31.1$ \\  
\noalign{\smallskip}
$\Sigma_2$ & $\Sigma_1$ &
    0.2 & $-47.6$ & 47.6 &  $-11.0$ & $-12.7$ & $-12.9$ \\
$\Sigma_2$ & $\Sigma_2$ &
    55.1 & 0.0  &  73.2  &   $-6.6$ & $+9.6$ & $-13.4$ \\
$\Sigma_2$ & $\Sigma_3$ &
    0.5  & $-18.7$ & 18.9 & $-4.8$ & $-4.6$ & $-4.8$ \\    
\noalign{\smallskip \noindent $\Sigma_{\rm zp}$: disk and halo \smallskip}
$\Sigma_{2+3}\hspace{-3mm}$ & $\Sigma$ &
    55.8 & $-32.9$ & 109. & $-14.4$ & $+1.1$ & $-22.3$ \\  
\noalign{\smallskip}
$\Sigma_{2+3}\hspace{-3mm}$ & $\Sigma_1$ & 
    0.2 & $-32.8$ & 32.9  & $-7.6$ & $-8.8$ & $-8.9$ \\
$\Sigma_{2+3}\hspace{-3mm}$ & $\Sigma_2$ &
    55.1 & $+9.1$ &  67.1 & $-4.4$ & $+12.0$ & $-10.9$   \\
$\Sigma_{2+3}\hspace{-3mm}$ & $\Sigma_3$ &    
    0.5  & $-9.1$ &  9.5  & $-2.4$ & $-2.2$  & $-2.4$    \\
\noalign{\smallskip \noindent $\Sigma_{\rm zp}$: halo \smallskip}
$\Sigma_3$ & $\Sigma$ &
    55.8 & $-1.2$ &  83.3 & $-6.8$ & $+9.3$ & $-13.9$   \\
\noalign{\smallskip}
$\Sigma_3$ & $\Sigma_1$ &    
    0.2  & $-18.9$ & 18.9 & $-4.3$ & $-5.0$ & $-5.2$  \\
$\Sigma_3$ & $\Sigma_2$ &      
    55.2 & $+17.7$ & 61.9 & $-2.4$ & $+14.3$ & $-8.6$ \\  
$\Sigma_3$ & $\Sigma_3$ &      
0.5 &  0.0  &  2.5  & $-0.1$  & $+0.1$ & $-0.1$  \\
\noalign{\smallskip\hrule}
\end{tabular}
\tablefoot{The table lists integrated polarization $X={Q_\phi,Q,P}$ and
  quadrant polarization values $Q_{\rm xxx}$ for different integration
  regions $\Sigma_{\rm int}$ of the PSF$_{\rm AO}$ convolved model, and for
  corresponding zp-corrected models using different zp-correction
  regions $\Sigma_{\rm zp}$. The parameters $\Sigma_{i'}X$ and $\Sigma_{i'}X_{xxx}$
  for the table columns stand for intrinsic values $X=X'$ and $Q_{xxx}=Q'_{xxx}$,
  for convolved values $X$ and $Q_{xxx}$, and for values $X^{{\rm z}i}$ and
  $Q_{xxx}^{{\rm z}i}$ with zp-corrections based on region $\Sigma_{zp}=\Sigma_i$,
  integrated in region $\Sigma_{i'}$
  (Table~\ref{ParXYTabzp}). All values are normalized to
  the intrinsic azimuthal polarization $\Sigma Q'_\phi$.} 
\end{table}

Table~\ref{Tab.ConvNorm} lists polarization parameters $Q_\phi$,
$Q$, $P$ and Stokes $Q$ quadrant values $Q_{000}$, $Q_{090}$, $Q_{180}$
for the RingI60 models with $r_0=100.8~$mas. Results
for many combination of used regions for the zp-correction
$\Sigma_{\rm zp}$ and integration
regions $\Sigma_{\rm int}$ are given as described
in Sects.~\ref{Sect.NormPolRing} and \ref{Sect.NormQuad} and partly shown in
Figs.~\ref{Fig.ConvNorm} and \ref{Fig.QuadNorm}.

The integration apertures are defined as follows: round apertures
$\Sigma$ with $r\leq 1.5''$ for the entire system and $\Sigma_1$
with $r\leq 0.027''$ for the star, and annular aperture $\Sigma_2$
with $0.027''< r\leq 0.2''$ for the disk region and $\Sigma_3$ with
$0.2''<r\leq 1.5''$ for the halo.
There is $\Sigma_1+\Sigma_2+\Sigma_3=\Sigma$. All polarization values in
Table~\ref{Tab.ConvNorm} are given relative to $\Sigma Q'_\phi$.

The first line gives values for the intrinsic model
which are identical to the first line in Table~\ref{TabDiski60},
where also the Stokes $U$ quadrant values $\Sigma U'_{045}$ and
$\Sigma U'_{135}$ are given.
For the intrinsic model the values for 
$\Sigma_2$ are identical to the system values $\Sigma$, because the
polarization signal in $\Sigma_1$ and $\Sigma_3$ are zero. 

The lines 2 to 5 in
Table~\ref{Tab.ConvNorm} list the polarization for the PSF$_{\rm AO}$
convolved RingI60 model for different integration regions $\Sigma_i$.
The values for $\Sigma$ are identical to entries in the fourth
last line of Table~\ref{TabDiski60}, which gives also the Stokes
$U$ quadrant values $\Sigma_2 U_{045}/\Sigma Q'_\phi=-1.2~\%$ and
$\Sigma_2 U_{135}/\Sigma Q'_\phi=+14.0~\%$. The values for the halo are
$\Sigma_3 U_{045}/\Sigma Q'_\phi=+0.2~\%$,
$\Sigma_3 U_{135}/\Sigma Q'_\phi=+0.3~\%$, while they are zero for
the $\Sigma_1$ region.
The values for $\Sigma_i=\Sigma_1$, $\Sigma_2$, and
$\Sigma_3$ show how the convolution changes the distribution
of the polarization radially.

The following lines in Table~\ref{Tab.ConvNorm} give the same parameters
as for the convolved model, but after zp-correction using different
reference regions $\Sigma_{\rm zp}=\Sigma$, $\Sigma_1$, $\Sigma_2$,
$\Sigma_{2+3}$ and $\Sigma_3$. It is obvious that the different
zp-correction offsets change strongly the values for $\Sigma_i Q$,
$\Sigma_i P$, and the Stokes quadrants $\Sigma_i Q_{xxx}$.

An important result is, that the azimuthal polarization
$\Sigma_i Q^{\rm z}_\phi$ is practically not changed by the applied correction
offsets. Also the Stokes $U$ quadrant values are not
changed because only offsets for the Stokes $Q$ polarization
component are applied in these simulations. The $Q$ quadrant values
$\Sigma_3 Q^{\rm z}_{000}$, $\Sigma_3 Q^{\rm z}_{090}$, and $\Sigma_3 Q^{\rm z}_{180}$
for the halo region
have after a zp-correction practically the same value, because
all quadrants are offset by the same amount. The same also applies
roughly for the Stokes $Q^{\rm z}$ quadrants integrated in the
stellar aperture $\Sigma_i=\Sigma_1$.

\subsection{ZP-corrected model RingI60 with $r_0=403.2$~mas}

\begin{table}
  \caption{Comparison for different
    zp-corrections for the extended model RingI60 with $r_0=403.2$~mas.}
\label{Tab.ConvNorm400}  
\begin{tabular}{lll ccc ccc }
\noalign{\hrule\smallskip}
$\Sigma_{\rm zp}$ & $\Sigma_{\rm int}$ &
               \multispan{3}{\hfil $\Sigma_{i'} X/\Sigma Q'_\phi$ [\%] \hfil} &
               \multispan{3}{\hfil $\Sigma_{i'} Q_{xxx}/\Sigma Q'_\phi$ \hfil}  \\
$\Sigma_i$ & $\Sigma_{i'}$ & $Q_\phi$ & $Q$ & $P$
                          & $Q_{000}$ & $Q_{090}$ & $Q_{180}$ \\
\noalign{\smallskip\hrule\smallskip}
\noalign{\smallskip \noindent PSF$_{\rm AO}$ convolved parameters \smallskip}
 & $\Sigma$ & 
     68.6 & $+40.9$ & 78.1 & $+2.2$ & $+22.4$ & $-6.1$  \\ 
\noalign{\smallskip}
 & $\Sigma_1$ &
      0.0  & $+0.2$ &  0.2  & $+0.1$ & 0.0 & 0.0 \\ 
 & $\Sigma_2$ &
     65.9 & $+31.9$ & 68.6  & $+0.5$ & $+19.5$ & $-7.7$ \\ 
 & $\Sigma_3$ &
      2.7 & $+8.9$  & 9.6   & $+1.7$ & $+2.8$  & $+1.6$ \\ 
      \noalign{\smallskip \noindent $\Sigma_{\rm int}=\Sigma_2$ parameters for
        different $\Sigma_{\rm zp}$ \smallskip}
$\Sigma$ & $\Sigma_2$ &      
    65.9 & $+26.0$ & 67.5 & $-0.9$ & $+18.1$ & $-8.2$ \\  
\noalign{\smallskip}
$\Sigma_1$ & $\Sigma_2$ &      
    65.9 & $+31.9$ & 68.6 & $+0.5$ & $+19.5$ & $-7.7$ \\
$\Sigma_2$ & $\Sigma_2$ &      
    65.9 & $0.0$ & 83.9 & $-7.2$ & $+11.6$ & $-16.0$ \\  
$\Sigma_{2+3}$ & $\Sigma_2$ &      
    65.9 & $+14.1$ & 74.2 & $-3.8$ & $+15.1$ & $-12.3$ \\  
$\Sigma_3$ & $\Sigma_2$ &      
    65.9 & $+25.0$ & 68.0 & $-1.1$ & $+17.8$ & $-9.5$ \\
\noalign{\smallskip\hrule}    
\end{tabular}
\tablefoot{The upper part gives results for the PSF$_{\rm AO}$ convolved
  model and the lower part the $\Sigma_2$ values after zp-correction based
  on different regions $\Sigma_{\rm zp}$ for the models shown in
  Fig.~\ref{Fig.ConvNormLarge}. All values are normalized
  relative to the intrinsic azimuthal polarization
  $\Sigma Q'_\phi$.}
\end{table}

Table~\ref{Tab.ConvNorm400} gives polarization parameters for the
very large RingI60 model with $r_0=403.2$~mas of
Fig.~\ref{Fig.ConvNormLarge} in Sect.~\ref{Sect.NormCoro}, and they
quantify the impact of different zp-correction for a disk with
$r_0$ similar to the radius of the PSF speckle ring of an AO-system.

The first line in Table~\ref{Tab.ConvNorm400} gives values
identical to the second last line of
Table~\ref{TabRingi60}, where also the $\Sigma U_{xxx}$ quadrant parameters
for this model are given. The
values for $\Sigma_{\rm int}=\Sigma_1$, $\Sigma_2$, and $\Sigma_3$ give
the radial distribution of the PSF$_{\rm AO}$-convolved polarization signal.
The integration regions are defined by the circular aperture
$r\leq 0.1''$ for $\Sigma_1$ representing the star or the coronagraphic mask,
and by the annular apertures $0.1''<r\leq 0.5''$ for $\Sigma_2$
and $0.5''<r\leq 1.5''$ for $\Sigma_3$. For the Stokes $U$ quadrants
there is $\Sigma_2 U_{045}/\Sigma Q'_\phi=-2.5$~\% and
$\Sigma_2 U_{135}/\Sigma Q'_\phi=+17.2$~\%, the corresponding $\Sigma_1$
values are zero and the $\Sigma_3$ values are the differences between the
$\Sigma$ and $\Sigma_2$ values. The distribution of the intensity
$\Sigma_i I/\Sigma I'_s$ in the PSF$_{\rm AO}$ convolved images
are 66.2~\%, 13.5~\% and 18.1~\% for the integration regions
$\Sigma_1$, $\Sigma_2$ and $\Sigma_3$, respectively. The intrinsic disk signal
is $\Sigma I'_d/\Sigma I'_s=1~\%$ and $\Sigma Q'_\phi/\Sigma I'_s=0.1~\%$.

The last five lines in Table~\ref{Tab.ConvNorm400} give the
polarization values for the disk aperture $\Sigma_i=\Sigma_2$
after zp-correction in different apertures $\Sigma_{\rm zp}$.
The $\Sigma_2 U_{xxx}$ values given above for the convolved
model are not changed by a Stokes $Q$ zp-correction.

The ideal case is the zp-correction using the unpolarized
star signal in $\Sigma_{\rm zp}=\Sigma_1$, which yields the same result
as the (initial) PSF$_{\rm AO}$ convolved model. For coronagraphic
observations a zp-correction including the $\Sigma_1$ region is not
possible. The $\Sigma_{\rm zp}=\Sigma_2$ correction sets the Stokes
$\Sigma_2 Q$ signal to zero and this introduces for this model
a very large polarization offset. Using the halo for the
zp-correction $\Sigma_{\rm zp}=\Sigma_3$ is practically equivalent to the
case $\Sigma_{\rm zp}=\Sigma$ for a total system correction, which
is a good approach to account for undefined instrumental polarization
effects, which may be re-calibrated with aperture polarimetry. 
A $\Sigma$ or system zp-correction means also
$\Sigma Q^{\rm z} = 0 \approx \Sigma_3 Q^{\rm z}$ and
$\Sigma_2 Q^{\rm z}\approx - \Sigma_1 Q^{\rm z}$ (Table~\ref{Tab.ConvNorm}).
Measuring the halo polarization signal can be problematic, if
the halo signal is noisy. A zp-correction based on a region including
disk and halo regions $\Sigma_{\rm zp}=\Sigma_{2+3}$ is also problematic because
this depends on the PSF-convolution.

\begin{table}
  \caption{Comparison of values for two compact DiskI60$\alpha$-2 models
     without and with two types of zp-correction.}
  \label{Tab.CentNorm}
  \begin{tabular}{l cc ccccc}
\noalign{\hrule\smallskip}
$\Sigma_{\rm int}$ &
               \multispan{2}{\hfil $\Sigma_{i'} X/Q'_{\phi,{\rm ref}}$ [\%] \hfil} &
               \multispan{5}{\hfil $\Sigma_{i'} X_{xxx}/Q'_{\phi,{\rm ref}}$ [\%] \hfil}  \\
$\Sigma_{i'}$ & $Q_\phi$ & $Q$       & $Q_{000}\hspace{-0.3mm}$ & $Q_{090}$
                                          & $\hspace{-0.3mm}Q_{180}$ 
                              & $U_{045}\hspace{-0.3mm}$ & $U_{135}$ \\
\noalign{\smallskip\hrule\smallskip}

\noalign{\smallskip \noindent cavity size $r_{\rm in}=0.125\,D_{\rm PSF}$ \smallskip}
\noalign{\smallskip \noindent intrinsic disk \smallskip}
$\Sigma$  & 167. & 73.7      & $-1.2\hspace{-3mm}$ & $+47.0$
                                            & $\hspace{-3mm}-22.3$  
                              & $-6.8$ & $\hspace{-3mm}+40.5$ \\ 
\noalign{\smallskip \noindent convolved disk \smallskip}
$\Sigma$   & 33.7 & 72.5      & $+12.6\hspace{-3mm}$ & $+25.0$
                                            & $\hspace{-3mm}+10.9$
                              & $+1.0\hspace{-3mm}$ & $+9.6$ \\ 
$\Sigma_{1+2}\hspace{-3mm}$ & 33.6 & 51.6 & 
                              $+7.4\hspace{-3mm}$ & $+19.7$
                                            & $\hspace{-3mm}+5.7$ 
                              & $+0.9\hspace{-3mm}$& $+9.5$ \\ 
$\Sigma_3$  &  0.1 & 20.9      & $+5.2\hspace{-3mm}$ &  $+5.3$
                                            & $\hspace{-3mm}+5.2$ 
                              & $+0.1\hspace{-3mm}$ & $+0.1$ \\ 
\noalign{\smallskip \noindent $\Sigma_{\rm zp}=\Sigma$: system corrected disk \smallskip}
$\Sigma$  & 33.6 & 0.0        &  $-4.6\hspace{-3mm}$ & $+6.0$
                                            & $\hspace{-3mm}-8.1$
                              & $+1.0\hspace{-3mm}$ & $+9.6$ \\ 
$\Sigma_{1+2}\hspace{-3mm}$ & 33.5 & 0.0     & $-4.6\hspace{-3mm}$ & $+6.0$
                                            & $\hspace{-3mm}-8.1$ 
                              & $+0.9\hspace{-3mm}$ & $+9.5$ \\ 
$\Sigma_3$  & 0.1    & 0.0    &  $0.0\hspace{-3mm}$ & $0.0$
                                          & $\hspace{-3mm} 0.0$
                              & $+0.1\hspace{-3mm}$ & $+0.1$ \\ 
\noalign{\smallskip \noindent $\Sigma_{\rm zp}=\Sigma_0$: center corrected disk \smallskip}
$\Sigma$  & 33.6 & 42.7 
                              & $+5.6\hspace{-3mm}$ & $+17.2$
                                             & $\hspace{-3mm}+3.1$ 
                              & $+1.3\hspace{-3mm}$ & $+9.9$ \\ 
$\Sigma_{1+2}\hspace{-3mm}$ & 33.5 & 30.4   & $+2.5\hspace{-3mm}$ & $+14.1$
                                            & $\hspace{-3mm} 0.0$ 
                              & $+1.1\hspace{-3mm}$ & $+9.7$ \\ 
$\Sigma_3$  &  0.1   &  12.3   & $+3.1\hspace{-3mm} $& $+3.1$
                                            & $\hspace{-3mm} +3.1$ 
                              & $+0.2\hspace{-3mm}$ & $+0.2$ \\ 

\noalign{\smallskip \hrule \smallskip}

\noalign{\smallskip \noindent cavity size $r_{\rm in}=0.5\,D_{\rm PSF}$ \smallskip}
\noalign{\smallskip \noindent intrinsic disk \smallskip}
$\Sigma$ & $Q'_{\phi,{\rm ref}}$
              & 42.1        & $-0.7\hspace{-3mm}$ & $+28.1$
                                             & $\hspace{-3mm}-13.4$ 
                              & $-4.1\hspace{-3mm}$ & $+24.3$ \\
\noalign{\smallskip \noindent convolved disk \smallskip}
$\Sigma$  & 31.5 & 41.4       & $+5.6\hspace{-3mm}$ & $+16.4$
                                             & $\hspace{-3mm}+3.2$ 
                              & $+0.4\hspace{-3mm}$ & $+8.5$ \\
$\Sigma_{1+2}\hspace{-3mm}$ & 31.4 & 29.4     
                              & $+2.6\hspace{-3mm}$ & $+13.4$
                                             & $\hspace{-3mm}+0.2$ 
                              & $+0.3\hspace{-3mm}$ & $+8.4$ \\
$\Sigma_3$  &  0.1   & 12.0    & $+3.0\hspace{-3mm}$ & $+3.0$
                                             & $\hspace{-3mm}+3.0$ 
                              & $+0.1\hspace{-3mm}$ & $+0.1$  \\

\noalign{\smallskip \noindent $\Sigma_{\rm zp}=\Sigma$: system corrected disk \smallskip}
$\Sigma$   & 31.4  & 0.0        &  $-4.2\hspace{-3mm}$ & $+5.6$
                                             & $\hspace{-3mm}-7.6$ 
                                & $+0.4\hspace{-3mm}$ & $+8.5$  \\

$\Sigma_{1+2}\hspace{-3mm}$ & 31.5 & 0.0    & $-4.2\hspace{-3mm}$ & $+5.6$
                                             & $\hspace{-3mm}-7.6$ 
                                & $+0.3\hspace{-3mm}$ & $+8.4$  \\

$\Sigma_3$  & -0.1  & $+0.1$    & $0.0\hspace{-3mm}$ & $0.0$
                                             & $\hspace{-3mm}0.0$ 
                                & $+0.1\hspace{-3mm}$ & $+0.1$  \\

\noalign{\smallskip \noindent $\Sigma_{\rm zp}=\Sigma_0$: center corrected disk \smallskip}
$\Sigma$  &  31.5   & 37.5      &  $+4.7\hspace{-3mm}$ & $+15.4$
                                             & $\hspace{-3mm} +2.2$ 
                                & $+0.5\hspace{-3mm}$ & $+8.7$  \\
$\Sigma_{1+2}\hspace{-3mm}$ & 31.4 & 26.6   & $+2.0\hspace{-3mm}$ & $+12.7$
                                             & $\hspace{-3mm}-0.5$ 
                                & $+0.4\hspace{-3mm}$ & $+8.5$  \\
$\Sigma_3$  &  0.1  &  10.9 & $+2.7\hspace{-3mm}$ & $+2.7$
                                             & $\hspace{-3mm} +2.7$ 
                                & $+0.1\hspace{-3mm}$ & $+0.2$  \\

\noalign{\smallskip\hrule}
\end{tabular}  
  \tablefoot{Result for models with
    $r_{\rm in}=0.125\, D_{\rm PSF}$ (upper part) and
    $r_{\rm in}=0.5\, D_{\rm PSF}$ (lower part) are given using
    different intergration regions $\Sigma_{\rm int}$. Values for the
    intrinsic and PSF$_{\rm AO}$ convolved models are given, and for models
    zp-corrected for the whole system and the center. 
    All values are normalized to $Q'_{\phi,{\rm ref}}=\Sigma Q'_\phi=100~\%$
    for the model with $r_{\rm in}=0.5\,D_{\rm PSF}$ (12.6~mas). The models for
    $r_{\rm in}=0.125\,D_{\rm PSF}$ (3.15 mas) are plotted in
    Fig.~\ref{Fig.NormDisk}.} 
\end{table}

For all $\Sigma_{\rm zp}$ regions practically the same azimuthal polarization
$\Sigma_2 Q_\phi$ is obtained, emphasizing the important invariance of
$\Sigma Q_\phi$ with respect to polarization offset. Of course, this
property is based on the use of an axisymmetric PSF and 
axisymmetric integration regions in the simulations. Contrary to this, the
integrated polarization $\Sigma_2 P$ changes by the use of different
integration region by up to about 20~\%, and this does not
include the problem of the noise bias, which might be very significant
for observations of extended, low SB scattering regions.

\subsection{Center corrected DiskI60$\alpha$-2 models}

The Table~\ref{Tab.CentNorm} gives numerical values
for the PSF$_{\rm AO}$ convolved DiskI60$\alpha$-2 models, and for
center corrected  $\Sigma_{\rm zp}=\Sigma_0$
and system corrected $\Sigma_{\rm zp}=\Sigma$ models. For the
size of the inner disk cavity the two cases
$r_{\rm in}=0.125~D_{\rm PSF}$ (3.15~mas) and $r_{\rm in}=0.5~D_{\rm PSF}$
(12.6~mas) were selected. The latter case is the reference
disk with $\Sigma Q'_\phi = Q_{\rm ref}=100~\%$ for the intrinsic
azimuthal polarization. For the model with the small cavity the
intrinsic polarization $\Sigma Q'_\phi=167~\%$ is much larger when
compared to $Q_{\rm ref}$, but most of the additional azimuthal
polarization is not resolved (see Table~\ref{TabDiski60}).
Also the $\Sigma Q$ signal is high for the $r_{\rm in}=0.125~D_{\rm PSF}$
model and Fig.~\ref{Fig.NormDisk} shows the PSF$_{\rm AO}$ convolved
polarization maps $X(x,y)$ for $X=\{Q,U,Q_\phi,U_\phi\}$, and the
resulting maps after a system zp-correction $X^{\rm z}$ and
a center zp-correction $X^{\rm ex}$.
The integration regions $\Sigma_i$ are defined by
round apertures $r\leq 1.5''$ for the whole system $\Sigma$,
and $r\leq 0.2''$ for the star and disk region $\Sigma_{1+2}$,
and by an annular aperture
$0.2''<r\leq 1.5''$ for the halo region $\Sigma_3$.

The system zp-correction $\Sigma_{\rm zp}=\Sigma$ sets the $\Sigma Q$ signal
to zero and this reduces for both $r_{\rm in}$ cases strongly the
$\Sigma Q_{xxx}$ quadrant values. The center corrected values
for the disk with the small cavity $r_{\rm in}=0.125 D_{\rm PSF}$ are
very similar (within 15~\%) to the case of the reference model with a
cavity size of $r_{\rm in}=0.5 D_{\rm PSF}$.
Thus, the center zp-correction accounts practically for the Stokes
$Q$ polarization offset introduced by the
unresolved central region.
This correction offset is
$-\langle q_0\rangle\, \Sigma I/Q'_{\phi,{\rm ref}}=29.8~\%$ for the
model with $r_{\rm in}=0.125 D_{\rm PSF}$ and 3.9~\% for
$r_{\rm in}=0.5 D_{\rm PSF}$. The quadrant values
for $U_{045}$ and $U_{135}$ are practically unchanged for a zp-correction
of a Stokes $Q$ polarization.

\end{document}